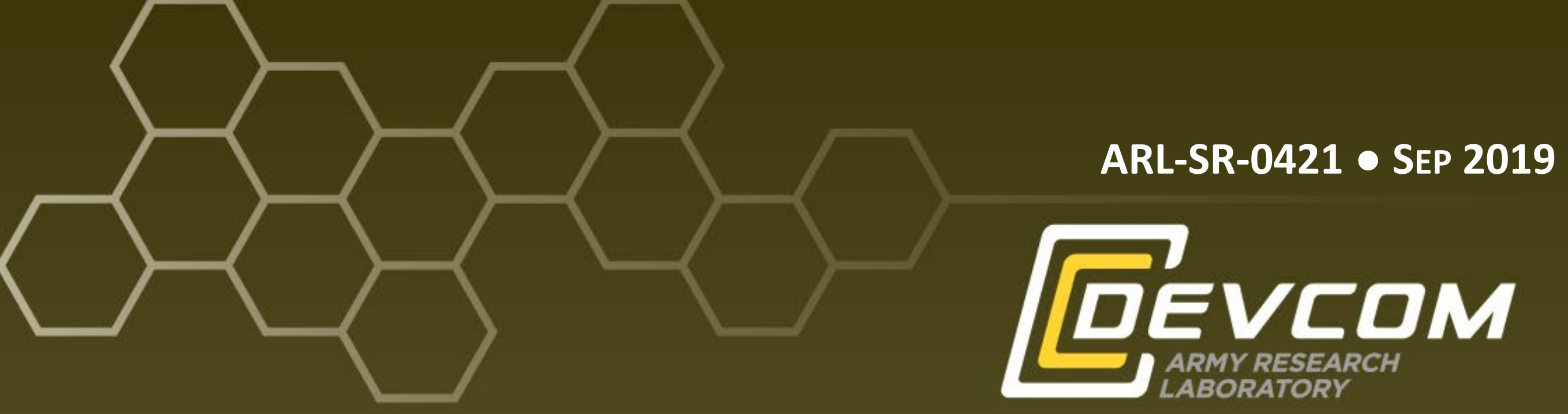

# Autonomous Intelligent Cyber-defense Agent (AICA) Reference Architecture Release 2.0

by Alexander Kott, Paul Théron, Martin Drašar, Edlira Dushku, Benoît LeBlanc, Paul Losiewicz, Alessandro Guarino, Luigi V Mancini, Agostino Panico, Mauno Pihelgas, Krzysztof Rzadca, and Fabio De Gaspari

## NOTICES

### Disclaimers

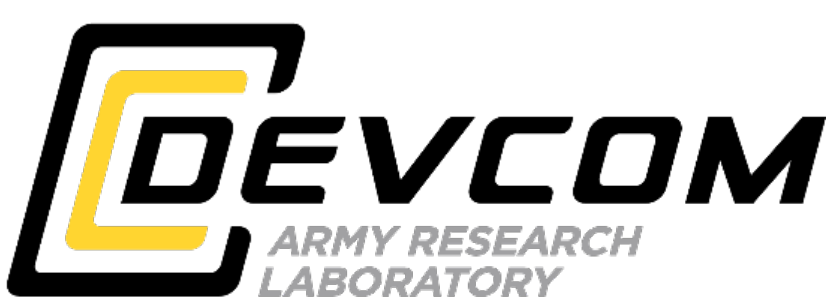

# Autonomous Intelligent Cyber-defense Agent (AICA) Reference Architecture Release 2.0

**by Alexander Kott,** ***Office of the Director, CCDC Army Research Laboratory***

**Paul Théron,** ***Thales, France***

**Luigi V Mancini, Edlira Dushku, Agostino Panico, and Fabio De Gaspari,** ***Sapienza Università di Roma, Italy***

**Martin Drašar,** ***Masaryk University, Brno, Czech Republic***

**Benoît LeBlanc,** ***Ecole Nationale Supérieure de Cognitique, Bordeaux, France***

**Paul Losiewicz,** ***Cybersecurity and Information Systems IAC, USA***

**Alessandro Guarino,** ***StAG Srl, Italy***

**Mauno Pihelgas,** ***NATO Cooperative Cyber-defense Centre of Excellence, Tallinn, Estonia***

**Krzysztof Rzadca,** ***Institute of Informatics, University of Warsaw, Warsaw, Poland***

| **REPORT DOCUMENTATION PAGE** | *Form Approved*<br>*OMB No. 0704-0188* |
|---|---|

| **1. REPORT DATE** *(DD-MM-YYYY)* | **2. REPORT TYPE** | **3. DATES COVERED (From - To)** |
|---|---|---|
| September 2019 | Special Report | 09 September 2016–31 October 2019 |

**4. TITLE AND SUBTITLE**

Autonomous Intelligent Cyber-defense Agent (AICA) Reference Architecture, Release 2.0

**5a. CONTRACT NUMBER**

**5b. GRANT NUMBER**

**5c. PROGRAM ELEMENT NUMBER**

**6. AUTHOR(S)**

Alexander Kott, Paul Théron, Martin Drašar, Edlira Dushku, Benoît LeBlanc, Paul Losiewicz, Alessandro Guarino, Luigi V Mancini, Agostino Panico, Mauno Pihelgas, Krzysztof Rzadca, and Fabio De Gaspari

**5d. PROJECT NUMBER**

**5e. TASK NUMBER**

**5f. WORK UNIT NUMBER**

**7. PERFORMING ORGANIZATION NAME(S) AND ADDRESS(ES)**

CCDC Army Research Laboratory
ATTN: FCDD-RLD
Adelphi, MD 20783-1138

**8. PERFORMING ORGANIZATION REPORT NUMBER**

ARL-SR-0421

**9. SPONSORING/MONITORING AGENCY NAME(S) AND ADDRESS(ES)**

**10. SPONSOR/MONITOR'S ACRONYM(S)**

**11. SPONSOR/MONITOR'S REPORT NUMBER(S)**

**12. DISTRIBUTION/AVAILABILITY STATEMENT**

Approved for public release; distribution is unlimited.

**13. SUPPLEMENTARY NOTES**

**14. ABSTRACT**

This report—a major revision of its previous release—describes a reference architecture for intelligent software agents performing active, largely autonomous cyber-defense actions on military networks of computing and communicating devices. The report is produced by the North Atlantic Treaty Organization (NATO) Research Task Group (RTG) IST-152 "Intelligent Autonomous Agents for Cyber Defense and Resilience". In a conflict with a technically sophisticated adversary, NATO military tactical networks will operate in a heavily contested battlefield. Enemy software cyber agents—malware—will infiltrate friendly networks and attack friendly command, control, communications, computers, intelligence, surveillance, and reconnaissance and computerized weapon systems. To fight them, NATO needs artificial cyber hunters—intelligent, autonomous, mobile agents specialized in active cyber defense. With this in mind, in 2016, NATO initiated RTG IST-152. Its objective has been to help accelerate the development and transition to practice of such software agents by producing a reference architecture and technical roadmap. This report presents the concept and architecture of an Autonomous Intelligent Cyber-defense Agent (AICA). We describe the rationale of the AICA concept, explain the methodology and purpose that drive the definition of the AICA Reference Architecture, and review some of the main features and challenges of AICAs.

**15. SUBJECT TERMS**

intelligent agent, autonomy, cyber warfare, cyber defense, agent architecture

| **16. SECURITY CLASSIFICATION OF:** | | | **17. LIMITATION OF ABSTRACT** | **18. NUMBER OF PAGES** | **19a. NAME OF RESPONSIBLE PERSON** |
|---|---|---|---|---|---|
| | | | | | Alexander Kott |
| **a. REPORT** | **b. ABSTRACT** | **c. THIS PAGE** | | | **19b. TELEPHONE NUMBER (Include area code)** |
| Unclassified | Unclassified | Unclassified | UU | 154 | (301) 394-1507 |

Standard Form 298 (Rev. 8/98)
Prescribed by ANSI Std. Z39.18

## Executive Summary

The North Atlantic Treaty Organization (NATO) Research Task Group IST-152 developed a concept and a reference architecture for intelligent software agents performing active, largely autonomous cyber-defense actions on military assets. In this report, which is an updated and extended version of its previous release, such an agent is referred to as an Autonomous Intelligent Cyber-defense Agent (AICA).

In a conflict with a technically sophisticated adversary, NATO military networks will operate in a heavily contested battlefield. Enemy malware will likely infiltrate and attack friendly networks and systems. Today's reliance on human cyber defenders will be untenable on the future battlefield. Instead, artificially intelligent agents such as AICAs will be necessary to defeat the enemy malware in an environment of potentially disrupted communications where human intervention may not be possible.

The IST-152 group identified specific capabilities of AICA. For example, AICA will have to be capable of autonomous planning and execution of complex multi-step activities for defeating or degrading sophisticated adversary malware, with anticipation and minimization of resulting side effects. It will have to be capable of adversarial reasoning to battle against a thinking, adaptive malware. Crucially, AICA will have to keep itself and its actions as undetectable as possible, and will have to use deceptions and camouflage.

The group identified the key functions, components, and their interactions for a potential reference architecture of such an agent, as well as a tentative roadmap toward the capabilities of AICA.

NATO should encourage the emerging interest in member nations' academia, industry, and governments toward the related research and development. AICAs are likely to become primary cyber fighters on the future battlefield, and NATO must not fall behind its adversaries in developing and deploying such technologies.

# 1. Introduction

Authors: Alexander Kott and Guido Gluschke

This report describes a reference architecture for intelligent software agents performing active, largely autonomous cyber-defense actions on military networks of computing and communicating devices. The report is produced by the North Atlantic Treaty Organization (NATO) Research Task Group (RTG) IST-152 "Intelligent Autonomous Agents for Cyber Defense and Resilience".

## 1.1 Objective

In a conflict with a technically sophisticated adversary, NATO military tactical networks will operate in a heavily contested battlefield. Enemy software cyber agents—malware—will likely infiltrate friendly networks and attack friendly command, control, communications, computers, intelligence, surveillance, and reconnaissance (C4ISR) and computerized weapon systems. To fight them, NATO needs an effective response. Artificial cyber defenders—intelligent, autonomous, mobile agents specialized in active cyber defense—are one form of adequate and effective response. The key roles of these agents will be to detect and defeat the enemy malware that infiltrated friendly systems and networks.

With this in mind, in 2016, NATO initiated RTG IST-152 "Intelligent Autonomous Agents for Cyber Defense and Resilience". Its objective is to help accelerate development and transition to practice of such software agents by producing a reference architecture and technical roadmap.

If such research is successful, it will lead to technologies that enable the following vision. Cyber-defense agents will stealthily monitor the networks, detect the enemy cyber activities while remaining concealed, and then destroy or degrade the enemy malware. They will do so mostly autonomously, because human cyber experts will be always scarce on the battlefield. They have to be capable of autonomous learning because enemy malware is constantly evolving. They have to be stealthy because the enemy malware will try to find and destroy them. At the time of this writing and to the best of our knowledge, autonomous agents with such capabilities remain unavailable. The IST-152 group performed focused technical analysis to produce a first-ever reference architecture and technical roadmap for autonomous cyber-defense agents. In addition, the RTG worked to identify selected elements of such capabilities that are beginning to appear in academic and industrial research.

The output of the RTG is a tangible starting point for acquisition activities by NATO nations. If based on a common reference architecture, software agents developed or purchased by different nations will be far more likely to be interoperable. Deployed on NATO networks, the autonomous cyber-defense agents will become a significant force multiplier: the agents will operate autonomously when it is necessary to augment the inevitably limited capabilities of human cyber defenders, and will work under the control of humans when ordered to do so and when conditions permit such a control.

With the help of autonomous, intelligent cyber-defense agents, NATO C4ISR will be more likely to survive an encounter with a determined, technically sophisticated enemy. To acquire and successfully deploy such agents, in an interoperable manner, NATO nations must have a common technical vision, including a reference architecture and a roadmap, of which this report is a beginning.

## 1.2 Fundamental Choices and Assumptions

A key assumption taken by this report is that in a conflict with a technically sophisticated adversary, NATO military tactical networks will operate in a heavily contested battlefield. Enemy software cyber agents—malware—will infiltrate NATO networks and attack NATO C4ISR and computerized weapon systems, with a significant probability that cannot be ignored.

To focus the attention of our research group, we have chosen to limit the scope of the problem as follows. We consider a single military platform, such as a vehicle, a vessel, or an unmanned aerial vehicle (UAV) with one or more computers residing on the platform, connected to sensors and actuators. Each computer contributes considerably to the operation of the platform or systems installed on the platform. One or more computers are assumed to have been compromised, where the compromise is either established as a fact or is suspected.

Due to the contested nature of the communications environment (e.g., the enemy is jamming the communications or radio silence is required to avoid detection by the enemy), communications between the vehicle and other elements of the friendly force are often limited and intermittent. At certain times and under some conditions, communications may be entirely impossible.

Given the constraints on communications, conventional centralized cyber defense (i.e., an architecture where local sensors send cyber-relevant information to a central location where highly capable cyber-defense systems and human analysts detect the presence of malware and initiate corrective actions remotely) is often infeasible. It is also unrealistic to expect that the human warfighters residing on the platform, for example, a vehicle, will have the necessary skills or time available to perform cyber-defense functions locally on the vehicle, even more so if the vehicle is unmanned.

Therefore, cyber defense of such a platform, including its computing devices, will have to be performed by an intelligent, autonomous software agent. The agent (or multiple agents per platform) will stealthily monitor the networks, detect the enemy agents while remaining concealed, and then destroy or degrade the enemy malware. The agent will have to do so mostly autonomously, without support or guidance by a human expert.

In most discussions in this report, the agent is considered as a monolithic piece of software. However, depending on the implementation, the agent's modules can be distributed over multiple processes or devices, or it could be implemented as a team of agents or subagents.

To fight the enemy malware that has infiltrated the friendly computer, the agent may have to take destructive actions, such as deleting or quarantining certain software. Such destructive actions are carefully controlled by the appropriate rules of engagement and are allowed only on the computer where the agent resides.

In most cases, the agent will not be able to stop the enemy from penetrating the platform's systems. However, it will be able to perform detection of, analysis, and response to a given threat. The actions of the agent, in general, cannot be guaranteed to preserve the availability or integrity of the functions and data of friendly computers. There is a risk that an action of the agent will "break" the friendly computer, disable important friendly software, or corrupt or delete important data. Developers of the agent will attempt to design its actions and planning capability to minimize the risk. This risk, in a military environment, has to be balanced against the death or destruction caused by the enemy if the agent's action is not taken.

Provisions will be made to enable a remote controller—human or automated cyber command and control (C2) node—to fully observe, direct, and modify the actions of the agent, and even to update the agent's software as needed. However, it is recognized that such a remote control is often impossible due to the difficulties of communicating between the agent and the control node. The agent, therefore, should be able to plan, analyze, and perform most or all of its actions autonomously.

Similarly, provisions should be made for the agent to collaborate with other agents (that reside on other computers); however, in many cases, because the communications are impaired or observed by the enemy, the agent has to eschew collaboration and operate alone.

The enemy malware, specifically, its capabilities and tactics, techniques, and procedures (TTPs), evolves rapidly. Therefore, the agent will be capable of autonomous learning. In case the enemy malware knows that the agent exists and is likely to be present on the computer, the enemy malware will seek to find and destroy the agent. Therefore, the agent will possess techniques and mechanisms for maintaining a certain degree of stealth, camouflage, and concealment. More generally, the agent takes measures that reduce the probability that the enemy malware will detect the agent. The agent is mindful of the need to exercise self-preservation and self-defense.

It is assumed here that the agent resides on a computer where it was originally installed by a human controller or an authorized process. We do envision a possibility that an agent may move itself (or move a replica of itself) to another computer. However, such propagation is assumed to occur only under exceptional and well-specified conditions, and takes place only within a friendly network—from one friendly computer to another friendly computer.

Here is a good place to mention the controversy about "good viruses". Such viruses have been proposed, criticized, and rejected earlier (Muttik 2008). These criticisms do not apply here. This agent is not a virus because it does not propagate except under explicit conditions within authorized and cooperative nodes. It is also used only in military environments, where the concerns listed in Muttik (2016) do not apply. As mentioned earlier, in a military environment, any drawbacks that might be associated with operations of an autonomous cyber-defense agent have to be balanced against the death or destruction caused by the enemy if the agent is not available.

Discussions of autonomous cyber-defense capabilities bring to mind the Defense Advanced Research Projects Agency (DARPA) Cyber Grand Challenge and the products that showed effective performance at that competition (e.g., Avgerinos et al. [2018]). However, unlike those products, the Autonomous Intelligent Cyber-defense Agent's (AICA's) purpose is not to find and fix vulnerabilities in friendly software, but rather to find and defeat the adversary's malware.

The field of collaborative intrusion detection (Zhou et al. 2010) is another topic that appears to be related to AICA. However, collaborative intrusion detection, while a possible useful capability for AICA, is not its central purpose. It is possible that extensive collaboration would not be possible for AICAs due to the need to maintain stealth.

Other related areas of research include software agents, multiagent systems, autonomous software, and host-based intrusion detection systems. Each of these areas is associated with voluminous literature. To our knowledge, an agent with AICA's purposes, capabilities, and architecture has not been discussed in the literature.

## 1.3 Basic Concepts and Terminology

In this report, the term "agent" denotes software or a collection of software that resides and operates on one or more computing devices, perceives its environment, and executes purposeful actions on the environment (and on itself) to achieve the agent's goals. We use the following acronyms: the agent is AICA and the architecture is the AICA Reference Architecture (AICARA).

The term "environment" here denotes everything that surrounds the agent and that the agent can perceive: the computer hardware and software where the agent operates, the vehicle, the enemy malware, the humans who communicate with the agent or with surrounding hardware and software, and other agents that this agent can find and with whom it can communicate.

The term "percept" denotes an element of information that the agent is able to obtain or receive; the percept reflects an attribute of the environment or a change in an attribute of the environment. The following are examples of percepts, partly inspired by De Gaspari et al. (2016):

- Report from Nmap probing
- Observation of a change to the file system
- A signal that someone has interacted with a fake webpage (honeypot page) or fake service

The term "action" denotes any action that a software agent can execute on its environment. It can include an impact on other software or data, or a communication to a human or another agent. The following are examples of actions (De Gaspari et al. 2016):

- Remap ports
- Check the integrity of the file system
- Create and deploy a fake password file, with an alarm mechanism activated when the file is accessed
- Create and deploy a fake webpage or web service
- Deposit a file with a "poison pill"
- Identify a suspicious file
- Sandbox a suspicious file
- Analyze the behavior of software in the sandbox

Examples of actions and situations in which the agent takes such actions are described in the Appendices.

The term "state" refers to a collection of values of the environment's attributes. Generally, the state is not known either fully or accurately, and the agent must infer it, at least in part.

The term "plan" here refers to a sequence or a directed graph of actions that the agent generates in order to transform the current state of the environment into a different state more desirable by the agent. The plan can be conditional (i.e., it includes intermediate decisions based on the perceived state) or temporal (i.e., it includes constraints on when the actions are performed).

## 1.4 Scope and Selected Requirements for AICA

For the purposes of describing its reference architecture, we assume that AICA resides on a physical military platform with the scope of ensuring availability and integrity of all relevant computerized functions of the platform against injected malicious code in order to ensure the correct behavior of the platform. Detecting abnormal functional behavior of the physical platform is not within the scope of the cyber-defense agent. This is assumed to be done by other operational monitor and control functions, manually or autonomously.

Taking a UAV as an example of a platform, the scope of AICA can be illustrated as in Fig. 1.

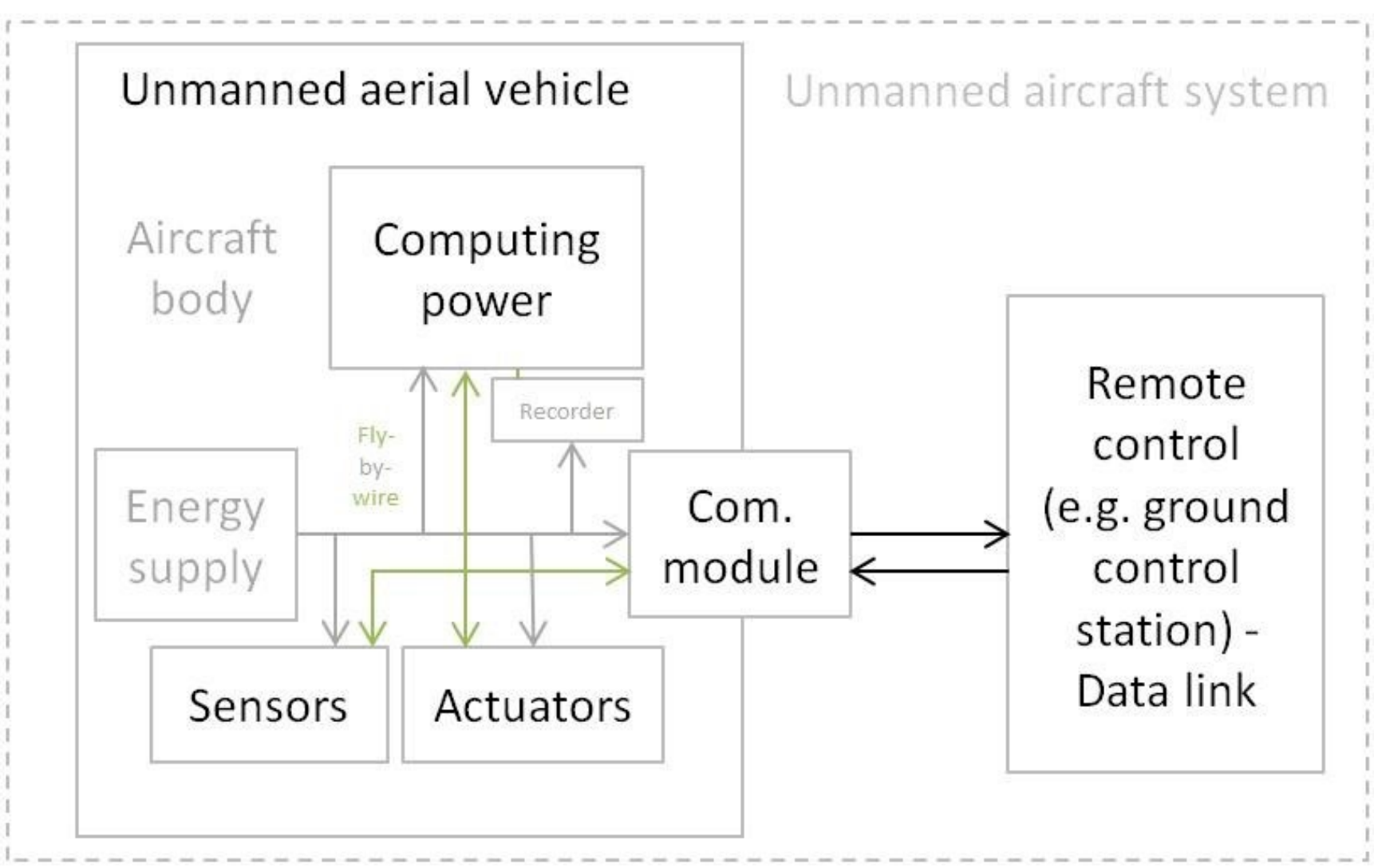

**Fig. 1 A possible scope of AICA implementation is illustrated in the context of a hypothetical UAV**

In the figure, computing power means the primary computers (one or multiple) that support the functions of the UAV. Actuators are physical devices for controlling the physical elements of the UAV. Here these devices are assumed to include computer processing, can be targets of cyberattacks, and therefore, should be protected by AICAs. The same argument applies to sensors and communication components. Therefore, in this example, the elements highlighted in the figure fall in the scope of AICA's responsibilities.

The following are some of the key requirements that can be seen as prerequisites for development of AICA's architecture:

- The agent shall reside on a military platform in a persistent and stealthy manner. Here, stealth refers to the agent's ability to minimize the probability that the adversary malware will detect and observe the agent's presence and activity.
- The agent shall be able to observe the state and activities within the elements within its scope of responsibilities, detect the enemy malware while remaining minimally observable to the malware, and destroy or degrade the enemy malware.
- The agent shall be capable of operating effectively in an environment compromised by an adversary malware.
- The agent shall be resistant to compromise.
- The agent shall be able to observe and understand the environment in which it is operating and for that it needs its own world model of the relevant environment.
- The agent shall be able to observe and influence all computational elements under its protection, including computational elements of all sensors and actuators of the platform.
- All relevant communications traffic shall be observable for the agent.
- The agent shall be able to function effectively when communications to other friendly elements or external controller are limited or unavailable.
- The agent shall function under specific circumstances, such as limited computing resources (memory, CPU, etc.) and special environmental conditions (e.g., temperature, air pressure, G-forces, size, and so on.)
- The agent shall function autonomously when necessary, that is, without depending on support of external friendly elements or an external controller. This implies that it has to be enabled to interact with all computational components of the platform, including the computational elements of sensors and actuators in real time; make its own decisions; and take the necessary actions.

- Provisions shall be made to enable a remote or local human controller to observe, direct, and modify the actions of the agent, when a need arises and circumstances permit.
- The agent shall be able to make nontrivial (and nonobvious to the adversary) plans in order to pursue a given goal and has to be able to execute defined actions resulted from the plan.
- The agent shall be able to take destructive actions, such as deleting or quarantining certain software and data, autonomously, while observing the specified rules of engagement. The agent shall have the means to assess the risk and benefits involved in such actions, and make its decisions accordingly.
- The agent should be able to collaborate with other friendly agents when a need arises and conditions permit. Collaboration schemes and negotiation mechanisms are needed for that.
- The agent should be able to perform autonomous learning, particularly regarding the capabilities, techniques, and procedures of the enemy malware. The learning should occur both offline and online, and the newly learned knowledge should be able to inform the agent during its operation.
- The agent, whenever requested, shall report data to the external controller that would enable the controller to make inferences about the trustworthiness of the agent.
- The agent should be able to self-propagate to a remote, friendly computing device. Self-propagation shall occur only under exceptional and well-specified conditions of military necessity.

The remainder of this report describes a proposed architecture that would meet such requirements.

Part A of the report provides the rationale and concept of operations of AICA, gives an overview of its architecture, and explains how the necessary data are stored and managed within the agent.

Part B offers a collection of exploratory discussions of possible approaches to implementing the key functions of the architecture. In this part, Section 5 describes how the agent acquires the information about its environment and determines the state of the environment. Section 6 discusses the means by which the agent plans its actions, including the prediction of actions' ramifications. Section 7 is about the ways in which the agent executes the actions it decided upon. Section 8 explains how the agent may collaborate with other agents. Section 9 outlines possible approaches to means by which the agent learns from its actions and observations.

# Part A. Presentation of the Architecture

# 2. Rationale of AICA and Scenario

Authors: Paul Losiewicz, Mauno Pihelgas, and Martin Drašar

## 2.1 Context

The threat of cyberattack on NATO-member military platforms cannot be underestimated. As described in a US General Accounting Office Report, GAO-19-128, *Weapon Systems Cybersecurity: DOD Just Beginning to Grapple with Scale of Vulnerabilities*, "Nearly all major acquisition programs that were operationally tested between 2012 and 2017 had mission-critical cyber vulnerabilities that adversaries could compromise" (GAO 2018).

In the wider cybersecurity domain, tactical targets for cyberattacks are termed "cyber–physical systems" (CPSs). The gateways for attacks are often the control systems (CSs) that manage the guidance and propulsion of a vehicle, the C2 of a system, or the operations of a system payload, such as weapon systems or intelligence, surveillance, and reconnaissance (ISR) sensor packages. CSs have been segregated into either facilities-related control systems (FRCSs) or platform control systems (often termed platform information technology [PIT]).

In this section, we describe an AICA employment strategy to defend against cyberattack in a notional platform using realistic threats in the military domain. It is understood that the modality of the vehicle (ground, aerial, surface, subsurface) will have different operational impacts in different contexts. In addition the targeted attacks can occur on the control systems of *manned, optionally manned,* and *unmanned* vehicles. We attempt to generalize as much as possible.

The following notional systems components (represented by boxes) are used in the vehicle (Fig. 2):

- **Bus (BUS)**: This box describes a component that is able to interconnect different devices, not relying on a specific technology. It includes all internal communication systems including intercoms that enable crew members who are physically separated to communicate within the vehicle, if the vehicle is crewed. There may be more than one, and busses can be repeated as needed.
- **Payload (PLD):** This is a symbolic box for a weapon, electronic warfare, ISR sensor package, or simply a cargo CS on the vehicle. There may be more than one, and the designation can be used as needed. The payload may include manned crew stations.

- **Communication system (COMMS)**: This box describes the communication systems between the vehicle and the external world (satellite communication, radio communication, etc.). There may be more than one, and the designation can be used as needed.

- **Vehicle Navigation System (VNS):** This box describes the internal position, navigation, and timing (PNT) system of the vehicle. The VNS receives input from PNT sensors on the vehicle or from offboard. The VNS provides input to the vehicle CS, either a pilot in case of a manned vehicle or the autopilot in case of an unmanned system.

- **Sensors (SENS):** This box represents the systems that can be used to provide input from the environment. The sensing function may be part of the VNS, vehicle CS, or a payload.

- **Vehicle Management System (VMS):** This box indicates the platform internal CS used to pilot the vehicle. It includes either a pilot in case of a manned vehicle or an autopilot and a contingency management system in case of an unmanned vehicle.

- **Battle Management System (BMS):** This box represents the system that is used by the operators, either onboard or offboard, to gather and send information about their tasking, platform status, and situation awareness of friends or foes. It is primarily used to report systems status and update mission tasking. It usually includes a bidirectional geo-information system that relies on information from the vehicle CS and navigation sensors, and payload sensors. It gets information from either a centralized battle manager or forms part of a distributed, noncentralized BMS.

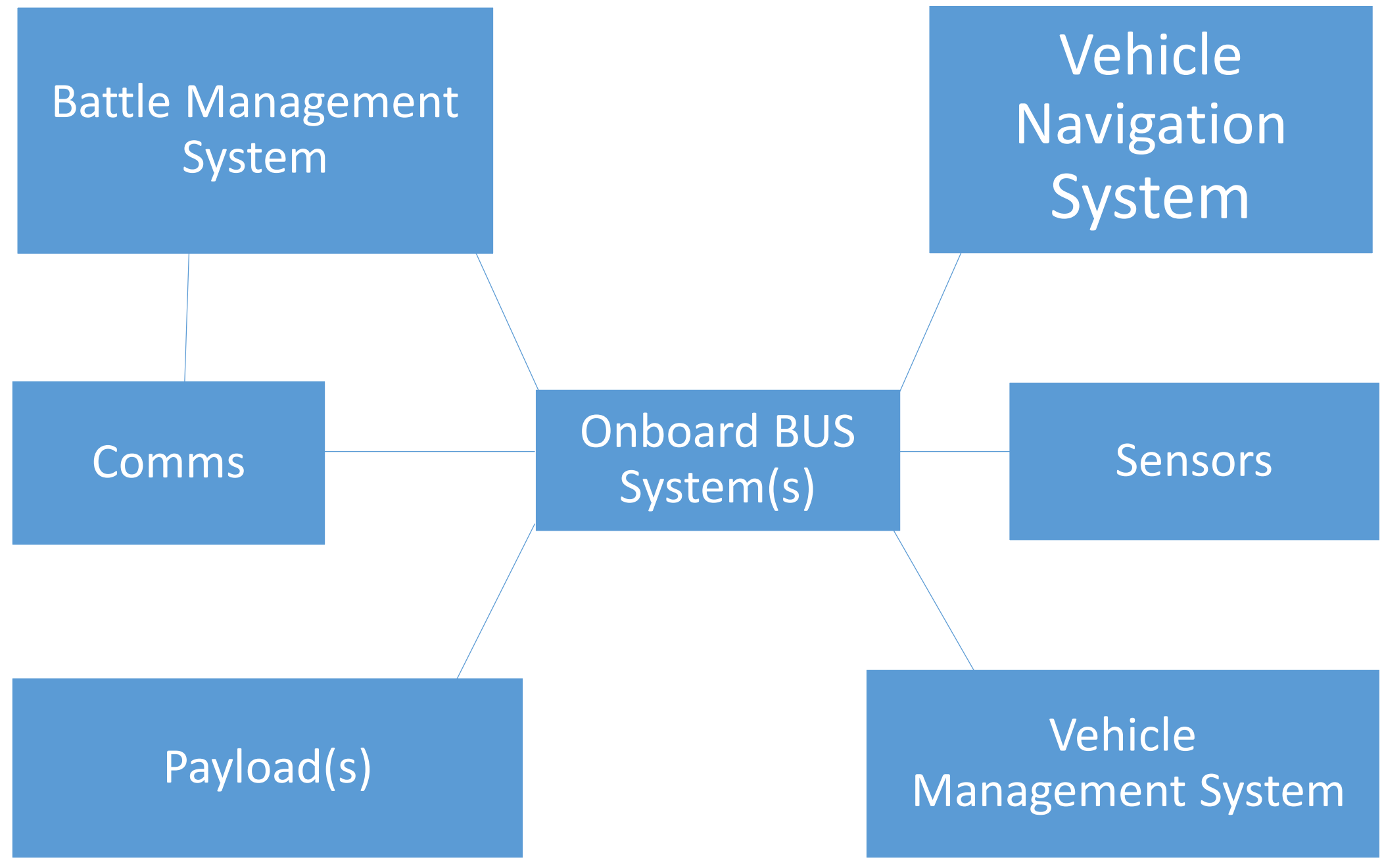

**Fig. 2 Vehicle systems and network structure**

## 2.2 Agent Deployment

Agents can be deployed in a centralized approach with master and client agents or as a distributed network of self-organizing agents (dotted lines indicate optional C2 configurations). The AICA we envision monitors internal component operations or communications bus traffic, which could include sensor readings, control signals, or TCP-IP packet data. For the autonomous CPS that may be operated remotely or operate with high degrees of autonomy, the data will consist of machine-to-machine (M2M) data passed over C2 or platform CS networks, using protocols and busses such as MIL-STD-1553, BACnet, MODBUS, ZigBee, IEBus, and ANSI/ISA-95. CSs are rapidly migrating to use of TCP-IP as well.

In a centralized approach, the evaluation of data and subsequent decision making is delegated to a master agent. The master agent controls the client agents and commands them to perform actions. The client agents, which have been installed on subsystem hardware can be very simple (e.g., scripts that send data and execute commands) or full replicas of the designated master agent that can be activated as needed.

In Fig. 3, the dotted lines indicate optional AICA communication schemes for either onboard or offboard C2.

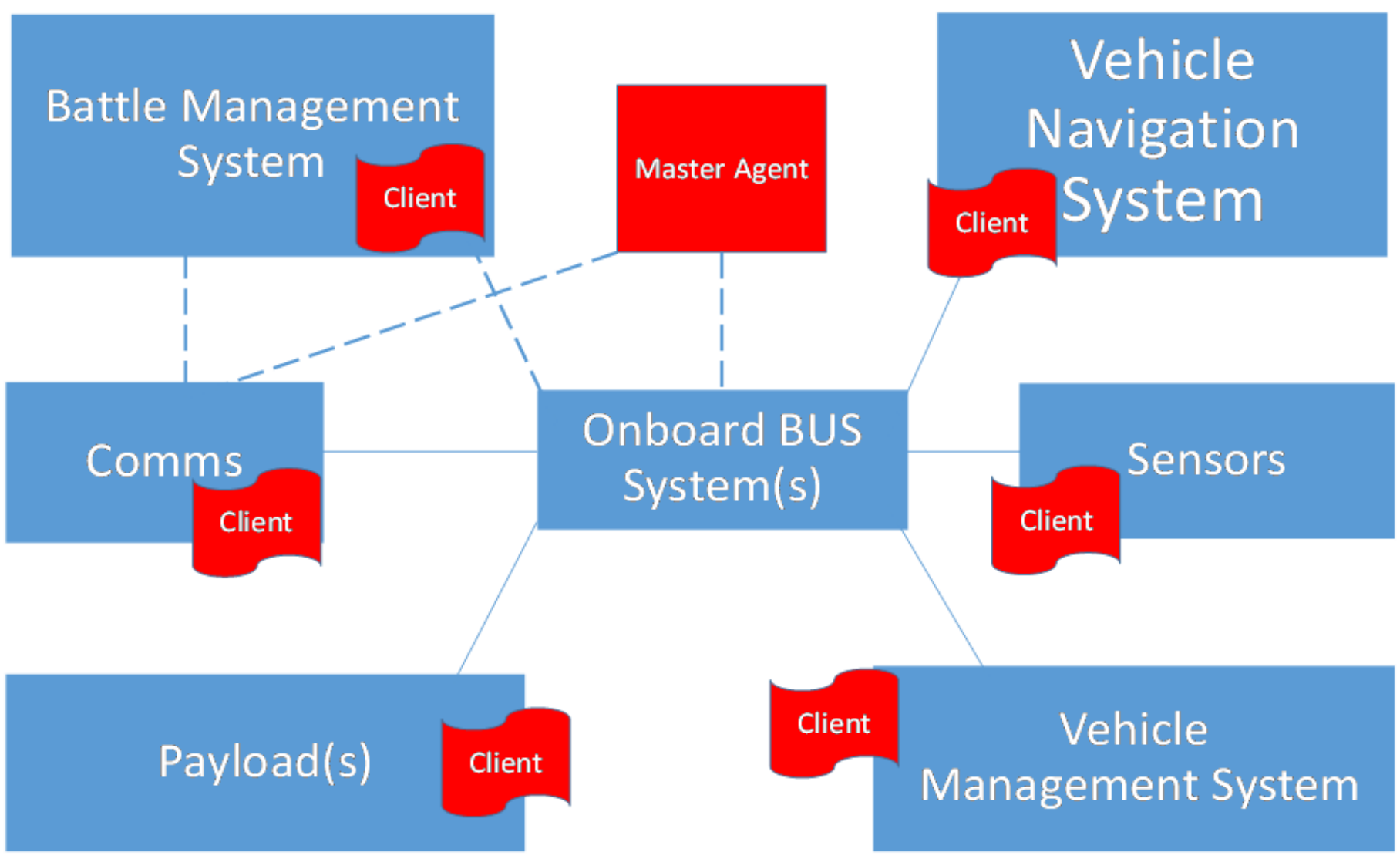

**Fig. 3 Centralized agent system**

A distributed network of self-organizing agents does not employ a centralized master agent, but uses more of a peer-to-peer (P2P) structure (Fig. 4). In the extreme case, the autonomous agents have to independently negotiate and coordinate tasking, attempt to maintain a common situational picture, and decide together about collaborative goals. This structure eliminates the master agent as a single point of failure and dramatically increases system resilience. Thus, even isolated or partitioned agents can continue to protect some portion of the entire system. This more resilient structure does come at the cost of more complexity in the maintenance of communications and the coordination of action and deliberation.

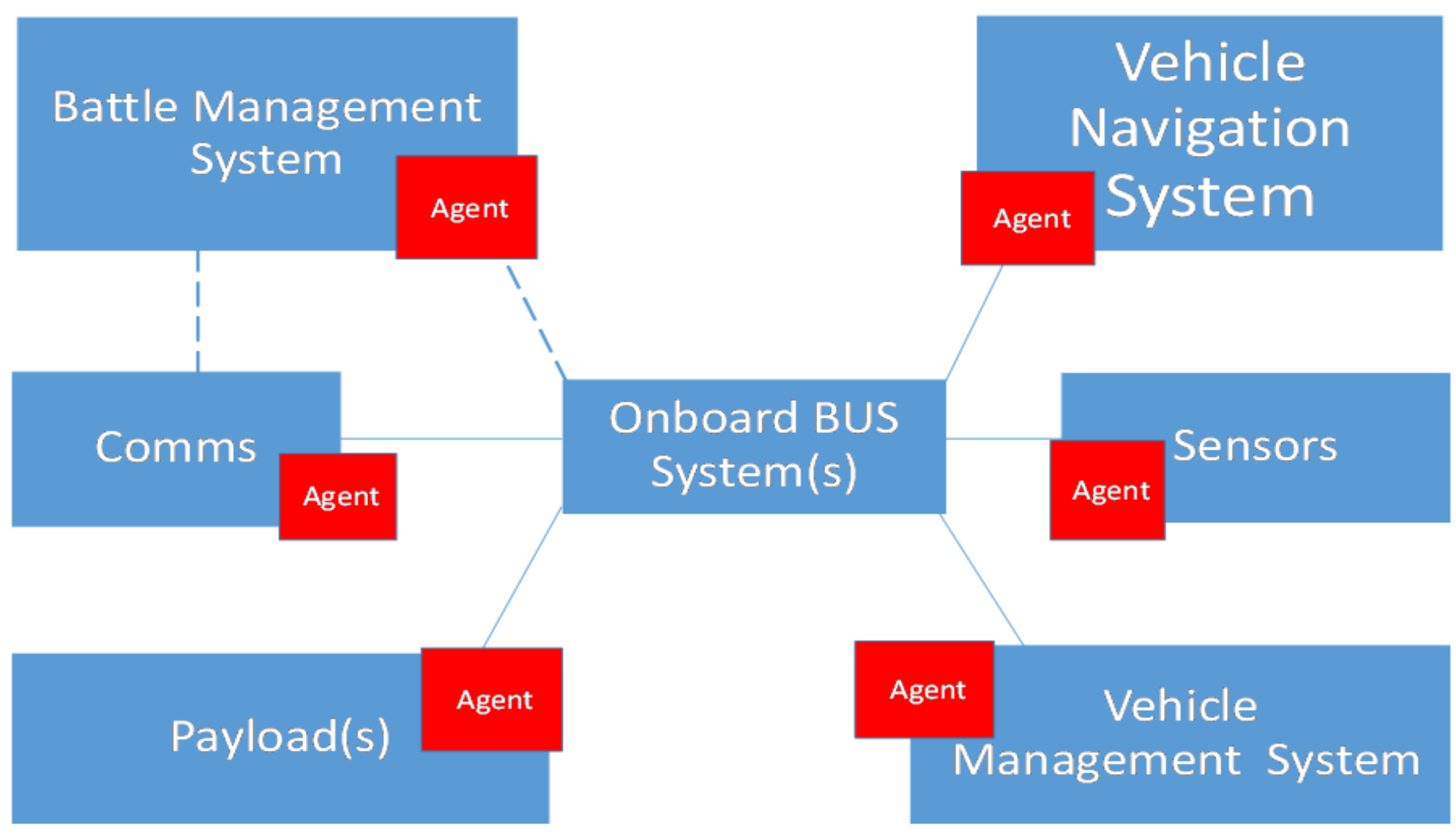

**Fig. 4 Distributed agent CS**

## 2.3 Cyber-defense Agent Concept of Operations (CONOPS)

CPSs operate over sensor or control links that can be degraded, denied, or corrupted by adversarial action in combat. The standard of performance of a cyber system is how well the system guarantees confidentiality, integrity, and availability of the data that are transmitted or stored on the system. For the discussion that follows, we assume a decentralized command structure of autonomous agents operating to mitigate interference with a BMS of a military vehicle. We assume use of TCP-IP. Note that use of a decentralized cyber-defense agent command structure in a vehicle does not imply that the command structure for the vehicle as a whole is decentralized.

## 2.4 An Attack Example

Let us assume a vehicle in laager undergoing routine maintenance. A maintenance management system is connected to a system bus to record system network traffic during static pre-operations test procedures.

**Stage 0: Primary Infection**

Prior to the maintenance procedure, malware is loaded onto the maintenance system hardware, which has now become the vector for the attack on the vehicle. When the maintenance hardware is connected to the vehicle, malicious code is allowed to migrate to the VMS.

However, there is an AICA resident in the VMS, which in this case logs the network activity under the maintenance procedure, which includes identification and access management (IDAM), file transfers, and configuration file changes.

**Stage 1: Reconnaissance**

Once the maintenance hardware is removed, the malware begins to log network traffic in order to identify the attached subsystems of the vehicle. The location of the BMS is not known to the malware a priori, so the malware starts probing for any open ports commonly used by a BMS. Once located, it scans the BMS for vulnerabilities it can exploit from within the VMS.

The AICA in the VMS detects the scanning activity going out over the network. It puts AICAs in the target systems on alert following the anomalous activity of an unanticipated port scan. The VMS is given an alert from its AICA that there is anomalous activity originating from the VMS. This information is then used in accordance with platform TTPs to initiate systems diagnostics. Systems diagnostics notifications are sent offboard via the BMS and COMMS to a C2 node that the vehicle operational readiness might be degraded.

**Stage 2: Attack**

The malware in the VMS recognizes that VMS diagnostics have started and commences a lateral movement into the file structures of the BMS. It selects an appropriate exploit and executes code.

The AICA in the BMS recognizes a subsequent change in a configuration file as a result of the code execution and that it was updated outside of a scheduled maintenance period or without system access authorization. That AICA notifies the BMS to commence systems diagnostics. Another diagnostics alert is sent offboard to the C2 system. An AICA at the central C2 system recognizes an increased frequency of diagnostics alerts and sends status queries back to the vehicle system's AICAs. The malware now needs to move laterally again due to systems diagnostics procedures being initiated and selects the COMMS after a port scan. The AICA in COMMS has already been alerted to anomalous behavior in two other subsystems and notifies COMMS that no configuration changes are to be accepted without the elevated privileges required by systems recovery procedures. The malware is now unable to execute a hijacking of a software-defined radio (SDR) communications channel and resumes port scanning to seek further lateral movement and further targets.

**Stage 3: Recovery**

It is at this point that the AICAs at the various vehicle control subsystems isolate their systems on the various control busses and initiate automated diagnostics and recovery procedures. The vehicle maintenance team is alerted and diagnostics and forensics begins, whereby the malware is discovered, and agent-based examination of systems logs discovers the chain of events and the likely vector. Luckily, in this example, our vehicle never made it out of laager before recovery. But we can envision many other scenarios where we encounter cyber–physical attacks while underway, and mitigation and recovery processes have to be carried out during a mission. In some of these instances, the deliberative actions of the embedded agents will have to include prosecuting a mission with degraded capabilities or autonomous recovery of a vehicle with minimal human intervention.

## 3. Architecture Overview

Author: Paul Théron

This section provides an overview of the AICARA. First, it presents the agents' functional architecture and its components as it is assumed today. Next, it identifies five high-level functions of agents and, for each of them, details their main features. Finally, recognizing that the components of the functional architecture will have dependencies, the last section presents what they are going to be.

Cyber-defense agents considered in the AICARA can essentially do the following:

- They can handle autonomously and in a trustworthy manner the cyberattacks affecting the perimeter they defend.
- They can cooperate, with one another, a cyber C2 system, or even a human operator, when and as required and feasible.

Each agent is implemented within or in attachment to one delimited system or device. Cooperation between agents is achieved through available communication channels. These communication channels must be as covert as feasible because agents must be as stealth as possible in order to protect themselves from attacks by enemy malware.

The AICARA, derived from Russell and Norvig (2010), is assumed to include the functional components outlined in Fig. 5.

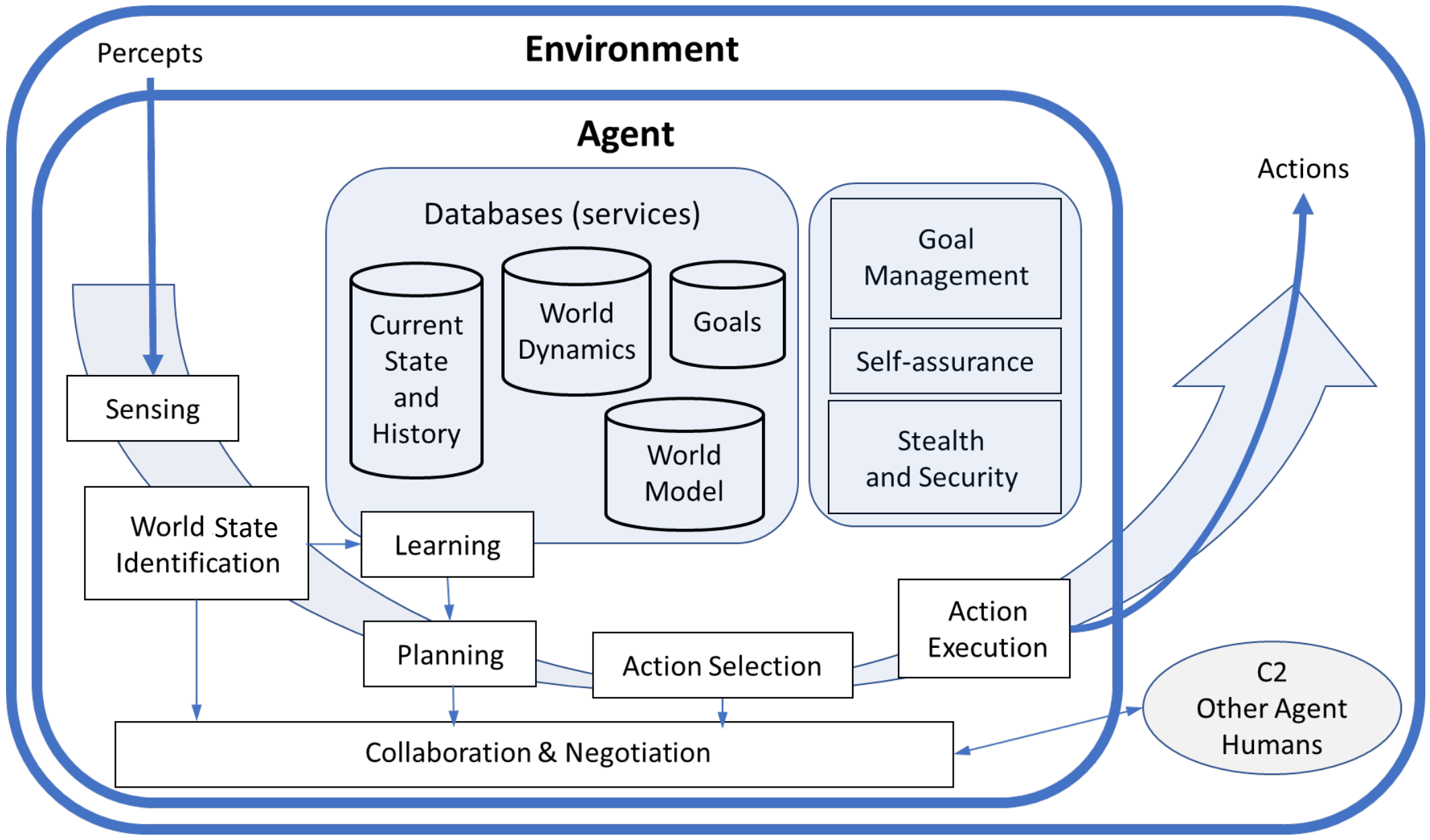

**Fig. 5 AICA's functional architecture, AICARA**

The agents' functional components belong in three classes as outlined in Table 1.

**Table 1 Classes of AICA components**

| Class of components | Functional components |
|---|---|
| Core components | Sensing<br>World State Identification<br>Planning<br>Action Selection<br>Action Execution |
| Support functions | Collaboration and Negotiation<br>Learning<br>Goals management<br>Self-assurance<br>Stealth and security |
| Data services | World model<br>Current state and history<br>World dynamics<br>Goals |

Note that at the time of publication, the AICARA stands as an initial assumption. It is discussed and exemplified in later sections.

AICAs can be implemented in three different ways, and each option would entail specific choices both in terms of technology and doctrine of use:

1) **A society of specialized agents**: This option refers to the distributed implementation of the reference agent's functional components (as in Fig. 5) as a group of specialized agents, each one owning/delivering one of the functions of the AICA presented, and the sum of the agents delivering the entire reference agent's cyber-defense capability. Major questions are the following: Where should these agents be implemented/located within the defended system? What happens if one specialist agent is disabled (for instance, out of an attack directed at it) and thus breaks a functional chain: would it be replaced, how would its own knowledge/working memory be preserved, how would the tasks it was performing before being attacked continue, and so on? Does this option allow or jeopardize agents' stealth and can covert communication channels hide a possibly intense traffic between specialist agents?

2) **A multiagent system**: This option refers to a swarm or cohort of fully functional agents the architecture of which would be as in the AICA model presented previously, each one being capable of executing all AICA functions, and the swarm as a whole being supposed to deliver a collective response to a cyberattack. Major questions are the following: What is the collective intelligence of the swarm and how does it emerge? Are multiagent systems less stealthy than option 1 and option 3?

3) **An autonomous collaborative agent**: This option refers to a fully functional agent, capable of performing full cyber-defense duty on its own territory and capable, when and as needed and circumstances permitting, of communicating with other agents. Major questions are the following: What is the purpose of communications/collaborations between agents? Where is this single agent implemented within a system? How different is this agent from cyber-detection agents currently under development or already available?

This report does not advocate any choice of an implementation option. The choice of an implementation option may be guided by criteria such as, but not limited to, the following:

- The type and level of classification of systems to defend
- Their architecture and topology
- Their technical capacities (computing power, memory, communications, etc.)
- Agents' performance and cost requirements

AICA contributes to the cyber defense of a military system or device through five main high-level functions (Fig. 6):

1) *Sensing and World State Identification*
2) *Planning and Action Selection*
3) *Collaboration and Negotiation*
4) *Action Execution*
5) *Learning*

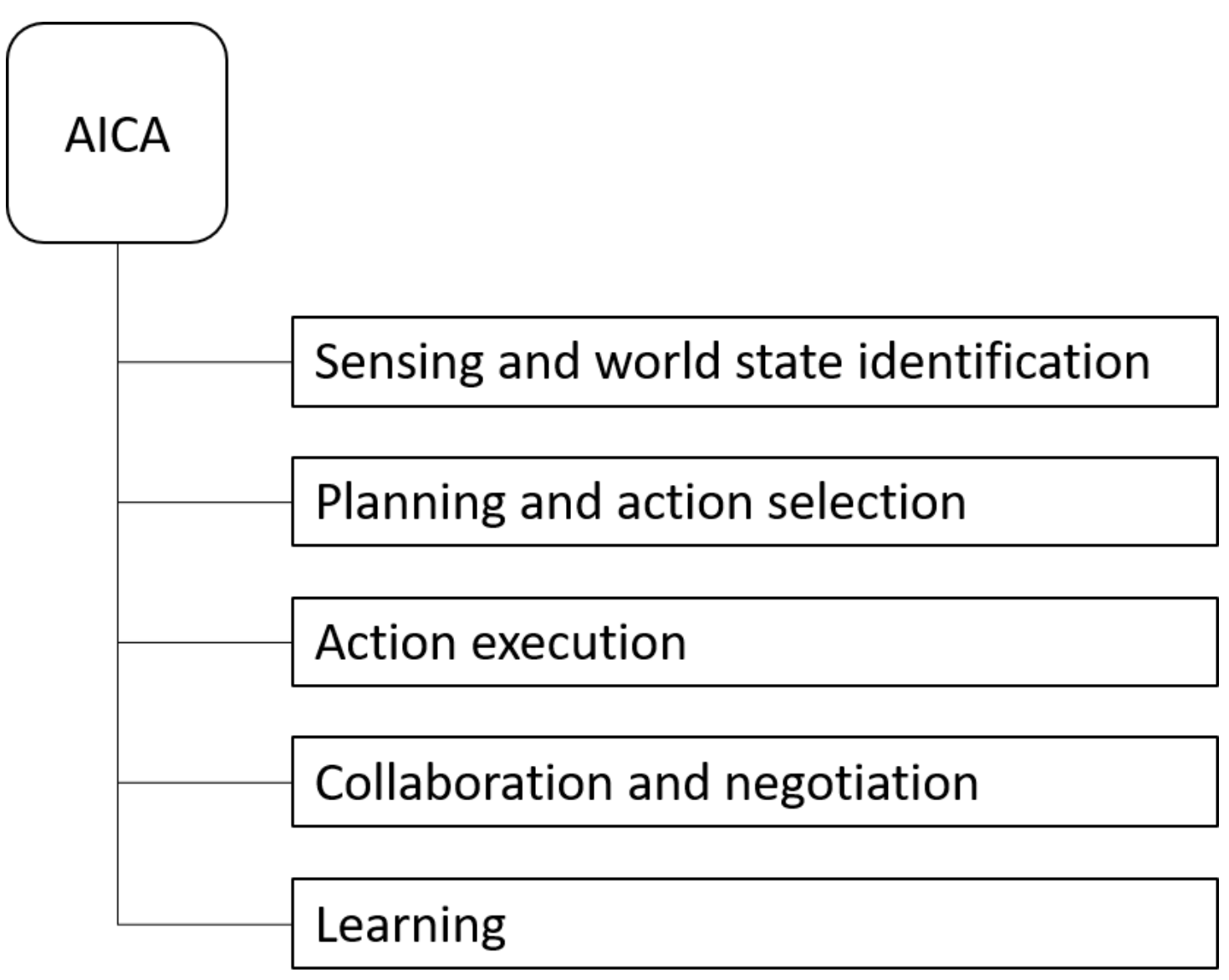

**Fig. 6 AICA's main five high-level functions**

## 3.1 Sensing and World State Identification

*Sensing and World State Identification* is the AICA high-level function that allows a cyber-defense agent to acquire data from the environment and systems in which it operates, as well as from itself, to reach an understanding of the current state of the world and, should it detect risks in it, trigger the *Planning and Action Selection* high-level function.

This high-level function relies upon the world model, current state and history, sensors, and world state identification components of the assumed functional architecture.

It includes the following two functions:

- Sensing
- World state identification

### 3.1.1 Sensing

Sensing operates from two types of data sources:

- External (system/device-related) current world state descriptors
- Internal (agent-related) current state descriptors

Current world state descriptors, both external and internal, are captured on the fly by the agent's sensing component. They may be double checked, formatted, or normalized for later use by the world state identification component (to create processed current state descriptors).

### 3.1.2 World State Identification

The world state identification function operates from two sources of data:

- Processed current state descriptors
- Learned world state patterns

Learned world state patterns are stored in the agent's world knowledge repository. Processed current state descriptors and learned world state patterns are compared to identify problematic current world state patterns (i.e., presenting an anomaly or a risk). When identifying a problematic current world state pattern, the world state identification function triggers the *Planning and Action Selection* high-level function.

## 3.2 Planning and Action Selection

*Planning and Action Selection* is the AICA high-level function that allows a cyber-defense agent to elaborate one to several action proposals and propose them to the action selection function, which decides the action or set of actions to execute to resolve the problematic world state pattern previously identified by the world state identification function.

This high-level function relies upon the world dynamics, actions and effects, goals, the actions' effect predictor, and action selection components of the assumed functional architecture.

It includes the following two functions:

- Planning
- Action selection

### 3.2.1 Planning

The planning function operates on the basis of two data sources:

- Problematic current world state pattern
- Repertoire of actions (response repertoire)

The problematic current world state pattern and repertoire of actions (response repertoire) are concurrently explored to determine the action or set of actions (proposed response plan) that can resolve the submitted problematic current world state pattern. The action or set of actions so determined are presented to the action selection.

It may be possible that the planning function requires some form of cooperation with other agents or a central cyber C2 to come up with an optimal set of actions forming a global response strategy. Such cooperation could be to either request from other agents or the cyber C2 complementary action proposals or delegate to the cyber C2 the responsibility of coordinating a global set of actions forming the wider response strategy. This aspect is not yet studied in the present release of the AICARA.

### 3.2.2 Action Selection

The action selection function operates on the basis of three data sources:

- Proposed response plans
- Agent's goals
- Execution constraints and requirements (e.g., environment's technical configuration, and so on)

The proposed response plan is analyzed by the action selection function in the light of the agent's current goals, and the execution constraints and requirements that may either be part of the world state descriptors gained through the *Sensing and World State Identification* high-level function or be stored in the agent's data repository and originated in the *Learning* high-level function. The proposed response plan is then trimmed from whatever element does not fit the situation at hand and augmented by prerequisite, preparatory, precautionary, or postexecution recommended complementary actions. The action selection thus produces an executable response plan, which is then submitted to the *Action Execution* high-level function.

Like with the planning function, it is possible that the action selection function is required to liaise with other agents or a central cyber C2 to come up with an optimal executable response plan forming part of and being in line with a global response strategy. Such cooperation could be to exchange and consolidate information with other agents or the central cyber C2, and then agree collectively on the assignment of responsibilities over the various parts of the execution of the global executable response plan to specific agents. Alternatively, it could be to delegate to the cyber C2 the responsibility of elaborating a consolidated executable response plan and then assign to specific agents the responsibility of executing part(s) of the overall plan within their dedicated perimeter. This aspect is not yet studied in the present release of the AICARA.

## 3.3 Action Execution

*Action Execution* is the AICA high-level function that allows a cyber-defense agent to effect the action selection function's decision about an executable response plan (or the part of a global executable response plan assigned to the agent), monitor its execution and its effects, and provide friendly agents with the means to adjust the execution of their own part of the response plan as and when needed.

This high-level function relies upon the goals and actuators components of the assumed functional architecture.

It includes the following four functions:

- Action activation
- Execution monitoring
- Effects monitoring
- Execution adjustment

### 3.3.1 Action Activation

The action activation function operates on the basis of two data sources:

- Executable response plan
- Environment's technical configuration

Taking into account the environment's technical configuration, the action activation function executes each planned action in the scheduled order.

### 3.3.2 Execution Monitoring

The execution monitoring operates on the basis of two data sources:

- Executable response plan
- Plan execution feedback and status

The execution monitoring function should be able to monitor (possibly through the sensing function) each action's execution status (for instance, done, not done, or wrongly done). Any status apart from "done" should trigger the execution adjustment function.

### 3.3.3 Effects Monitoring

The effects monitoring function operates on the basis of two data sources:

- Executable response plan
- Environment's change feedback and status

It should be able to capture (possibly through the sensing function) any modification occurring in the plan execution's environment. The associated data set should be analyzed/explored. The result of such data exploration might (should) provide a positive (satisfactory) or negative (unsatisfactory) environment change status. Should this status be negative, this should trigger the execution adjustment function.

### 3.3.4 Execution Adjustment

The execution adjustment function operates on the basis of three data sources:

- Executable response plan
- Plan execution feedback and status
- Environment's change feedback and status

The execution adjustment function should explore the correspondence between the three data sets to find alarming associations between the implementation of the executable response plan and its effects. Should warning signs be identified, the execution adjustment function should either adapt the actions' implementation to circumstances or trigger a tactical revision/adaptation to the plan.

The update of the response plan in the course of its execution is not studied in the current release of the AICARA. It presents issues that require further research work such as the need for collaboration and negotiation between agents. A notion of tactical superiority can be envisaged but is not studied in this report.

## 3.4 Collaboration and Negotiation

*Collaboration and Negotiation* is the AICA high-level function that allows a cyber-defense agent to 1) exchange information with other agents or a central cyber C2, or possibly with a human operator, for instance, when one of the agent's functional components is not capable on its own of reaching satisfactory conclusions or usable results; and 2) negotiate with its partners the elaboration of a consolidated conclusion or result.

This high-level function relies upon the collaboration and negotiation component of the assumed functional architecture.

It includes, at the present stage, one function:

- Collaboration and negotiation

The collaboration and negotiation function operates on the basis of three data sources:

- Internal, outgoing data sets (i.e., sent to other agents, a cyber C2, or human operator)
- External, incoming data sets (i.e., received from other agents, a central cyber C2, or human operator)
- The agents' own knowledge (i.e., consolidated through the *Learning* high-level function).

When an agent's *World State Identification, Planning*, or *Action Selection* high-level function (or potentially any other functional component) needs it, the agent's collaboration and negotiation function is activated. Depending on collaboration policies memorized in the agent's stealth and security component, ad hoc data are sent to authorized agents or a central cyber C2, possibly to a human operator. The receiver(s) may negotiate with the emitting agent or may not be able to elaborate further on the basis of the data received through their own collaboration and negotiation function. When agents (including possibly a central cyber C2 or human operator) have elaborated and reached shared conclusions, agent(s) will spark the next function within their own decision-making process.

When the agent's own security is threatened, the agent's collaboration and negotiation function should at least help warn other agents (or a central cyber C2 or possibly a human operator) of this state.

This release of the AICARA does not describe the agent's security monitoring and management.

Furthermore, the agent's collaboration and negotiation function may be used to receive warnings from other agents that may trigger in the agent a higher state of alarm.

Finally, the agent's collaboration and negotiation function should help agents discover other agents and establish links with them.

This release of the AICARA does not describe nor specify the exchange protocol and the negotiation process, nor the alarm-raising mechanism and the agent discovery mechanism. These are issues to be further studied in later research.

## 3.5 Learning

*Learning* is the AICA high-level function that allows a cyber-defense agent to use the agent's experience to improve progressively its efficiency with regard to all other functions.

This high-level function relies upon the learning and knowledge improvement components of the assumed functional architecture.

It includes two functions:

- Learning
- Knowledge improvement

### 3.5.1 Learning

The learning function operates on the basis of three data sources:

- Feedback data from the agent's environment changes
- Feedback data from the agent's functioning
- Feedback data from the agent's actions

The learning function collects and analyzes the corresponding data sets possibly in conjunction with the reward function of the agent (or distance between goals and achievements). Results feed the knowledge improvement function.

### 3.5.2 Knowledge Improvement

The knowledge improvement function operates on the basis of two data sources:

- Results (propositions) from the learning function
- Current elements of the agent's knowledge

The knowledge improvement function merges results (propositions) from the learning function and the current elements of the agent's knowledge.

The current release of the AICARA provides only a basic description or examples of the *Learning* high-level function and of the role of artificial intelligence in this context.

## 3.6 Agents' Generic Process Flow

The overall functioning of an agent is summarized in the following graph that shows the agent's generic process flow (Fig. 7).

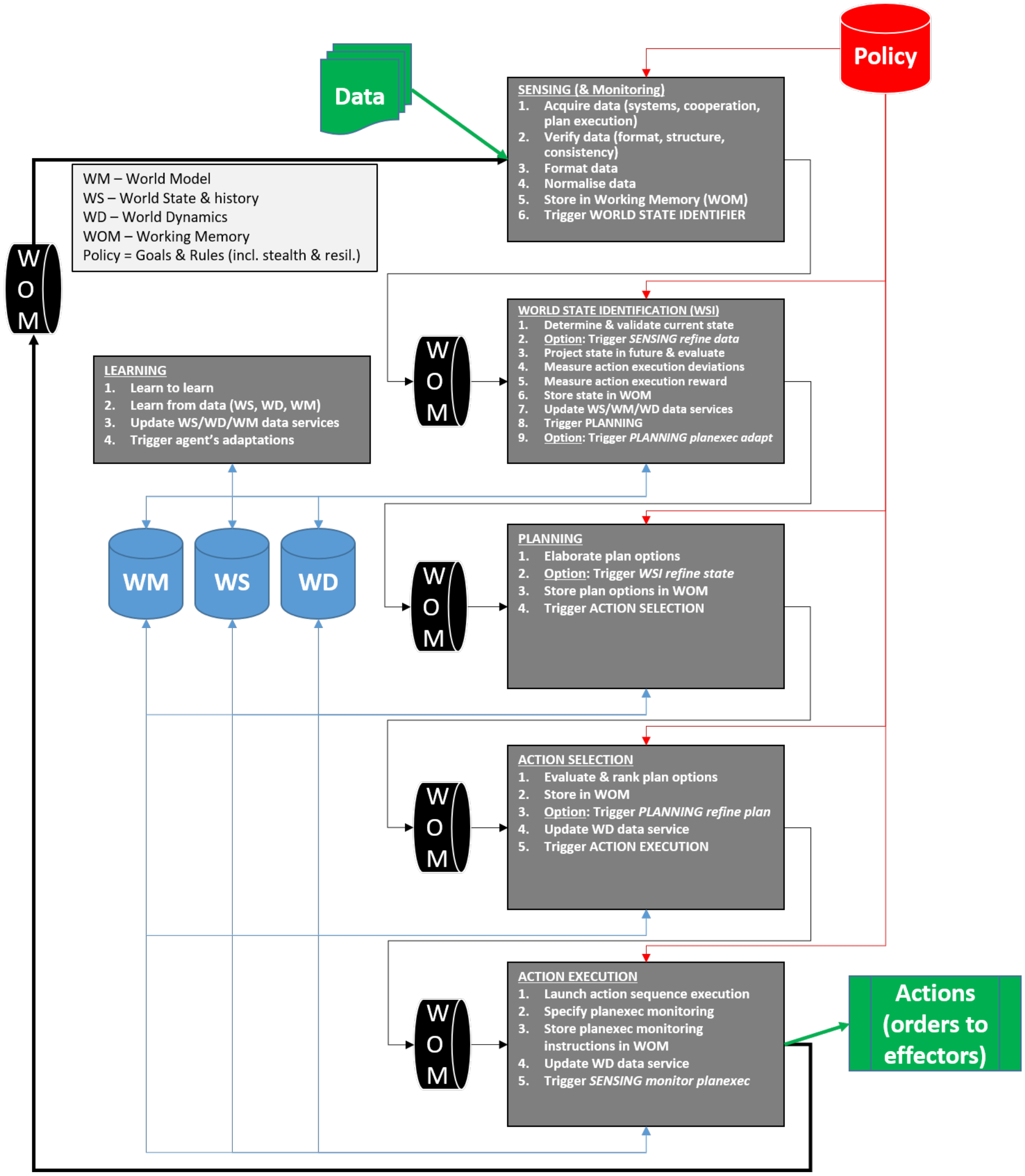

**Fig. 7 AICA's generic process flow**

In this diagram, each component of the AICARA details its principal tasks.

A working memory, which could be implemented either as temporary stacks or as a shared blackboard or common work area, will help with passing data on from component to component.

Besides the acquisition of data relating to the systems falling in the agent's scope and of data exchanged with other entities (other agents, a central cyber C2 system, or human operators via an ad hoc human–computer cooperation mechanism), the sensing component has also a monitoring function that covers the execution of action plans launched by the action execution component.

The world state identification component does the following:

- May ask the sensing component for further data if it cannot compute the current state of the environment in the agent's remit.
- Computes how good or poor the performance of previously launched plans of actions is, and if poor or inadequate, it triggers the planning component for a revision/tactical adaptation of these plans in order to better match the attacker's action.
- Updates, when possible/appropriate, the world current state and history, world dynamics, and world model databases.

The planning component does the following:

- Elaborates a number of options of action (countermeasure) plans in response to the current state identified previously.
- May ask the world state identification component for a refinement of the computation of the current state if it lacks elements to elaborate a plan of countermeasures.

The action selection component does the following:

- Evaluates and ranks (in terms for instance of cost, time to deliver effects, risks, etc.) the plan options presented by the planning component.
- May ask the planning component to refine its plan options.
- Updates the world dynamics database component when it has made a clear choice of a plan and associated it with the current state found by the world state identification component.

The action execution component does the following:

- Launches the orders corresponding to the plan and sends them to the ad hoc effectors across the system defended by the agent.
- Specifies what the sensing component must monitor to supervise the execution of the action plan, and it stores those elements in the working memory to pass them on to the sensing component.

- Updates the world dynamics database components with these complementary elements of information.

The learning component does the following:

- Has a generic learning mechanism that reinforces itself with experience.
- Learns on the fly from the data acquired and stored by the agent.
- Updates the database components with new elements of knowledge.
- Should trigger the ad hoc adaptations of the agent's internals to improve the latter's performance.

## 4. Data Services within Agents

Author: Paul Théron

This section describes the initial assumptions made about the following AICA data services:

- World model
- World current state and history
- World dynamics knowledge

These modules of the agent are not just mere data repositories but producers of processed data (i.e., "information"). They embark on an intelligence of their own or rely on external sources to produce information, possibly cooperating with other agent services for higher-order intelligence or support. Their communication with other agent data services and functional components implies the definition of internal protocols. Their data must be protected. The agent's data services are built in a way similar to that of the diagram in Fig. 8.

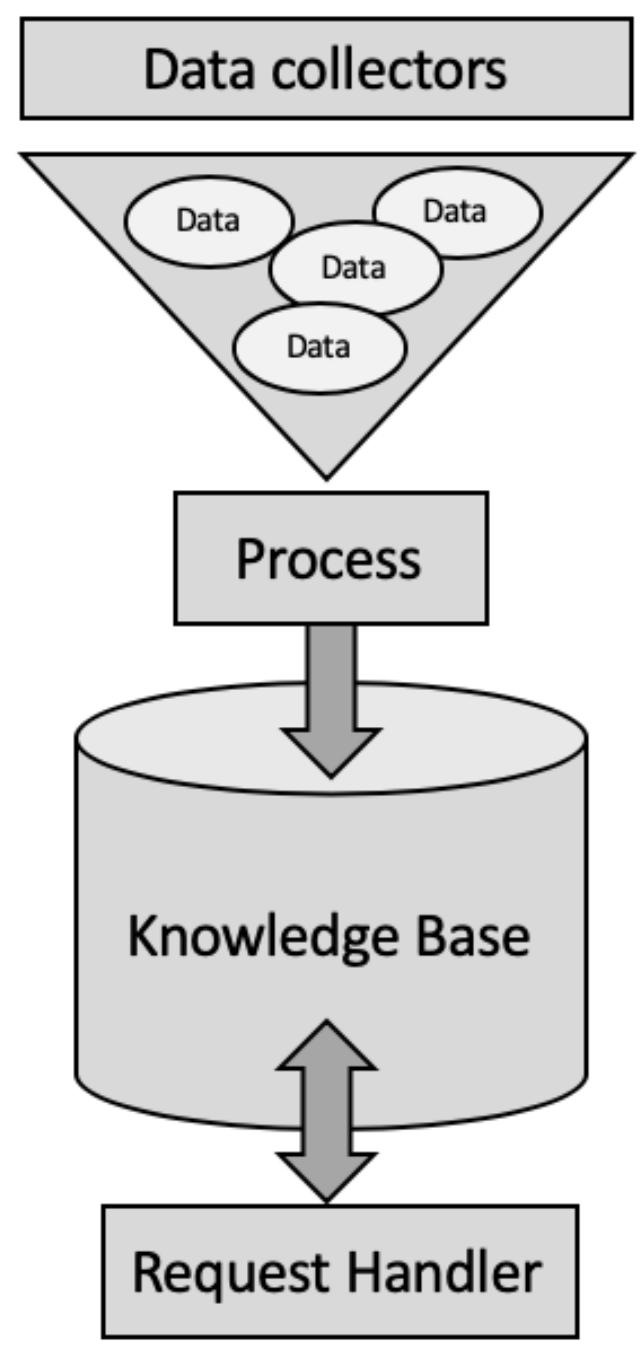

**Fig. 8 General architecture of the data services**

At the present stage, many options are open. We hypothesize the following ones:

- Data collectors accept incoming data records and check their compliance to formatting and consistency rules.

- Once verified, data records are processed. Processing may be limited to mere storage instructions or the data service module may have to perform data normalization/consolidation/aggregation functions as well as exploratory data analysis and exploratory factor analysis operations.

- Data records and elaborated information can be requested by the agent's other components. In this case, the data service's request handler should be designed to check the request against validity and security rules (according to agent design options and security policies), and then data are extracted, sorted, grouped, and bundled into an appropriate data container and returned to the requesting module.

## 4.1 World Model Data Service

### 4.1.1 Definition

We hypothesize that a world model is the following:

- A formal descriptor of the elements it supplies to the agent's other components:
    - The nominal and degraded ontology or configuration of the agent
    - The nominal and degraded ontology or configuration of the system and environment (systems and threat) to defend
    - The nominal and degraded ontology or configuration of cyber threats against the system and environment to defend and against the agent itself
    - The nominal and degraded patterns of the world's state (agent + environment + threat). Patterns express the agent, the system or its environment's static and dynamic relations, and the concurrency of their configurations.
- It is based on the following:
    - A theory of world models in the context of the cyber defense of military systems
    - A formal descriptive language
    - Validated algorithms transforming inputs into descriptors
- Embedded into the agent, the model is determined by one of the following:

- Calculated by the agent (which inflates the agent’s size and requires computing power) or
    - Loaded from external sources (which requires periodic or occasional downloads of the agent’s data and updates/uploads into the agent of data produced by external sources).

### 4.1.2 Inputs

We hypothesize that the world model data service may take the following classes of data as input:

- Data about the agent:
    - Architecture, modules, and functions
    - Communication
    - Collaboration links
    - Processes and protocols
    - Performance descriptors
- Data about the defended system:
    - Identified vulnerabilities
    - Security devices and barriers
    - Topology of friendly agents network
    - Connection components problems
    - Hardware components problems
    - Firmware components problems
    - Operating system (OS) components problems
    - Middleware components problems
    - Applicative components problems
- Data about cyber threats:
    - Cyberattack, cyber vulnerabilities, cybersecurity, and cyber-defense state-of-the-art technologies

- MITRE and other useful classifications (Common Attack Pattern Enumeration and Classification [CAPEC], Common Vulnerabilities and Exposures [CVEs], etc.)
- Kill chain-like models

- Data about the defended system's environment:
  - Sources of threats and attack C2 and tools
  - Threat and vulnerability patterns (CAPEC, CVEs, etc.)
  - Indicators of compromise (IoCs) (OpenIOC, Malware Information Sharing Platform [MISP], etc.)
  - Cybersecurity and cyber-defense dispositions and their topology
  - Available cyber-defense resources
  - Surrounding systems and their cybersecurity and cyber-defense dispositions and topologies

The data sources would then be the following:

- Cyber-threat intelligence sources
- System descriptors (Simple Network Management Protocol [SNMP] data, packet-based switching [PBS], topology, configuration, etc.)
- The world state and history data service

### 4.1.3 Process

There are two ways to produce ontologies and patterns of the world state:

- They can be created within the agent.
- They can be uploaded into the agent's database.

When created within the agent, input data are processed in the following ways:

- Collected through the agent's sensor (a standard format is required).
- Verified and preprocessed by the world model data service (e.g., normalized, formatted, and so on).
- Associated by the world model data service to form ontologies and patterns (ad hoc functionalities are required).

- Stored in the world model data service's database (a standard format is required).

### 4.1.4 Outputs

The hypothesized outputs of the agent's world data service are the following:

- Domain ontologies
    - Agent
    - System
    - Environment
    - Communication
    - Threat
    - Nominal and degraded
- World patterns
    - Cross-domain patterns
    - Domain-specific patterns
    - Nominal and degraded

### 4.1.5 Current Issues and Lines of Research

Several issues can be identified at the present stage:

- The data classes required as input and the exact nature of output information
- The data formats of input data, data exchange protocols, and output information
- The algorithms for preprocessing, creating, and indexing data
- The implementation option of agents
- The risks to the agent's stealth due to the required memory size, processing power, and communication needs

## 4.2 World Current State and History

### 4.2.1 Definition

We hypothesize that the world current state is the evaluated distance between the world as it is and what it should be (based, for instance, on set goals or standards). Pieces of information such as the following may be required to form world state vectors describing the agent's world and that can be used by the world state identification component of AICA:

- Nominal and degraded states of reference of agents and their cohort, defended systems, their environment and connections, and threats, including the current state and the track record of past states
- Memory of cyber-defense actions and their impacts on the state of the world (current and past)
- Current data about agents and their cohort, defended systems, their environment and connections, and threats

The world state identification module can then be hypothesized to do the following:

- Calculate the current world state data vector.
- Measure the deviation of the current world state data vector from the nominal world state data vector.
- Interpret (meaning) the measure of the deviation (based on history, actions in progress, etc.).
- Appraise the deviation (i.e., determine the positive or negative valence of the deviation).

The current world state data vector is a formal descriptor of the appraised world's state at a given point in time and circumstances, usable by the world state identification module.

The world state history is the chronological track record of world state descriptors.

### 4.2.2 Inputs

The world current state and history data service takes world state records from the world state identification module.

### 4.2.3 Process

The world current state and history data service labels (with metadata) and stores the new world state data vector provided by the state identification module into its database.

### 4.2.4 Outputs

World state descriptor records are stored in the world current state and history's database.

### 4.2.5 Current Issues and Lines of Research

The issues identified at the present stage are the following, among possible others:

- The specification of world state data descriptors/vectors
- The computation of world states, both nominal and degraded
- How the historical records of world state descriptors are used by the world state identification
- The size of the world current state and history's database

## 4.3 World Dynamics Data Service

### 4.3.1 Definition

Given that the world can be defined as a collection of interrelated objects, we hypothesize that world dynamics are the following:

- An agent's behavioral rules and related expected states (nominal and degraded) in given circumstances
- Defended systems and other world objects' behavioral rules and related expected states (nominal and degraded) in given circumstances

They can be measured in the following four manners, summarized in Table 2:

- A measure of how the world changes given its own event parameters (state changes, actions, events); the world includes agents, defended systems, those systems' environment and connections, and the threats on agents, the defended systems, and their environments.
- A measure of how the world changes given agents' event parameters (state changes, actions).
- A measure of how agents change given events in the world.

- A measure of how agents change given their own event parameters (state changes, actions).

**Table 2 Measuring the world dynamics data**

| **Factors of change ➔**<br><br>**Changing object 🡻** | **World's events** | **Agents' events** |
|---|---|---|
| World as a whole | 1 | 2 |
| Agents and agent cohort | 3 | 4 |

Those laws of world's dynamics can be computed out of the following data:

- States of reference (nominal and degraded) descriptors for agents, systems, environments, and threats
- Agents and world entities' events and actions descriptors
- Agents' and world entities' initial and final state descriptors

The world dynamics data service computes state transition patterns. Confidence estimators are associated with state transition patterns. State transition patterns and confidence estimators can be applied to identified initial states of world entities or agents to predict their likely end states.

### 4.3.2 Inputs

We hypothesize that the world dynamics data service requires the following classes of data as input:

- Data from the world model data service
- Data about cyber-threat dynamics:
  - Patterns of behavior of malware
  - {cyber threat; targeted world entities and topologies}
  - {cyberattack patterns; expected defense responses}
  - {initial state; end state} and their factors
  - Circumstances/context of cyber threats
- Data about defended systems and cyber-defense dynamics:
  - Monitored and surrounding system(s)

- Patterns of behavior of world events
        - {world events; expected world retroaction}
        - {world's initial state; world's end state} and their factors
        - Circumstances/context of world changes
    - Agent itself and other friendly agents inside/outside agent's cohort
        - Patterns of behavior of agent events
        - {agent events; expected agent retroaction}
        - {agent's initial state; agent's end state} and their factors
        - Circumstances/context of agent changes
    - Incident response mechanisms (IRM) dynamics
        - Patterns of behavior of IRM events
        - {IRM events; expected IRM retroaction}
        - {IRM's initial state; IRM's end state} and their factors
        - Circumstances/context of IRM changes

### 4.3.3 Process

There are two possible ways to compute world state transition patterns and associated confidence estimators out of input data:

- Data are processed live by the world dynamics data service.
- Data are uploaded into the world dynamics data service's database from externally provided records.

### 4.3.4 Outputs

The world dynamics data service computes state transition patterns and associated confidence estimators.

### 4.3.5 Current Issues and Lines of Research

The complexity of the world (and even of the agent, as it is internally dynamic and adjusts to the world's changes) poses computational challenges. The second kind of technical challenges is related to the memory size and computation power required to compute state transition patterns. The third challenge is associated with using state transition patterns and their confidence estimators.

## Part B. Discussion of the Architecture's Main Functions

# 5. Sensing and World State Identification

Authors: Martin Drašar, Mauno Pihelgas, Markus Kont, and Benoît Leblanc

## 5.1 Overview

To interact with the world, the agent has to perceive and understand itself and its surroundings. This is accomplished by two functional components of the AICARA: *Sensing and World State Identification*:

- The *Sensing* function provides the agent with data about itself and its environment.
- The *World State Identification* function interprets the collected data as the following:
    - The current state of the world and of the agent itself
    - Changes in this context since last observations were made
    - Adversarial or suspicious events
    - Anomalies in collected data

The *Sensing* function addresses the question "What do I observe?" and the *World State Identification* addresses the question "What is the situation?". Their combined actions participate to agent's situational awareness.

## 5.2 Sensing

The *Sensing* function (Fig. 9) can contain a number of subsystems:

- **Self**: Collects data about the agent's memory and functions to ensure the agent's integrity.
- **System**: Collects data about the defended system's resources like memory, file system, and so on. It also monitors results of actions performed by the agent. It can either be part of a monolithic agent or function as a separate module that feeds the agent data.
- **Environment**: Used for monitoring data coming from outside the agent. Can either be part of a monolithic agent or function as a separate module that feeds the agent data.

- For the sensory data to be useful to the rest of the system, they should be properly normalized, correlated, fused, and deduplicated, so that only unique and relevant bits are passed on.

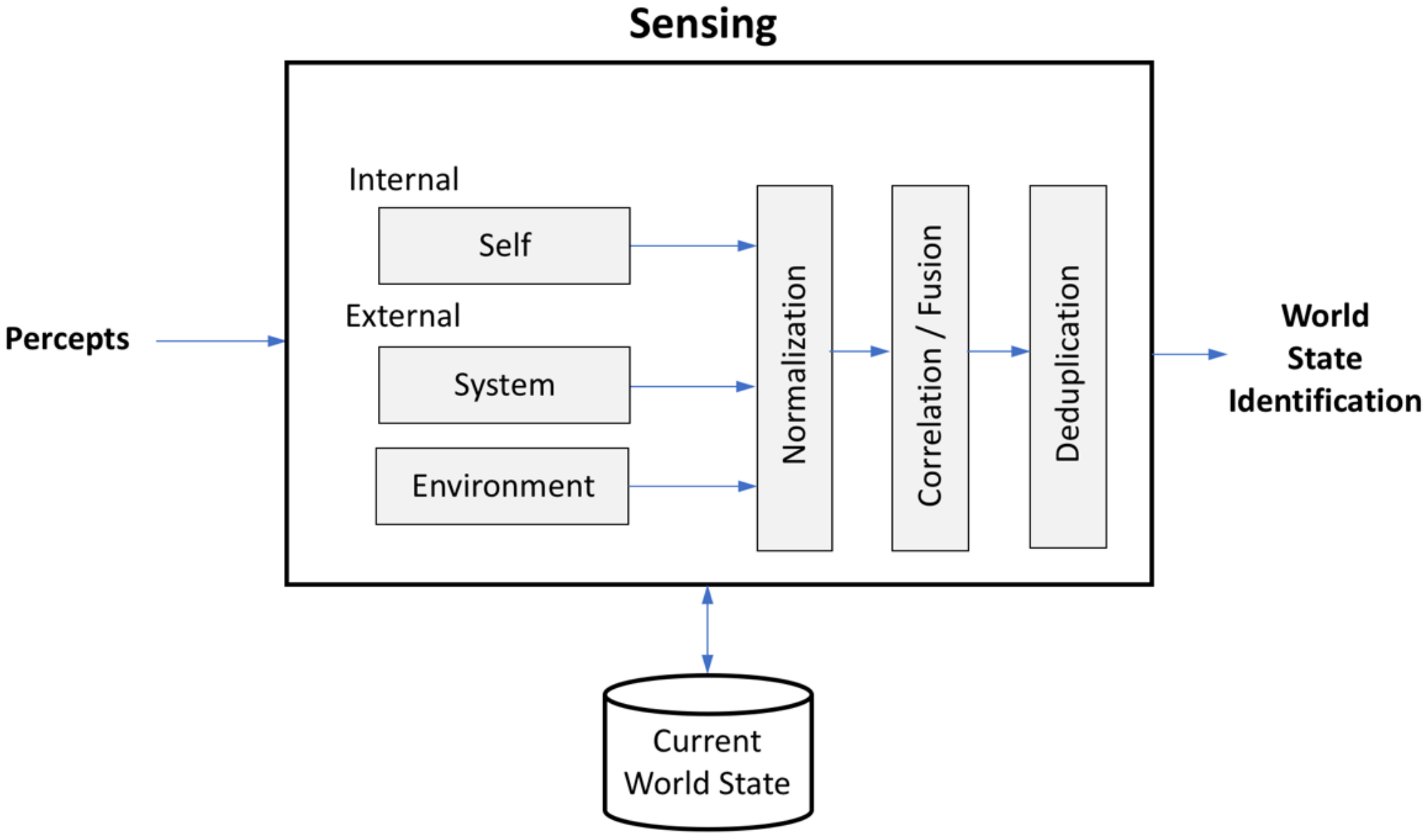

**Fig. 9** ***Sensing*** **component**

The dataflow of the *Sensing* system includes the following traits:

- *Sensing* is the function responsible for gathering and processing data from both external and internal sources.

- To ensure continued operation, the *Sensing* function monitors its internal health by collecting runtime statistics and checking their integrity.

- Furthermore, the input modules of the *Sensing* function fulfil the typical roles of system and network monitoring tools. The *Sensing* function collects logs and metrics from the other internal systems of the agent, the underlying host system (i.e., the OS), and relevant applications running on the host. *Sensing* is also capable of capturing network traffic from the host network interfaces. Alternatively, in the case of a centralized agent, it is possible to capture traffic from a dedicated test access point (TAP) device or monitoring port.

- The data from the input modules always go through the input sanitation (normalization, correlation, fusion, and deduplication) process, which ensures that the data can be processed by other functions of the agent. This also applies to data received from other agents and C2, because the adversary may try to inject malicious or garbage data into the agent.

- The *Sensing* function passes its data on to the *World State Identification* function.
- The *Sensing* function can also deal with the communication component to be able to ask question or discuss what is observed with other agents, cyber C2 or humans, via a secure channel.

It should be noted that when agent's stealth is a concern, all *Sensing* operations should be done on-demand. Unlike in the physical world, it is nearly impossible (excluding specific side channels) for an agent to do a truly passive reading of a sensor, because the system calls needed may by intercepted by an adversarial agent. It should also be noted that by applying this policy an agent can severe itself from any external orders, which may substantially diminish its usability,

## 5.3 Current World State Identification

The *World State Identification* component (Fig. 10) processes data given from sensing to assess the state the world is in with respect to the world model. It consists of up to four processes:

- **Environment identification:** Based on the sensing data and the knowledge of expected world state, it identifies the environment the agent is running on. This process is mostly needed to distinguish running inside a virtual machine or inside a debugger to limit the adversary's ability to reverse engineer the agent.
- **Friend or foe identification:** Used mainly for identification and tagging of processes and files on the system. It is a prerequisite for offensive and defensive actions against adversaries as well as correct strategy planning.
- **Anomaly identification:** Used for detecting anomalies in data from the *Sensing* function. The baseline for anomaly identification is encoded into the world state. The detection can be rule- or pattern-based, or based on behavioral detection.
- **World State Update:** Transforms sensor data and data from environment identification and friend/foe identification into a world model and world dynamics update.

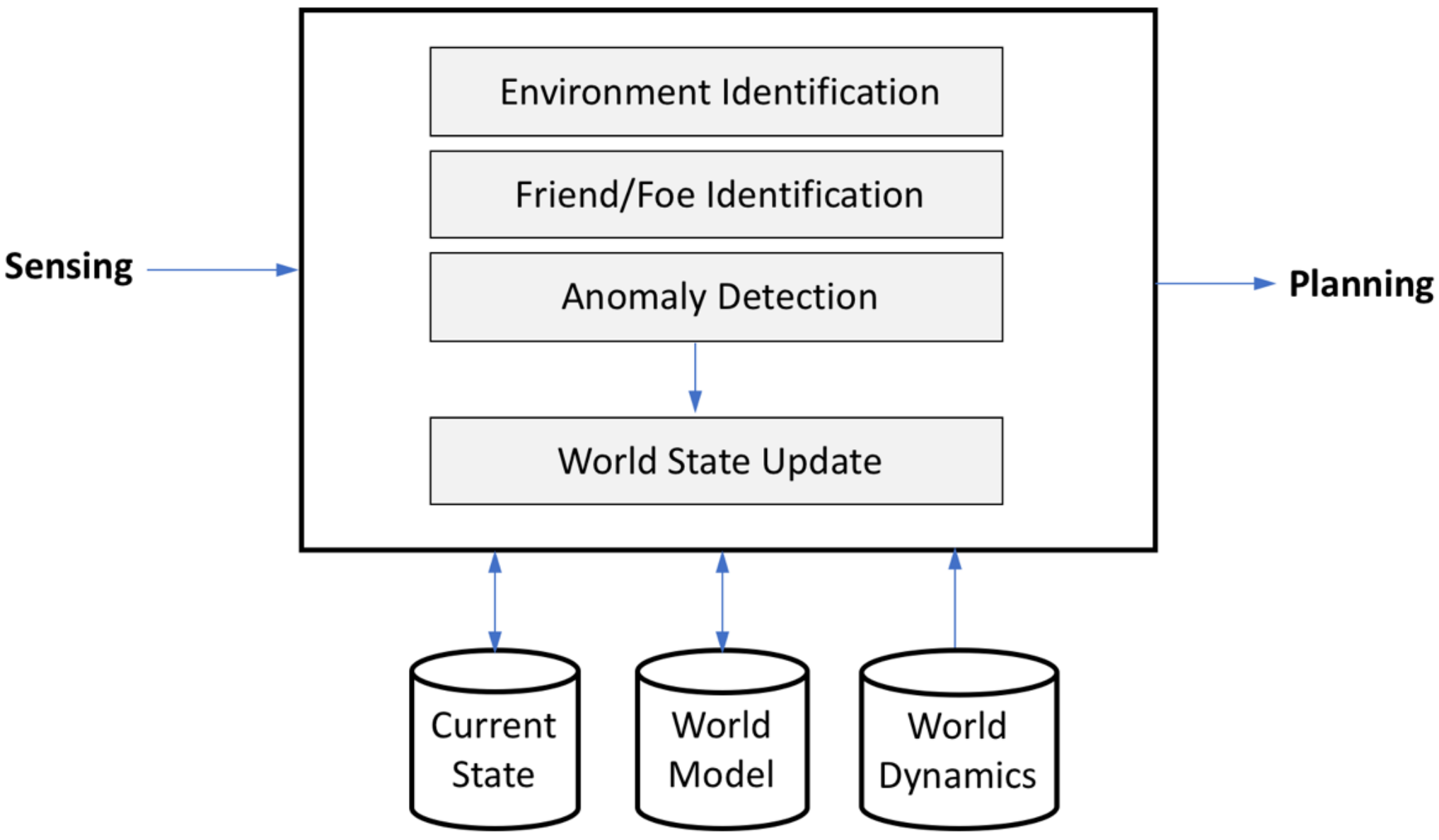

**Fig. 10** ***World State Identification*** **component**

The dataflow of the full-featured system includes the following:

- Updates on self and environment changes are received from the *Sensing* function.
- The world model database, the current state and history database and the world dynamics database are queried to have a baseline for sensing data processing.
- The environment identification component assesses any changes in the agent's environment.
- The friend/foe identification component identifies any potential adversaries and produces IoCs.
- The anomaly detection component estimates potentially anomalous behavior in sensing data and produces IoCs.
- Findings from the previous three components are combined with input sensing data and transformed to a world state update. This update is propagated to a world state database.

## 5.4 Anticipation of the Future World State

This section deals with world state anticipation. This is based on the knowledge of the actual state of the world and the knowledge of the laws of a regular behavior.

A world model is an abstraction of reality that provides a semantic meaning to perceived data. Its actual representation is strongly dependent on the implementation of the agent. In the optimal case, an agent is using data services to process, store, and employ sensory information transformed into the world model and world dynamics knowledge. These data services conform to the general description provided in Section 4 and are built with their own sets of constraints, which dictate their structure and capabilities.

In this section, we present a set of recommendations for a minimalistic world model and world dynamics structures, which are required for successful operation of an in-vehicle AICA. Given the large amount of data the agent could be processing and the number of different states the agent could be in, the following should be satisfied:

- The model should use features based on the properties of the machines and network, which are normal during non-anomalous operation. Provided that most Army systems have precisely defined operation parameters, establishing a model as a baseline should be attainable.
- The model should encode explicit IoCs.
- The goals of the agent should be expressible as a function of a world state.
- Both the current state and world dynamics are also highly dependent on the agent's implementation and the design decisions for the model. Nevertheless, given the expected operational parameters of the agent, we suggest that the model used for the world and the current state should contain the components listed in Table 3. The world dynamics knowledge should be computed on the fly from the world model and the current state and history.

**Table 3 Components of the world and current state and history models**

| Component | Model | Description |
|---|---|---|
| Flow database | Current state and history | Record of network flows, which can be augmented by full traffic traces where allowed by space constraints. |
| Log stash | Current state and history | Collection of system and application logs, preferably in a unified form suited for quick searching and analysis. |
| System metrics | Current state and history | Performance and operation characteristics of an agent and the system it is running on. |
| Whitelists | World model | Policies and baselines of normal behavior derived beforehand from the knowledge of the agent's environment. |
| Entity description | World model<br>Current state and history | Both the description of entities in the agent's proximity and their current operational status as viewed by an agent (e.g., probability of compromise). |

## 5.5 Use Case

To illustrate possible relations among the *Sensing* function, *World State Identification* function, and world state, we present a scenario where an AICA is deployed in a vehicle. In this scenario, malicious code was inserted during maintenance to the VMS and manifests on the battlefield, propagating to the BMS and then to the COMMS.

The use case timeline and events are as follow:

1) The VMS gets infected during maintenance.
    - Sensing (S): No information.
    - World state identification (I): No information.
    - World state (W): No change.
2) Malware activates and attempts to infiltrate the BMS.
    - S: Detected connection between the VMS and BMS.
    - I: Identified an anomalous connection and produced an IoC.
    - W: Updated with the IoC; the VMS and BMS are flagged as anomalously acting systems with potential to compromise.

3) BMS successfully compromised.

    - S: The BMS supervising process identifies an integrity violation and logs the information.
    - I: Logged information is transformed into an IoC.
    - W: Updated with the IoC; the BMS is flagged as a potentially compromised system with higher confidence.

4) Malware attempts to infiltrate COMMS.

    - S: Detected connection between the VMS and COMMS.
    - I: Identified an anomalous connection and produced an IoC.
    - W: Updated with the IoC; the VMS, and COMMS are flagged as anomalously acting systems with potential to compromise, the VMS with higher confidence.

5) COMMS successfully compromised.

    - S: No information.
    - I: No information.
    - W: No change.

6) The VMS is functionally affected.

    - S: Detected anomalies in vehicle responses.
    - I: Anomaly report is converted into an IoC.
    - W: Updated with the IoC; the VMS is flagged as compromised with the highest probability.

## 6. Planning and Action Selection

Authors: Benoît LeBlanc and Krzysztof Rzadca

### 6.1 Overview

We propose to decompose the part of the decision-making process that decides the actions to be performed into two components: the *Planning* function and *Action Selection* function.

The goal of *Planning* is to create a set of possible plans of actions that lead from the current world state to some interested future world states. As a new plan is computed, *Planning* sends it to *Action Selection*, in a semi-continuous process. Then, the *Planning* function continues to compute alternative plans and proposes them, one by one. Some of them are real new plans, some others are just new versions or adaptations of previous plans. Finally, it answers the question: "What could be done?"

The *Action Selection* receives continuously proposed plans of actions leading to some future world states of interest. Because of enemy actions or the world dynamics, there might be multiple future world states stemming from a single action in the first step. Aware of the goals of the agent, the *Action Selection* function choses an action plan that leads to the most desirable future world states and then sends it to the *Action Execution* component. Finally, *Action Selection* function answers the question: "What must be done?"

*Action Selection* may ask *Planning* for a more precise or a fitter plan if needed.

*Planning* produces plans and *Action Selection* is able to negotiate details of these plans in an iterative cycle between the two components. This is a major part of the decision-making process and it eventually produces a "to do list" of actions.

### 6.2 Planning

The *Planning* function has access to a database representing a repertoire of actions: a kind of a dictionary of all possible actions, including preconditions and prerequisites for each action. Two functions are used by *Planning* to implement a tree exploration. The first one is called function $f_a$. It is in charge of proposing actions. It maps the current world state (given by the world state identification [WSI] component) to a set of feasible actions (i.e., a subset of the discussed database). The second is called function $f_w$. It maps a world state and an action to a set of future world states (possibly with some information on the probability of individual states) (Fig. 11).

Planning

World State Identification

Current World State

Repertoire of actions

Proposed Plan

fa
-> Set of possible plans of actions

fw
-> Set of proposed plans of actions

Action Selection

Current State

World Model

World Dynamics

**Fig. 11** ***Planning*** **component**

Starting with the current state, *Planning* uses $f_a$ to produce a set of alternative actions. For instance, and greatly simplifying the situation, if the current state given by the WSI module is (*vehicle engaged in combat*; and *a system file with changed SHA-1 hash* and *previously unseen radio transmission detected*), the result of $f_a$, the set of feasible actions might be {no action, shut down COMM radio Y, shut down the entire computer system}.

Then, on each of these actions, *Planning* uses $f_w$, leading to a future world state. For instance, the previous world system state combined with the action "shut down COMM radio Y" may lead to the world state (*vehicle engaged in combat*; *a system file with changed SHA-1 hash* and *weapon system Z malfunction*). The process of invoking $f_a$ and $f_w$ is continued, resulting in a tree; a leaf of this tree is a set of future world states (or a probabilistic distribution over this set).

When a path in the tree is determinate, it leads to a transmission to the *Action Selection* component of a plan such as "(#plan, (action, state), (action, state), etc.)".

Functions $f_a$ and $f_w$ can be implemented by trained neural networks or rule-based systems. Both should be precomputed and implemented in the system. Updates could be done during the vehicle overhaul. If the agent has a *Learning* component (see Section 9 about learning), then this function can be updated continuously based on learning from experience.

Function $f_a$ is a service using the database "actions and effects": receiving the world state and returning a set of actions. Function $f_w$ is a service of world dynamics knowledge, but we stress that $f_w$ must consider not only the internal evolution of the world, but also the effects of a concrete action.

The *Planning* function's algorithm effectively builds a complete search tree of the future actions and world states, and we acknowledge that such a tree probably must be pruned because of a possibly exponential search space.

An alternative to building a tree is to use a trained neural network directly. The input to the network is the current world state; the network produces future world states. Thus, the network implicitly implements $f_a$ and $f_w$.

After all, given basic rules (such as "avoid RF propagation" or "keep the initiative"), given criticality analysis of the assets, given a topology (which facilitates circulation of information) and security architecture (which curbs circulation of information), given resources that can be mobilized (such as a sandbox, a file cleaning tool), and given possible enemy tactical movements (such as expected expectations of the opponent), *Planning* combines consistency preservation and combinations of resources to produce possible plans of actions. Such plans are lists of proposed actions representing options of several sets of things to do.

## 6.3 Action Selection

Based on the current world state, *Action Selection* decides whether the situation is urgent, for instance, when the vehicle is engaged in combat. In the urgent decision-making mode, one of the first action plans suggested by *Planning* is chosen; otherwise, *Action Selection* may wait for a longer time for *Planning* to send more actions plans.

The *Action Selection* function chooses an action plan based on how the predicted future world states match with the goals of the mission. A goal can be expressed through a function of the (future) world state, mapping the state into the degree this particular goal is fulfilled. We acknowledge that there might be multiple goals and that their relative importance might change depending on the current state of the mission. For instance, one goal might be to maintain the information integrity of the vehicle, another to keep the crew as safe as possible, and yet another to achieve the mission's principal, tactical objective.

The *Action Selection* (Fig. 12) works to propose a multicriteria analysis of proposed plans of actions. Efficacy, rapidity, and assumed risks are the main criteria that must be considered. Policies, expressed in term of rules or goals, lead the choice of actions.

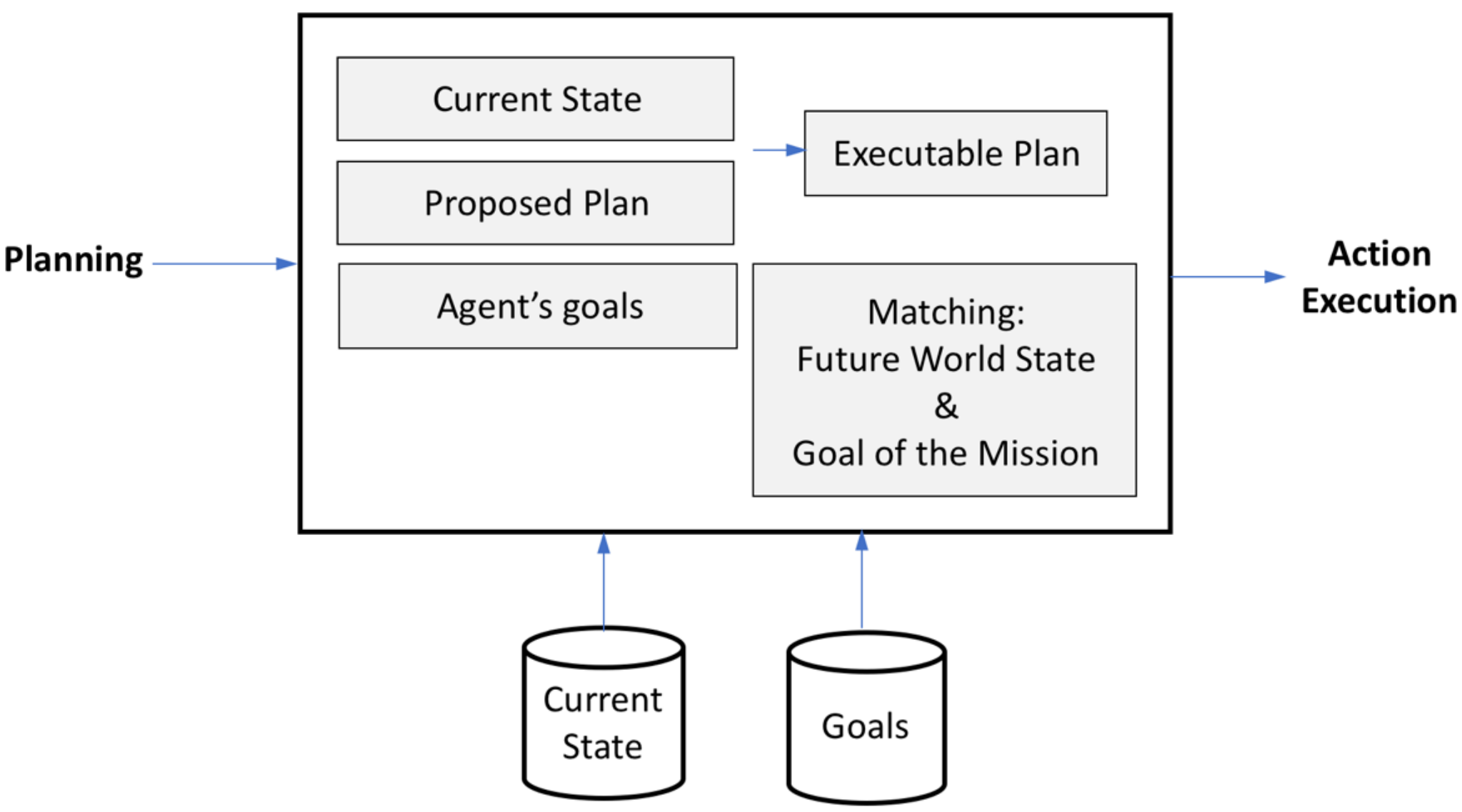

**Fig. 12** ***Action Selection*** **component**

The *Action Selection* component picks an action plan and then sends it to execution, beginning with the first action of the plan. *Action Selection* expects that, in the future, *Planning* will send an updated action plan, taking into account the chosen action and the actual change observed in the world state. *Action Selection* must inform *Planning* that it has chosen an action and sent it for execution. As the world state is then no longer valid, *Planning* should stop generating new action plans.

We assume that *Action Selection* can provide its analysis plan to *Planning*, for clarification, refinement, or modification of its own work. This does not lead to an infinite loop; it just envisages the case in which a complement is needed.

An alternative design is that *Action Selection* chooses a plan, and then sends autonomously the actions to be executed, without consulting *Planning* (or alternatively it can interrupt a plan with another unrelated plan, proposed in the future by *Planning*). However, this solution does not allow for refinement of a plan by *Planning*: for instance, an action might lead to three different world states; when the *Action Execution* component has executed an action, *Planning* can refine the plan based on the actual observed new world state.

## 6.4 Example

We present here an example of the operation of the *Planning* and *Action Selection* functions in the context of a military vehicle in which several machines are connected by an internal network. We consider three different situations.

### 6.4.1 A Common Cyberattack

Here a common cyberattack is defined as a known attack that, with high probability, would not lead to negative effects. For instance, it is an attack that tries to use a known bug that is patched in the version of the OS used by all the machines in the vehicle. Furthermore, we assume that there is no time urgency—for instance, the vehicle is parked at the base.

The world state catches the symptoms of the attack by the network sensors (e.g., flow of data or connection attempts to a certain port). The *Planning* function constructs an action plan leading to a future world state in which the attack is attributed. The plan starts with planting false information on the machine (e.g., the first action is to create a mock-up password file and another is to create a file with a name suggesting classified content). In a future world state, after creating a mock-up password file, this password file is either accessed or not. If there is an access, the next action is to generate more mock-up password files. If the file is ignored, the next action is to generate a file that pretends to contain classified information. However, the *Planning* function also proposes other action plans unrelated to the current attack (such as proposing actions executing orders from the C2) or doing nothing.

As the vehicle is parked, the goal of attributing the attack is the most important. Thus, *Action Selection* choses the action "small mock-up information" and send it to be executed (i.e., to generate a mock-up password file).

### 6.4.2 An Unexpected Cyberattack

An unexpected cyberattack is detected through its results, rather than by intercepting the attack as it happens. For instance, a routine file system check may detect a changed hash value of a system file. There might be also time urgency: the vehicle might be engaged in active combat. As such a situation is a threat to the integrity of the system, *Planning* and *Action Selection* must act quickly. The *Planning* component suggests a short plan of, for example, restoring the changed file from a backup; *Action Selection* chooses this action based on the goal of maintaining integrity.

### 6.4.3 Cyber Exploration

If a vehicle is parked and the world state model does not detect an attack, AICA is in a kind of "exercise situation". Depending on what network is protected by AICA, it will be possible or not to activate some deliberated actions and observe regular results on agent and/or on environment. This use case is certainly inappropriate in most part of military situations, but it could be considered in specific cases as an intermediary mode between sandboxes and real-life activities. In this cyber-exploration opportunity, the *Planning* and *Action Selection* components might use the chance to provide new data for the learning module collected in real and controlled situation. As the number of monitored characteristics of the world are vast, one of the important goals of the system is to be able to automatically distinguish a rare threat from a large number of normal, acceptable states. Similarly, given a large number of possible actions (closing a communication port, restoring a file, creating a file, etc.), the system must be able to learn the effects of the intended consequences of actions (e.g., closing TCP port 12345 would shut down the internal communication system XYZ).

During the cyber-exploration scenario, *Planning* would create action plans consisting of steps of basic actions (close TCP port 12345, open TCP port 12345) and observe their effects on the integrity of the vehicle.

# 7. Action Execution

Authors: Fabio De Gaspari, Luigi Mancini, and Agostino Panico

## 7.1 Purpose

The overall purpose of the *Action Execution* component is to execute the executable response plan of actions that the *Planning* and *Action Selection* components have chosen according to the mission goals and the situation. *Action Execution* works as an actuator that provides the following functionalities: 1) action activation, 2) execution monitoring, 3) effects monitoring, and 4) execution adjustment. The architecture overview of the *Action Execution* component is shown in Fig. 13. The *Action Execution* component has administrative privileges to execute the actions, and it should be able to perform all the actions required to accomplish the typical tasks of a system administrator, including the security analysis of the system. To guarantee complete execution of the actions, the *Action Execution* component should run only atomic actions, either all the operations are completed or nothing occurs. To this end protocols such as OpenC2 (OpenC2 Forum 2015), the de-facto standard for C2 functions, can be used by the agent to communicate actions to the actuators.

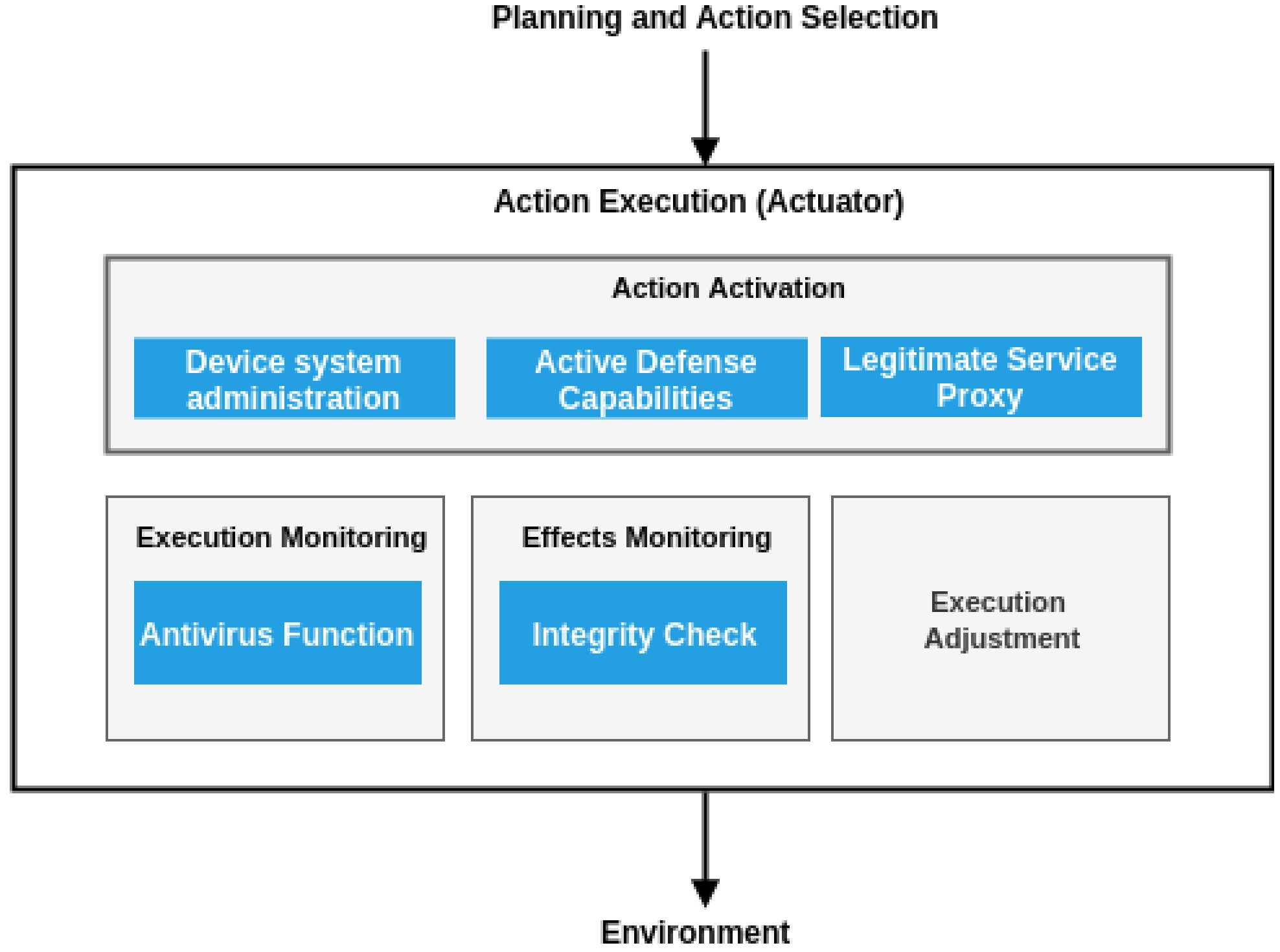

**Fig. 13 Overview of the *Action Execution* functionalities**

The *Action Execution* component is also connected with two other components: *Sensing and World State Identification* and *Learning*. As an actuator, the *Action Execution* component executes actions that produce valuable data that can be sensed by *Sensing and World State Identification.* Consider for instance the scenario when *Sensing and World State Identification* detects an anomalous situation and requires the execution of a customized antivirus function to perform a detailed verification of the current anomalous behavior of the system. In this case, *Action Execution* should be able to run a customized antivirus functionality, which will generate data that can allow the *Sensing and World State Identification* component to identify risky current world state patterns. In addition, the *Action Execution* component continuously updates the internal rules and conditions with the feedback provided by the *Learning* component.

In the following sections, we describe the scope and conditions for each of the functionalities of the *Action Execution* component.

### 7.1.1 Action Activation

Action activation takes as input an executable response plan from *Planning and Action Selection* component and environment's technical configuration and then proceeds with the execution of the planned actions according to the scheduled order. This function outputs the response of the execution, which can be a message that confirms the successful execution or provides some details about the reason why the operation failed (e.g., if an action "Delete a file" fails, then the agent should provide to other agents some details like "The file cannot be deleted. The requested file does not exist." or "The file cannot be deleted. The requested file is protected."). In order to guarantee a secure execution of the actions, the action activation function should have the capability to perform five main types of actions: 1) device system administration, 2) customize the antivirus function, 3) integrity check, 4) active defense actions, and 5) legitimate service proxy.

### 7.1.2 Device System Administration

Device system administration deals with normal system operations, incident handling, and root cause analysis (RCA). Overall, device system administration includes a set of actions that can be summarized as follows:

- Install/remove software application
- Software update
- Registry modification
- User management

- Log access
- Baseline creation and periodic check sent to sensors

Based on the feedback that derives from the *Learning* component, device system administration should be able to dynamically integrate new rules for each of the aforementioned set of actions.

### 7.1.3 Antivirus Function

The *Action Execution* component should cover the antivirus function. This means that this component should be able to behave as an antivirus software, perform analysis, and not impact system functionality. For instance, this function should be able to execute the action "perform a full scan". The antivirus actions that the actuator should perform are the following:

- Executable analysis
- Complete device scan
- Basic malware analysis heuristics

The deployment of this set of actions in the actuator aims to enable the antivirus functionality of the machine and reduce the installation of antivirus software on the device itself. This means that when the device does not use an endpoint protection solution, the agent should be still able to guarantee the defense of the device. However, in the case when the device uses an endpoint protection solution, then the agent should be able to communicate and interact with this solution to take the necessary steps to quarantine, delete, or report the infected items. In this case, the antivirus function serves as a sensing function.

### 7.1.4 Integrity Check

Integrity check function evaluates the changes of the machine's state by periodically checking the stability and integrity of critical files that must not be changed without proper authorization. Depending on the configuration of the integrity check and the need of the action selection, the *Action Execution* component should check the integrity of the file system against both a whitelist and a blacklist. Since the integrity check is an action, it should be executed by an actuator, which then sends the data to the *Sensing and World State Identification* to perform further analysis.

### 7.1.5 Active Defense Capabilities

Active defense is a popular defense technique based on systems that hinder an attacker's progress by design, rather than reactively responding to an attack only after its detection. Since the goal of active defense systems is to reduce the risk of a compromised system, in some cases, active defense can be used as a measure against lateral movements. Note that the purpose of active defense is not to defend or prevent the attacker from performing some actions. Instead, its goal is to slow the attacker down and allow optimal operation of the traditional defense systems. To this end, active defense tools can be integrated with preexisting security information and event management (SIEM) systems. This feedback allows response teams, or the autonomous agent in our case, to refine detection criteria for traditional security systems, as well as provides useful intelligence on how to react to the ongoing attack.

The *Action Execution* component should be able to implement active defense capabilities to have the ability to perform annoyance, attribution, and, under some circumstances, even attack. The range of active defense actions can be described as follows:

- Port remapping
- Fake files
- Fake services and network port
- Fake web services
- Fake supervisory control and data acquisition (SCADA) services
- Attribution capabilities
- Building a covert communication channel

The deployment of this set of actions enables a machine to act and react to the actions of an attacker, or an abnormal behavior of a legitimate user, by slowing the adversary down with annoyance and attribution, and eventually, attack.

### 7.1.6 Legitimate Services Proxy

The legitimate services proxy should be implemented according to the active defense capabilities of *Action Execution* and should be able to proxy any legitimate service of the host machine. The idea of legitimate services proxy is to use the actuator as a frontend interface toward the external environment, and then perform a security analysis of the incoming and outgoing traffic through the proxy. To support flexible active defense strategies, every legitimate port of the host machine should be bound to a port of the actuator, so that such port could be redirected to another port according to the active defense objectives. In other words, the proxy function should be able to expose a legitimate service to a nonstandard port without modifying the host machine.

### 7.1.7 Execution Monitoring

Execution monitoring function aims to observe the real-time execution of action activation. This function should be able to check the execution traces triggered by a given input and should have prior knowledge regarding the expected legitimate state of the software running during *Action Execution*. For instance, execution monitoring can trace logical statements (invariants) to identify properties of a running software, which can be helpful to check the legitimate execution traces.

Execution tracing is also important to identify the cases when the task action activation is not completed successfully. In this case, execution monitoring will activate execution adjustment to adjust the action implementation.

### 7.1.8 Effects Monitoring

Effects monitoring function should be able to monitor the environmental changes caused as result of *Action Execution*. In particular, effects monitoring should monitor data, access control, and the integrity of the components that are expected to be effected from *Action Execution*. Note that some of these functionalities also be covered also by the *Sensing and World State Identification* component. However, the effects monitoring aims to monitor only the effects of *Action Execution* and provide a very detailed analysis of particular software components, which may not be always detected by the general execution of *Sensing and World State Identification* for the entire system**.**

### 7.1.9 Execution Adjustment

Execution adjustment should be able to handle the cases when the execution of a plan of actions produces security-critical effects. In this case, this function should have the ability to adjust the action implementation to the environmental setting. When the adjustment is not possible (e.g., due to technical or security reasons), execution adjustment function will trigger the *Collaboration and Negotiation* component in order to interact with other agents, C2, or human operators for agreeing on changing the plan of actions.

## 7.2 Use Cases

In this section, we discuss some use cases, providing concrete examples of how the *Action Execution* component works in realistic scenarios.

### 7.2.1 Anomalous Behavior of a Military Vehicle

This attack scenario (Fig. 14) considers a compromised device that tries to probe the environment for information gathering. This behavior can result in detection by a number of active defense tools, providing early indication of compromise. For instance, the compromised device might interact with fake services exposed by another component or might access fake files triggering immediate detection. Upon detection, neighbor agents react based on the intelligence provided, deploying appropriate active defense tools and reconfiguring classic security tools based on the behavior of the compromised component. For example, the neighbor agents can remap service ports or launch new fake services that are interesting for the attacker in order to profile them. At the same time, the neighbor agents perform a scan on their local file system to check for suspicious executables and share the findings among themselves to build a real-time IoC. This operation involves a set of actions that should be executed on the infected device with administrator privileges and aims to restore the infected machine back to the normal state.

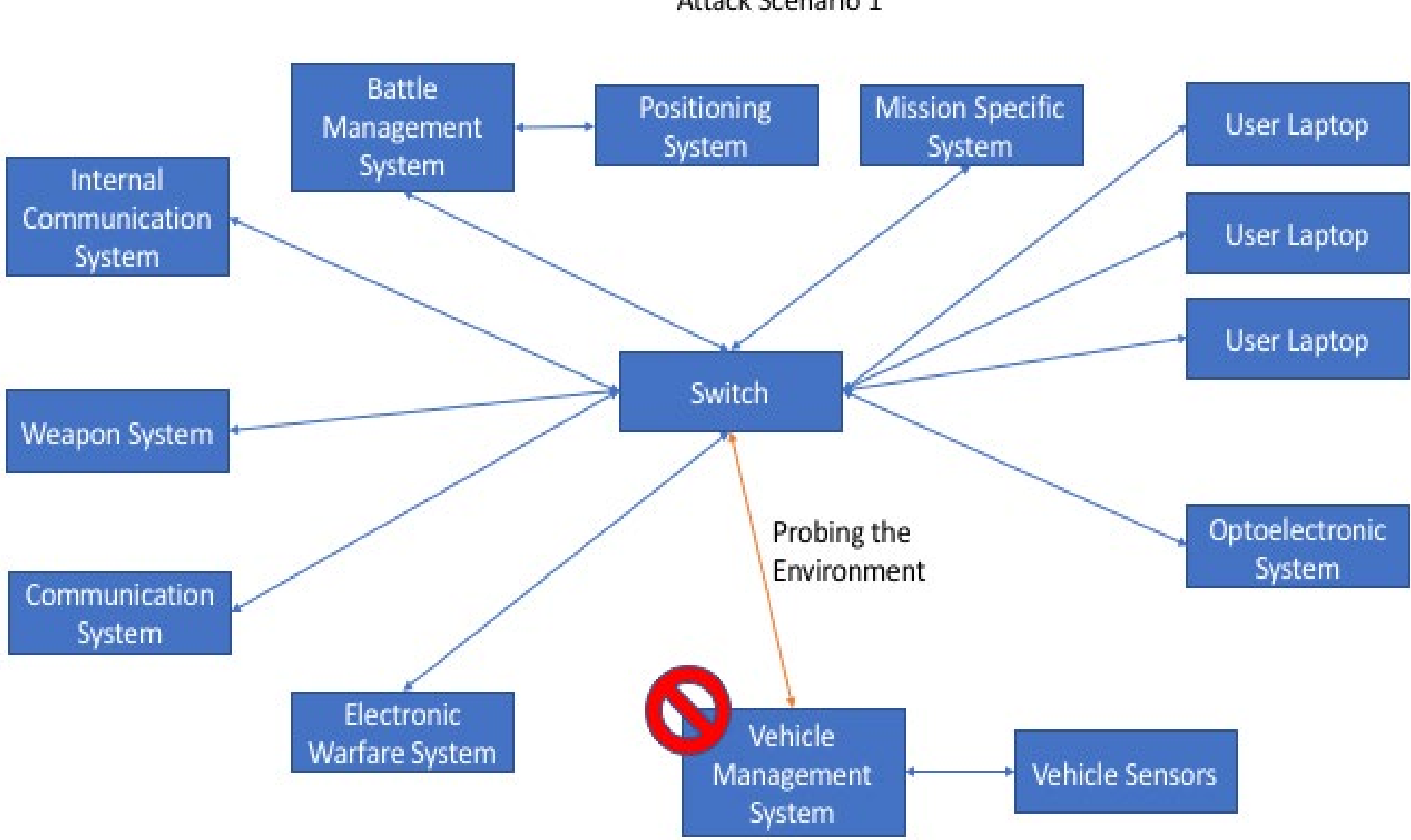

**Fig. 14** **Anomalous behavior of a military vehicle scenario**

With active defense capabilities, agents are able to perform early compromise detection based on abnormal behavior, reducing the risk of being persistently compromised by the attacker. Furthermore, deception techniques slow down ongoing attacks, providing the agent with time to understand the attacker's goal and devise a defense strategy.

### 7.2.2 Battle Management System, Vehicle Management System, and Communication System Compromised

In this scenario (Fig. 15), an agent that detects a compromised component creates a covert channel with other noncompromised agents, allowing them to communicate without being detected by the compromised devices. The goal of this communication is to alert the other agents of the network not to trust the data that are coming from the compromised vehicle, avoid the transmission of sensitive data to such vehicle, and agree on a plan to recover the compromised devices. Noncompromised agents can also use the covert channel to define a service remapping strategy and potentially which fake services to expose in their stead, as well as start a general integrity check to verify that no other systems are compromised.

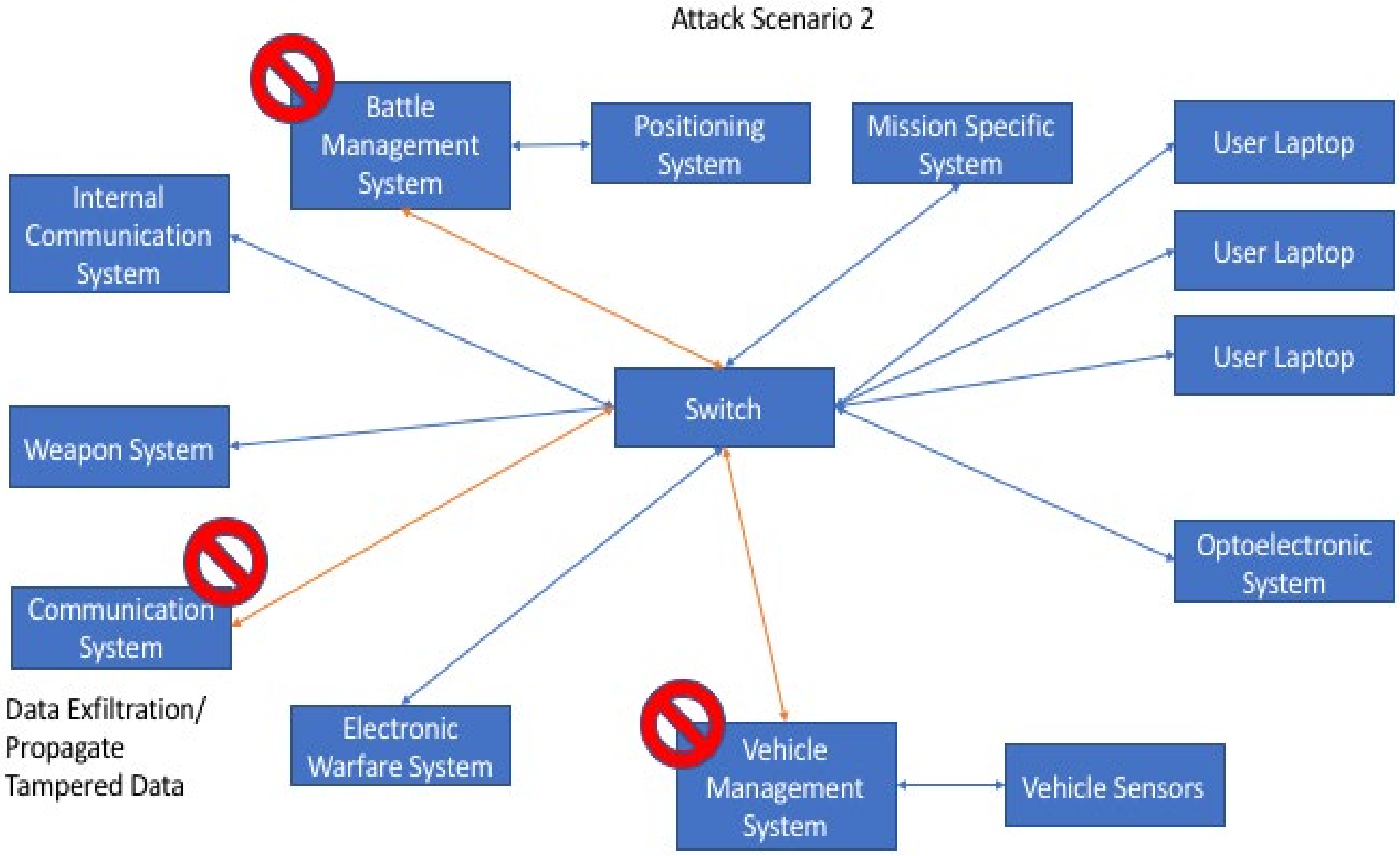

**Fig. 15 BMS, VMS, and COMMS compromised scenario**

# 8. Collaboration and Negotiation

Authors: Edlira Dushku and Luigi V Mancini

## 8.1 Overall Purpose

Battlefield operations are characterized by an unreliable communications infrastructure, limited network coverage and also the presence of enemy forces that intend to compromise these operations. Considering these limitations, an intelligent agent that operates in a battlefield environment should be able to plan its own actions and possibly perform them in an autonomous way. However, under some conditions, a group of autonomous intelligent agents may need to collectively decide a joint plan of actions that solves a set of common goals. In this context, the collaborative agent model emerges as an effective approach that allows autonomous agents to collaborate and negotiate among themselves to accomplish their mission-critical goals and confront adversarial actions.

In the collaborative model of AICA, an agent can individually perform one or multiple tasks and also choose to cooperate with other agents to perform coordinated actions. Different from multiagent systems that aim to solve problems that are difficult or impossible for an individual agent to solve, in the collaborative agent system of AICA, each individual agent should be able to solve the problem autonomously and only start to collaborate with other agents to extend the individual capacities of *World State Identification*, *Planning*, or *Action Selection*. In general, the interoperation between autonomous intelligent agents in AICA intends to improve the active defense capabilities of the contested battlefields by enabling a collaborative decision-making process and improving the goal execution capabilities of individual agents.

Agent interoperability in AICA is enabled by the *Collaboration and Negotiation* component, which coordinates the interactions agent–agent, agent–C2, and agent–human. In AICA, the *Collaboration and Negotiation* component can be initialized by one of the following components: *World State Identification*, *Planning*, or *Action Selection.* Overall, the *Collaboration and Negotiation* component consists of three functions:

1) **Collaboration:** The collaboration function allows an individual agent A to interact with other agents (or C2 or human operators) to make agent A's plan of actions more effective or solve a task that is beyond agent A's capabilities. The sensitive information that agent A perceives about the world state should remain local. The agents involved in a collaboration process should be able to exchange only the relevant information that is required for the collaboration.

2) **Negotiation:** The goal of the negotiation is to reach an agreement within a set of agents regarding a goal or a plan execution. During the negotiating process, agents agree on performing some tasks that are beyond the capabilities of the individual agents of the friendly forces. Most importantly, the agents agree on coordinating their plan of actions to reach a common goal.

3) **Agreement:** The agreement defines the conditions that the agents agreed during the negotiation procedure. The agreement first registers the plan of actions that the agents agreed during the negotiation and then returns a response to the AICA components that triggered the initialization of the *Collaboration and Negotiation* component about an updated action that should be executed.

## 8.2 Architecture of the Collaboration and Negotiation Component

The *Collaboration and Negotiation* function (Fig. 16) in the agent's structure should provide these fundamental services: 1) agent inquiry and discovery, 2) name discovery, 3) authentication, and 4) service and capacity discovery (SCD).

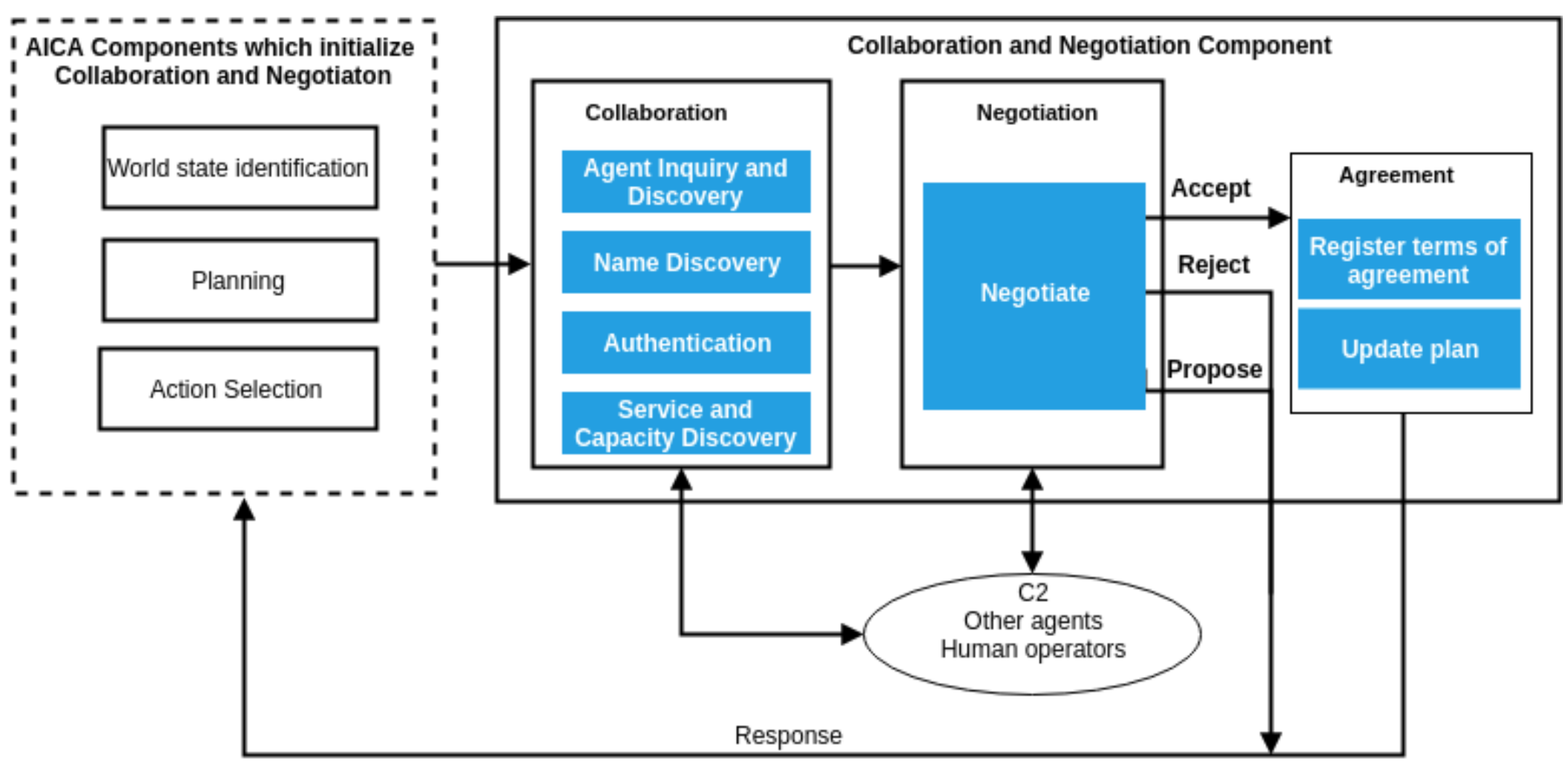

**Fig. 16** **Architecture overview of the *Collaboration and Negotiation* component**

### 8.2.1 Agent Inquiry and Discovery

Agent inquiry and discovery is a procedure that allows an agent to be discovered by friendly forces. When a new agent joins the network, its presence can be detected by other agents and they can start collaborating. To guarantee a stealth communication in under-attack situations, agent inquiry and discovery should allow agents to enable a covert channel communication among themselves. In this case, the data conveyed by the covert channel should be encrypted to prevent revealing of secret or sensitive information to unauthorized entities. Optionally, under high-risk conditions, an agent can use this service to make a choice whether to be discoverable in the network or not. Likewise, when severe attacks are detected on the battlefield, the C2 unit can call this service to make agents undiscoverable from other agents.

### 8.2.2 Name Discovery

A procedure for retrieving the name of a connectable agent. Friendly forces should share a common name taxonomy and should have some pre-shared cryptographic keys. The name of the agent should be connected to some configurations that agents know about each other. The agent process should be resilient to a Sybil attack. The agents that participate in the decision-making process should have the identity of the friendly forces. The enemy should not be able to influence the common goals.

### 8.2.3 Authentication

The collaboration function must enforce the authentication of the agents and protect the confidentiality and the integrity of the communications between agents by supporting standards such as AES-256, TCG Opal, and so on. The authentication procedure should comprise the required security mechanisms that should be applied when an agent initiates a collaboration request to a remote agent and when an agent receives a service collaboration from a remote agent.

The trust establishment process is a prerequisite of the authentication procedure.

### 8.2.4 Service and Capacity Discovery

SCD involves a set of procedures for querying and browsing the services offered by or through another agent. SCD does not define methods for accessing services; once services are discovered with SCD, they can be accessed in various ways, depending upon the service. After communication between two agents is established, they start exchanging information and computation. They also declare their capacities (memory, storage, CPU), which is very important in the later decision of allocating tasks to other agents.

The negotiation function provides the negotiate service, which returns as output 1) accept, 2) reject, and 3) propose. The agreement function consists of two services: 1) register terms of agreement and 2) update plan.

### 8.2.5 Collaborative Planning

Each agent has planning capabilities and can autonomously execute its local plan. An agent can extend its own capabilities by interacting with other agents to collaboratively construct a joint plan to accomplish their common mission goals.

When there is a new task/goal that the agent should achieve, the agent can choose to do the following:

1) Fulfill the task autonomously. In this case, the agent does not communicate with other agents and does not influence the goals of the other agents.

2) Distribute information about the task among the agents to reach a common plan of actions. This case requires several interactions among agents until they reach a common plan of actions. Since agents differ in capabilities and knowledge, they have different views regarding the task that should be fulfilled. During the interoperation, the data that an agent make accessible to other agents should present only the relevant information that is required for the collaborative planning and should not reveal sensitive information of the agent. This is important because an adversary, which could take control over an agent, should not be able to gain access to the sensitive information of other agents.

### 8.2.6 Communication Protocols

Agents can exchange information by using the following application protocols:

1) client-server:
   Simple Object Access Protocol (SOAP), Restful HTTP/Constrained Application Protocol (COAP)

2) publish-subscribe
   Message Queuing Telemetry Transport (MQTT), Advanced Message Queuing Protocol (AMQP), Requested Power To Send (RPTS)

Agents should use protocols that guarantee the confidentiality and integrity of communications, for example, the basic security protocols such as Transport Layer Security (TLS)/Datagram Transport Layer Security (DTLS).

## 8.3 Collaboration Process

The *Collaboration and Negotiation* component allows an agent to discover other agents in the network and start to collaborate with them. Figure 17 captures the process flow of the activities in the *Collaboration and Negotiation* component while interacting with the *Planning* component.

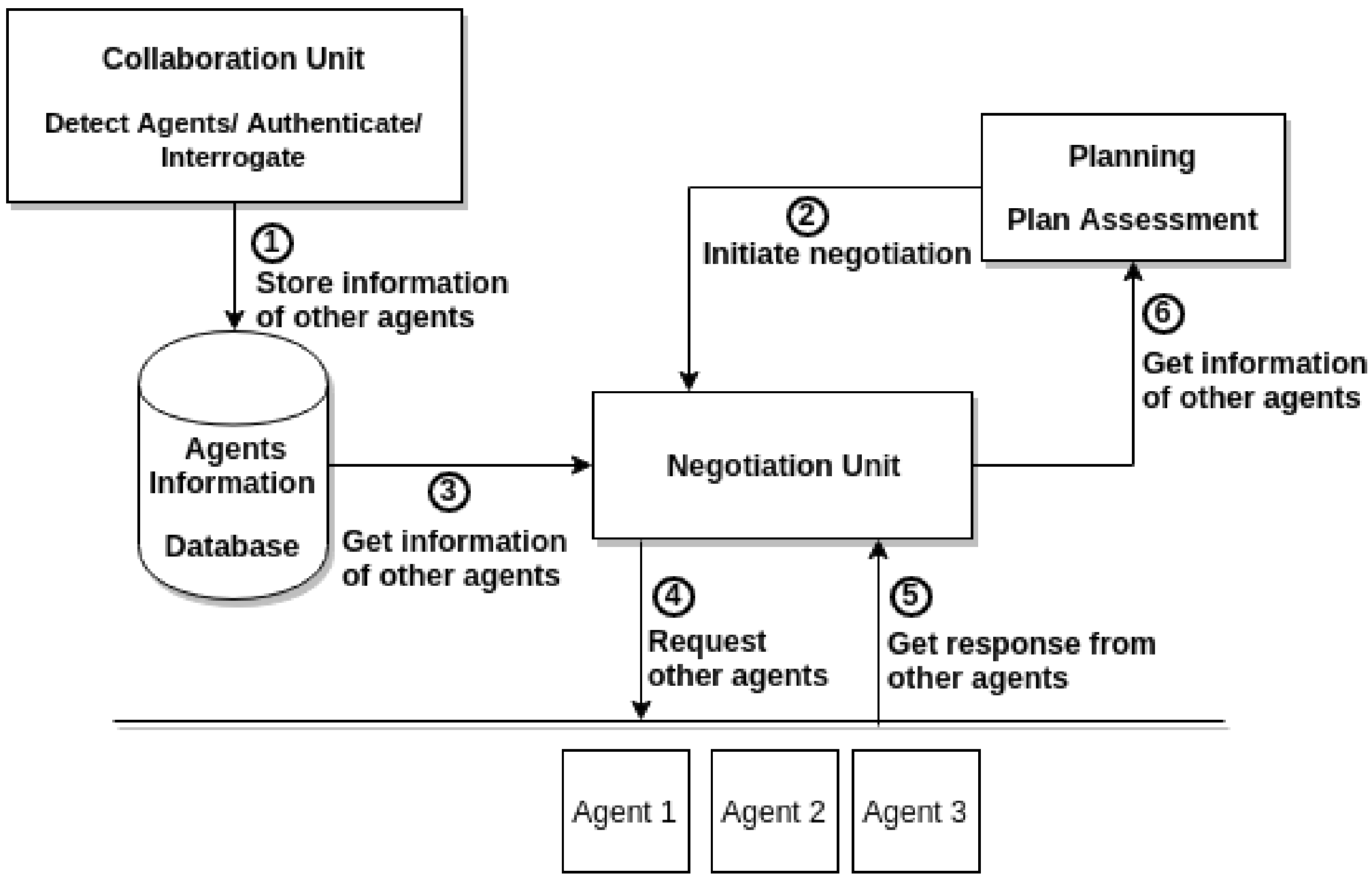

**Fig. 17 Process flow in the *Collaboration and Negotiation* component**

The process start with an agent that discovers other agents in the network. The collaboration between agents starts with a P2P authentication process. After the authentication, neighboring agents interrogate among each other about the services and the capacities that they offer. Each agents saves in a storage all the information related to the other agents, as shown in step 1 in Fig. 17.

The negotiation among agents is instantiated by the *Planning* component. When the planned action is a complex task that requires more resource capacities than the autonomous agent can handle or the planned action affects the common plan of actions, then the planning unit decides to negotiate the plan of actions with other agents (step 2). The negotiation unit retrieves all the agents' information from the database (step 3) and then, based on the services and the resources that each agent offers, the negotiation unit requests the agent that satisfy the requirements of the planned action that should be executed (step 4). Obviously, reasoning which agents can work on a planned action is a crucial factor for an effective collaboration among autonomous agents.

After sending the negotiation requests to some agents, the negotiation unit handles the responses that come from these agents (step 5) and then forwards them to the planning unit (step 6) for elaborating the next step of the plan execution.

## 8.4 Use Cases

1) An Agent A coordinates with C2:
    a) Agent A detects a condition where C2's decision is needed (e.g., Application X1 in the sandbox behaves suspiciously).
    b) Agent A sends a question to C2 (e.g., should I delete/kill application X1? Or else?).
    c) C2 may reply or not.
        i) If C2 sends an authenticated reply to the agent, then Agent A performs the following steps:
            1) The agent receives the command sent from C2 (e.g., uninstall the application).
            2) The agent checks the feasibility of the execution (e.g., checks for permissions).
            3) The agent responds to C2 (e.g., "will do" or "cannot do, explain why").
            4) If "will do", Agent A generates plan and executes the plan (e.g., agent has to make a plan that checking all the dependencies, if there are some necessary services that are critical to be deleted, generate concealment plan).
            5) The agent sends the resulting state to C2 (e.g., respond "success" or "action failed, explain why").
        ii) If C2 does not give a response, then Agent A performs a local decision (e.g., agent decides what to do).

The flowchart of the interactions between Agent A and C2 is depicted in Fig. 18.

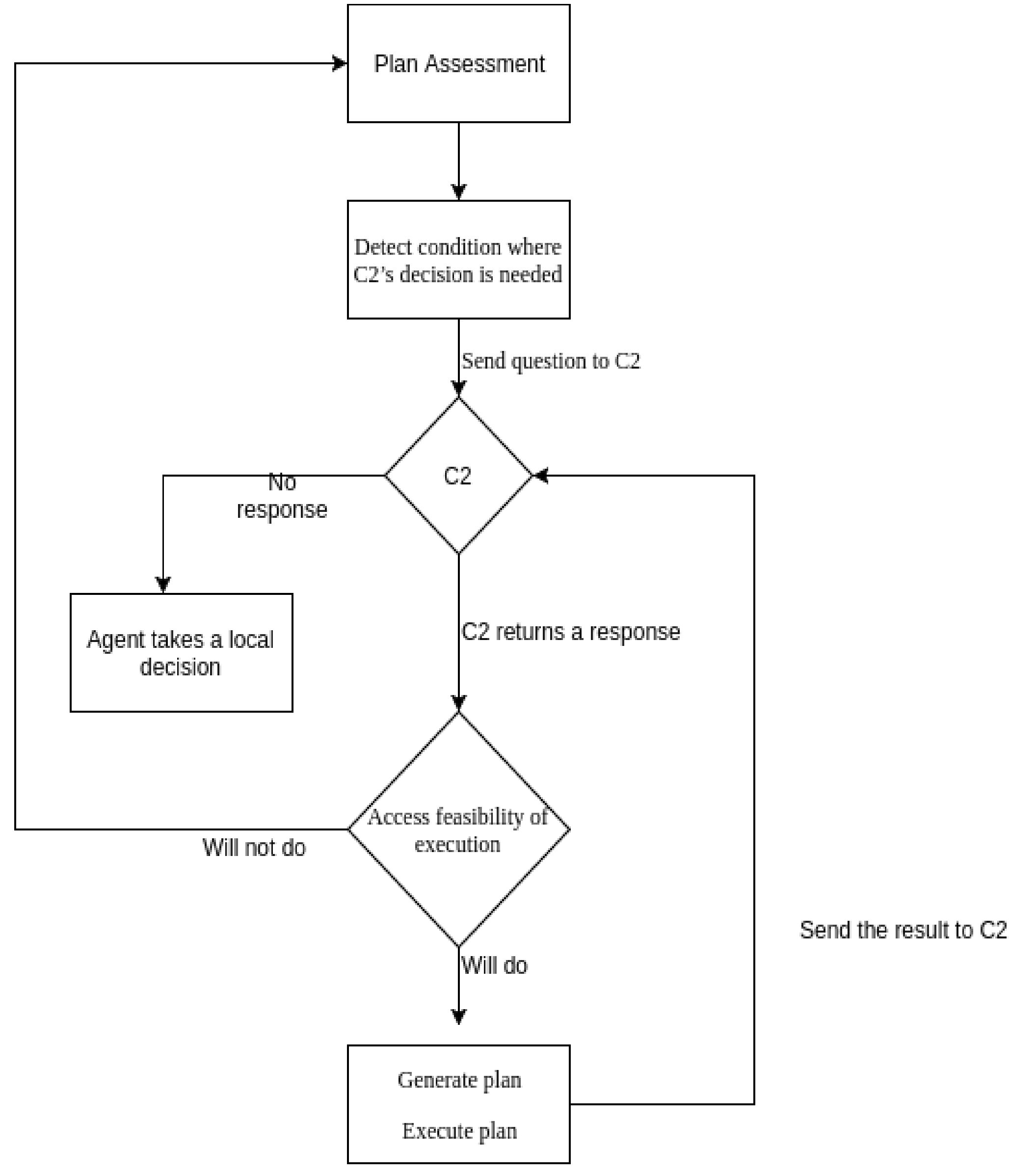

**Fig. 18 Flowchart of interaction Agent–C2**

2) Agent A collaborates with other agents:

   a) An agent communicates with other agents to improve the common plan of actions. For instance, if an Agent A identifies a malicious behavior, Agent A notifies other agents and agree on changing their individual plans.

      i) Agent A identifies anomalous traffic caused from a malicious service S1.

      ii) Agent A detects the presence of Agent B nearby.

iii) Agents A and B establish communication and authentication, and declare services.

iv) Agent A notice that Agent B provides the same service S1.

v) Agent A notifies Agent B about the risk.

vi) Agent B gets the alert from Agent A.

vii) Agent B may perform one of the following actions:

1) Agrees to kill immediately service S1 that is running on the machine and evaluates again its local plan of actions considering the non-availability of S1.

2) Completes the execution of the current plan and then kills S1.

3) Agrees to kill S1 if Agent A accepts to perform one of the action plan that Agent B must do.

4) Ignores the alert sent from A and continues its local plan.

b) In a similar way as the scenario explained previously, Agent A communicates with other agents to extend its local capacities in executing its individual plan of actions. For example, if a task needs to be executed and the resources are beyond the capacities of a single agent, then the task can be scaled to a group of agents:

i) Agent A realizes that the execution of an action X2 is taking a lot of time.

ii) Agent A detects the presence of Agent B nearby.

iii) Agents A and B establish communication and authentication, and declare services and capacities.

iv) Agent A notice that Agent B has the required capacities to perform the same action as A is running.

v) Agent A requests the Agent B to perform the action X2.

vi) Agent B gets the request from Agent A.

vii) Agent B may perform one of the following actions:

1) Agrees to immediately run X2.

2) Completes the execution of the current plan and then executes X2.
3) Rejects the request.

## 8.5 Conclusions

We presented the collaborative model of AICA and summarized the main properties of the *Collaboration and Negotiation* component. We emphasize that each AICA should be able to perform their individual plan of actions autonomously and only start to collaborate with other agents to extend their individual capabilities and improve the common plan of actions. The current version of the AICARA describes the functionality of the collaboration component, while the future work will focus on providing details of the operation of the negotiation component and elaborating the functionality of agreement for the communication among many autonomous agents.

Furthermore, the future works includes designing a secure collaborative model of AICARA. In particular, a trusted collaboration between autonomous intelligent agents will require further discussion regarding trust establishment between autonomous agents in battlefields. In this context, it is crucial to exploit the key management protocols and trustworthy information delivery in other domains to gain insight into the most relevant pre-shared key schemes that are suitable for heterogeneous battlefield environments, as well as communication protocols that could assure the confidentiality and integrity of the exchanged messages.

# 9. Learning

Authors: Alexander Kott and Alessandro Guarino

The environment in which an agent operates can change rapidly, especially (but not exclusively) due to enemy action. In addition, the enemy malware, its capabilities, and TTPs could evolve rapidly. Therefore, the agent must be capable of autonomous learning that could help it adapt to a changing environment and enemy. Numerous approaches to learning and purposes of learning are possible. In this section, we offer merely a few illustrative, highly simplified sketches of how the agent's learning could be implemented and for what purposes learning might be used. The discussion here is heavily influenced by the concepts of Reinforcement Learning (RL; Sutton et al. 1998) and Partially Observable Markov Decision Process (POMDP) formalism, but is arranged so that the reader does not need to be familiar with RL or POMDP.

The reasoning capabilities (such as planning, prediction of effects, state identification, etc.) of the agent rely on its knowledge (which could include various models such as world state model, etc.). The purpose of the learning function(s) is to modify the knowledge of the agent in a way that enhances the success of the agent's actions. The success of an agent, or in other words its level of performance, will be measured as the distance from the goal, in some sense. This of course implies close collaboration with the WSI function and access to the world state, world dynamics, and goals databases.

The agent learns from its experiences. These experiences could be acquired when the agent engages in an actual confrontation with the enemy malware, or in exercises or simulations where the agent performs against a threat in a simulated or cyber-range environment. It is conceivable that full-fledged AICAs will need to undergo a substantial period of "training" before being deployed. This necessity implies a huge challenge to be met before employing them: the building and maintaining of appropriate testing and simulation platform and—probably most importantly—the standardization of the training procedures. Since the AICA that will emerge from the training period with new knowledge is different from the one that entered it, we need new ways to certify and validate them.

A general cycle of agent learning from its experiences is the following:

- The agent has a knowledge.
- The agent uses the knowledge to perform actions and also makes observations (receives percepts). The ensemble of actions and observations constitute the agent's experience.

- The agent uses this experience to learn the desirable modifications to the knowledge.
- The agent modifies the knowledge.
- Repeat.

Many different types of resulting knowledge could be obtained. The agent may learn the world dynamics model (Sections 4 and 6), the mapping of sensed percepts to states (Section 5), or predictions of results of planned actions (Section 6), and so on. For the purposes of further discussion on this section, we focus on how AICA can learn recommendation(s) of a suitable action(s) (i.e., a plan, possibly as a function of state, or of a sequence of prior actions and observations, regardless of prior sequence). This knowledge too could be used in the processes discussed in Section 6.

Connected to those described previously, AICAs have the opportunity to learn such information that could lead to altering the original world state goal. The opportunity of designing AICAs with this capability, as well as the extent to which such a capability should be allowed is an open question. This function looks like one of the most appropriate parts of the AICARA in which to incorporate ethical and legal guidelines.

In the following sections, we explore details and illustrative examples of this cycle of learning from experience.

## 9.1 Representation of the Agent's Experience

Let's explore a simple sketch of how experience could be represented in AICA.

At any time t, the agent performs action *a*, which could be a NULL action (i.e., there was no action); and perceives percept *e*, which also could be NULL. Note that generally it is impossible to obtain a percept without performing an action, even if the action is as simple as reading data. If the percept *e*, in conjunction with any prior information that the agent has, provides the agent with sufficient information to determine the value of the state of the environment (i.e., the "goodness" of the situation), then the agent may also be able to determine the value of the state V and consequently evaluate the current distance from the desired goal. Otherwise, the reward is NULL. (We return to the topic of reward later.)

Therefore, all the experiences of the agent can be represented with this sequence:

(t1, a1, e1, V1) (t2, a2, NULL, NULL) (t3, NULL, e3, V3) … (tn, an, en, Vn).

Here t1 is the time when the agent starts to record an experience and tn is the moment "now".

To make the representation more compact and useful, we can divide it into shorter chunks; the length of the chunk is implementation dependent. We call such a chunk an episode. Episode Ej is a sequence of pairs {a1, ei}, and the resulting state value Vj:

$$Ej = (\{ai, ei\}, Vj).$$

The following is an example of a short episode:

- a1 checks file system integrity.
- e1 finds unexpected file.
- a2 deletes file.
- e2 file gone.
- a3 NULL.
- e3 observes enemy C2 traffic.
- Value is (–0.09).

Here is another example of an episode:

- a1 checks file system integrity.
- e1 finds unexpected file.
- a2 creates poisoned password file.
- e2 NULL.
- a3 NULL.
- e3 receives alert from node 237.
- Value is (–0.57).

## 9.2 Approach Example 1: Case-Based Reasoning

In this approach (Fig. 19), the learning is largely implicit. The agent collects its experiences in a collection of experiences and augments that collection by determining values for those states, using a function called "state assessor". When the agent wants to determine a plan of action, it looks at its most recent actions and matches them to the experiences. If a well-matching episode is found in its experiences and the resulting value for that episode was sufficiently high, the agent uses that episode as its plan for future actions. This and related classes of approaches are studied in research areas such as case-based reasoning (Kolodner 2014), learning from demonstration (Abbeel and Ng 2004), and inverse reinforcement learning (Argall et al. 2009). Here we offer a very simple sketch of the idea.

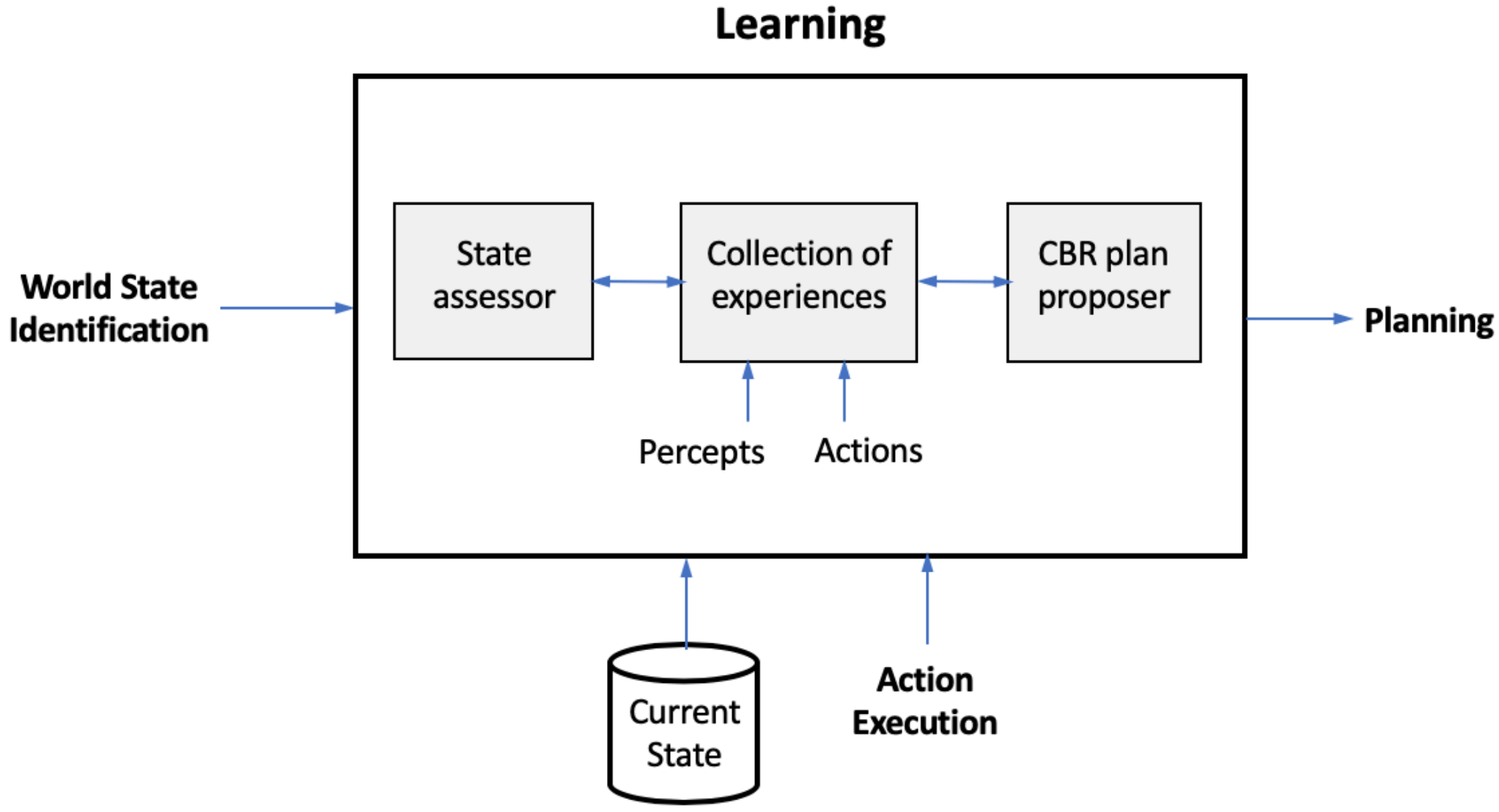

**Fig. 19 Approach example 1: case-based reasoning**

Consider the following, highly simplified illustration. Suppose the agent most recently took actions a13 and a76. The agent wants to formulate a plan of its next actions. The agent wants to make sure that the value of the state that would result from its future actions should be at least 0.75. The agent accesses its collection of experiences and finds there the following episode: a13, a76, a06, a52, V = 0.83. The first two actions of that episode match the most recent actions taken by the agent. The resulting value is very good, higher than the 0.75. The agent, therefore, takes the remaining actions of that episode as its plan: it will proceed to execute actions a06 and a52.

Let us consider what, in this particular example, are inputs and outputs of the learning module.

Inputs include the following:

- Actions (each with a timestamp) that are provided most likely by the *Action Execution* module.
- Percepts (each with a timestamp), each of which is likely to be a change of state, arriving from the state model database.

Outputs include the following:

- Updates to the collection of experiences, which serves here as the form of knowledge. It can be made available to other modules, either directly or via the mediation of data services.
- Episode and the associated reward provided to the *Action Selection* and *Planning* modules.

Alternatively, if the agent has a separate *Planning* function that generates plans, it can use its collection of experiences to predict the values of states that would result from executing that plan. For example, again suppose the agent most recently took actions a13 and a76. The *Planning* function proposed a plan to execute actions a06 and a52.

The agent wants to know what will be the values of states resulting from executing that plan. The agent accesses its collection of experiences and finds there the following episode: a13, a76, a06, a52, V = 0.83. The episode matches its past actions and the proposed future actions. Now the agent knows the value if the proposed plan is executed: V= 0.83.

In this case, the inputs and outputs differ partially from the ones mentioned previously.

Inputs include the following:

- Actions (each with a timestamp) that are provided most likely by the *Action Execution* module.
- Percepts (each with a timestamp), each of which is likely to be a change of state, arriving from the state model database.
- Plan provided by the *Action Selection* and *Planning* modules.

Outputs are the following:

- Updates to the collection of experiences, which serves as the primary knowledge base (KB). It can be made available to other modules, either directly or via the mediation of data services.
- Value of the state expected to result from the proposed plan, provided to the *Action Selection* and *Planning* modules.

Let's add a few words about the state assessor function. How could this function determine value of a state V? Here is a very simple, illustrative (not really workable) way to do this: let subject-matter experts assign each percept a number that characterizes the degree to which the percept indicates the strength of adversarial activity. Then, for each episode, add up such numbers. The sum would constitute a negative "value". Needless to say, other approaches are possible.

Of course, this highly simplified illustration eschews many critical details: we did not mention anything about the percepts and states, and we did not discuss what to do when the match is not perfect. Nevertheless, the gist of the approach should be clear.

## 9.3 Approach Example 2: Deep Neural Network to Learn the Reward for the Next Action

This approach is inspired by the successes of deep reinforcement learning such as described in Mnih et al. (2013). Here our agent uses the collection of experiences to train a neural network. The inputs are the actions and percepts for a number of previous time points. The outputs are, for each possible action of the agent, the value associated with taking that action as the next action. Once the neural network is trained, it is used at each time point to determine the next action—the one with highest value.

To explain what the neural network might look like, consider a highly simplified example. Suppose, at any given time, the agent can take one of only three actions: a1, a2, and a3. (In a practical implementation, there could be thousands of possible actions.) At any given time, it can receive one of only four percepts: e1, e2, e3, and e4. (In practical implementations, there could be thousands of possible percepts.) In our neural network, we consider only two time points: the most recent time an action was taken and the previous time point. (In a practical implementation, multiple time points could be considered.) Figure 20 depicts the neural network after it has been trained. At the most recent time, the agent has performed action a2 and received percept e3. Right before that, it performed a3 and perceived e1. These are the data that go into the input layer. The neural network uses these inputs to produce the outputs: if the next action taken by the agent is a1, the reward will 0.07, if the next action is a2, the value will be 0.023, and if the next action is a3, the value will be 0.79. Naturally, the agent will select a3, the one with the highest value.

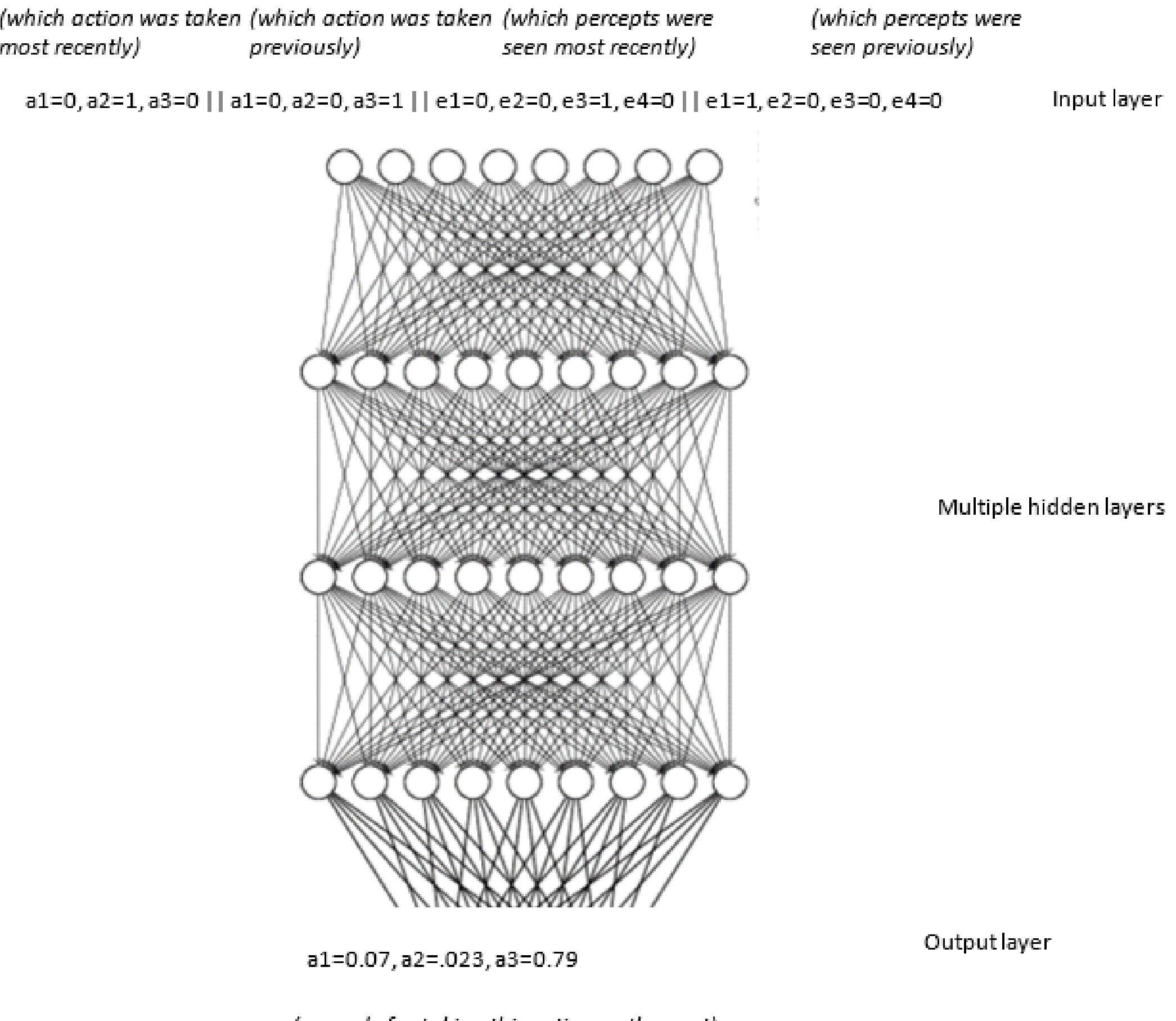

**Fig. 20 Neural network after it has been trained**

The architecture of this approach is illustrated in Fig. 21.

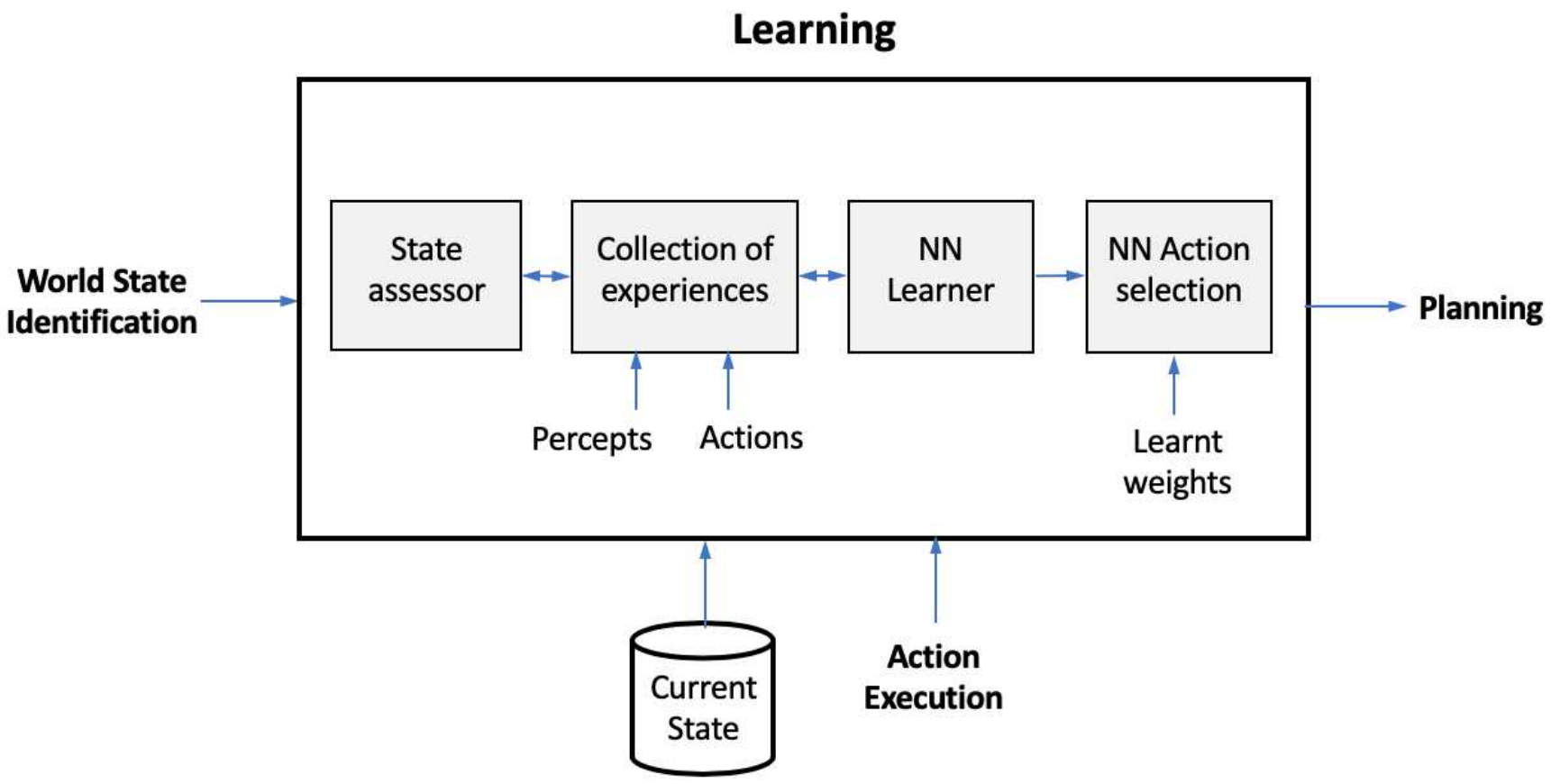

**Fig. 21 Approach example 2: deep neural network to learn the reward for the next action**

Let us consider what, in this particular example, are the inputs and outputs of the *Learning* module.

Inputs include the following:

- Actions (each with a timestamp) that are provided most likely by the *Action Execution* module.
- Percepts (each with a timestamp), each of which is likely to be a change of state, arriving from the state model database.

Outputs include the following:

- Updates to the weights of the neural net, which serve as the primary KB.
- The best next action and the associated value provided to the *Action Selection* and *Planning* modules.

Note that Mnih et al. (2013) used a deep neural network as a component of a Q-learning approach (Watkins and Dayan 1992). Indeed, Q-learning could be appropriate in such problems as ours. It is extremely unlikely that a sufficiently complete model (i.e., probabilities of state transitions given an action) can be constructed for operations of a computer or a network of computers. Therefore, an attractive option is to pursue some form of model-free reinforcement learning. This could mean Q-learning, that is, action-value learning. On the other hand, there is some evidence that for multistep agents or plans with complex time dependencies, Q-learning may not train well. Alternatives might include policy optimizing algorithms or variants of LSTM neural networks and their combinations.

## 9.4 Approach Example 3: Learning the Reward Function

In the classical approach to reinforcement learning the reward function is a given input to the agent and is immutable. The reward function embodies in part the goals and objectives for the agent; featuring the capability of learning a better one is a big step toward complete autonomy. Of course, introducing such a capability comes not only with an opportunity but with risks as well. The behavior of AICA gains a degree of flexibility that could not be appropriate in all real-world use cases and scenarios.

This approach is particularly useful in the training phase, where AICA has the opportunity to learn a compact formulation of the experience, goal, and task to be performed, in the form of a reward function that will be used during the mission. If an adequate level of resources is available (computing power, memory…), this approach could be usefully employed to model adversarial software agents against which AICA is posed, for immediate use in combat or for later reporting.

This approach presents several challenges (see, for instance, Ng and Russell [2000]) but would be important for agents that operate in complex environments where the optimal reward function is not easily formalized. One limitation to be considered is that AICA may need a "teacher", whether another agent or a human expert to show examples of valuable behavior.

## 9.5 Always Learning?

One of the assumption of the reference architecture presented in this work is that AICAs will possibly operate in an environment where limited computing resources and capabilities are available. If this holds true, real-world AICAs could implement a number of strategies in deploying the *Learning* function.

In the most extreme case, no learning at all happens during the actual mission and AICA relies on preloaded databases, including mission-specific packages (e.g., including the topology and details of the networks to protect). In a second scenario, the *Learning* functions are active but with no bearing on the agent's actual behavior. They are only used to gather information on enemy malware and attacks, if encountered, for later or real-time reporting. In the third scenario, full *Learning* capabilities are active in real time during the mission as well as in training and databases and the agent's own goals and policies are changing and responding to the evolving environment.

## 10. Conclusions

Authors: Alexander Kott, Paul Théron, Benoît LeBlanc, Alessandro Guarino, Martin Drašar, and Paul Losiewicz

There is a strong rationale for pursuing the development of intelligent autonomous agents of the kind we describe in this report as AICA. In a conflict with a technically sophisticated adversary, NATO military tactical networks will operate in a heavily contested battlefield. Enemy software cyber agents—malware—will likely infiltrate friendly networks and attack friendly C4ISR and computerized weapon systems. Bonware—intelligent, autonomous agents specialized in cyber defense, such as AICA—will be necessary to detect defeat the enemy malware.

The autonomy of AICAs, and the artificial intelligence underpinning their autonomy, is a necessity. Due to the contested nature of the communications environment (e.g., the adversary is jamming the communications or radio silence is required to avoid detection by the adversary), communications between any friendly battlefield asset, and other elements of the friendly force can be limited and intermittent at best. Given the constraints on communications, conventional centralized cyber defense is often infeasible. It is also unrealistic to expect that human warfighters will be commonly available and able to perform cyber-defense functions.

In general, today's reliance on human cyber defenders will be untenable in the future. The proliferation of intelligent agents is the emerging reality of warfare, and they will form an ever-growing fraction of total military assets. The sheer quantity of targetable friendly assets, the complexity and diversity of the overall network of entities and events, the fast tempo of robotic-heavy battle, the difficulties of centralized defense in a communications-contested environment, the relative scarcity of human warfighters in highly dispersed operations, and the high cognitive load imposed on them by activities other than cyber defense all make an intelligent, autonomous cyber-defense agent a necessity on the battlefield of the future.

Illustrative scenarios, a few of which are discussed in this report, spell out the need for specific capabilities and other requirements of AICAs. In particular, to highlight just a few, AICAs will have to be capable of planning and executing complex multi-step activities for defeating or degrading sophisticated adversary malware, with anticipation and minimization of resulting side effects. It will be capable of adversarial reasoning to conduct a dynamic, strategically minded battle of actions, reactions, and counteractions against a thinking, adaptive malware. It will be able to collaborate on planning and coordinating actions with friendly agents. Crucially,

AICAs will have to keep themselves and their actions as undetectable as possible, and will have to create and use deceptions and camouflage.

Our initial exploration—reflected in this report—identified the key functions, components, and their interactions for a potential reference architecture of such an agent. To mention just a few examples, *Sensing and World State Identification* is the AICA high-level decision-making function that allows a cyber-defense agent to acquire data from the environment and systems in which it operates, as well as from itself, to reach an understanding of the current state of the world. *Planning and Action Selection* is the AICA high-level decision-making function that allows a cyber-defense agent to elaborate one to several action proposals (*Planning*) and propose them to the *Action Selection* function that decides the action or set of actions to execute. *Learning* is the AICA high-level function that allows a cyber-defense agent to use the agent's experience to improve progressively its efficiency with regard to all other functions. For these and other high-level functions of AICA, our initial analysis suggests that the required technical approaches do not seem to be far beyond the current state of research.

The sum of challenges (Table 4) presented by the AICA concept appears, today, very substantial. Still, an empirical research program and collaboration of multiple teams should be able to produce significant results and solutions for a robust, effective intelligent agent. Based on the analysis of the proposed AICARA and available technological foundation, we envision a roadmap toward initial yet viable capabilities.

**Table 4 AICA research challenges**

| AICA component | Research challenges |
|---|---|
| Sensing | - definition of agents' sensing perimeter<br>- distribution of sensing goals and perimeters between agents in a swarm<br>- specialization or generality of the *Sensing* function or of agents in sensing |
| World State Identification | - embarking cyber-defense analysis tools (binary analysis, etc.), analytics and state estimator algorithms into an autonomous agent that must stay small and stealthy<br>- multiagent collaboration toward attack pattern identification<br>- computing possible future situations that may result from the current state of the world through, for example, attack path analysis and multiagent collaboration |
| Planning | - embarking in a small stealthy autonomous agent an AI + game theory + risk/criticality + efficiency-based response planning process + knowledge<br>- experience - routines<br>- multiagent collective optimized response plans |

**Table 4 AICA research challenges (continued)**

| AICA component | Research challenges |
|---|---|
| Action Selection | - evaluation of proposed reaction plans through simulation vs. dynamic risk analysis, tactical reasoning, multiple criteria analysis<br>- multiagent collaboration toward selecting optimal attack response plans |
| Action Execution | - tactical C2 of executing response plans by fully autonomous agents<br>- multiagent tactical C2 of response plans during their execution |
| Collaboration and Negotiation | - CONOPS for collaboration/warnings between agents, agent–cyber C2, agent–human operator<br>- agent discovery and identification (friend or foe), with/without an agents directory<br>- modification of swarms' composition (new entrants, defectors, connectivity issues)<br>- man-machine interface and working protocols for collaboration with humans<br>- inter-agent negotiation protocols, processes, policies, ontologies<br>- technical substrate for inter-agent / agent–cyber C2/agent–human collaboration<br>- inter-agent covert communication channels<br>- trust in inter-agent/agent–cyber C2/agent–human collaboration and negotiation |
| Learning | - learning on the fly or back-office learning?<br>- is learning an individual agent's task or is it a collective/swarm task? |
| Goals management | - what are the impacts of different contexts (combat, motion, idleness) and modes (fighting, fail-safe, isolated, etc.) of the agent/agent swarm on agent goals definition (missions + rules)?<br>- how and why to overrule autonomous agents' goals in specified circumstances? |
| Self-assurance | - definition and a theory of agents' self-assurance |
| Stealth and security | - technologies, processes and rules for autonomous (multi) agents' stealth<br>- interdependence between stealth and security (of agent, agents' communication…)<br>- cyber resilience of isolated agents and of multi agents swarms/cohorts |
| World model | - a theory and formal language of agents' world models (scope, nature, ontology, use, predictive power)<br>- algorithms for preprocessing, creating, and indexing data and machine learning based computation of world models<br>- should the world models of agents in a cohort be consolidated within each agent? |
| Current world state and history | - a theory and formal language of agents' world states (scope, nature, ontology, use, predictive power)<br>- algorithms for computing world states, both nominal and degraded and the role of AI in this process |

**Table 4 AICA research challenges (continued)**

| AICA component | Research challenges |
| --- | --- |
| World dynamics | - a theory and formal language of world's laws of dynamics applied to world objects and agents themselves<br>- world state transition patterns and confidence estimators<br>- can a single isolated/fully autonomous agent compute/learn its world's laws of dynamics? |
| Goals | - how to frame agents' behavior through goals<br>- formal language to express and compute goals and deviations from goals<br>- human–computer interaction and methods for defining agents' goals and embedding them into agents<br>- formal control of deviations from goals and alerts on deviations<br>- operational and ethical aspects of self-definition of goals on the battlefield |
| Development | - agents' fail-safe process: circumstances, process, and other features<br>- agent's database size vs constraints from host platforms' capacities<br>- risks to the agent's stealth due to agents' memory size, processing power, and communication requirements |
| Verification and validation (V&V) | - simulation as a way to validate agents' design and associated confidence estimators<br>- is V&V applicable to agents' knowledge?<br>- how to measure and validate the efficiency/pertinence of agents and swarms' outcome before pronouncing them fit for service? |
| Maintenance | - maintenance of agents through their entire lifecycle<br>- impacts of one agent's maintenance on other agents |
| Internal agent process flow | - agent's process orchestration<br>- optimization of agent's and inter-agent processes and performance |

The first phase of the roadmap, which could be lasting perhaps on the order of 2 years, will include the development of knowledge-based planning of actions, the execution functionality, elements of resilient operations under attack, and adaptation of the prototype agent for execution of a small computing device. This phase would culminate in a series of Turing-like experiments that would evaluate the capability of the agent to produce plans for remediating a compromise, as compared to experienced human cyber defenders.

The second phase, which could last about 3 years, would focus on adaptive learning, the development of a structured world model, and mechanisms for dealing with explicitly defined, multiple, and potentially conflicting goals. At this stage, the prototype agent should demonstrate the capability, in a few self-learning attempts, to return the defended system to acceptable performance after a significant change in the adversary malware behavior or techniques and procedures.

The third phase, potentially about 3–4 years, would delve into issues of multiagent collaboration, human interactions, and ensuring both the stealth and trustworthiness of the agent. Cyber–physical challenges may need to be addressed as well. This phase would be completed when the prototype agents are able to successfully resolve a cyber compromise that could not be handled by any individual agent.

NATO cyber defense would benefit from active encouragement of AICA development efforts. Relevant research in academia and in some government and industry research organizations is growing, and should be supported. It appears that academic institutions already begun work toward AICA-like capabilities, and results are beginning to be available for transition to industry. NATO defense agencies should query the cybersecurity software vendors about availability of AICA-like products. Creating a multi-stakeholder working group engaging industry, academia, and governments could help facilitate the development of AICA technologies. NATO must not fall behind its adversaries in developing and deploying such capabilities.

# Appendix A. Twenty-Eight Seconds in the Life of an AICA

Authors: Michael J De Lucia, Allison Newcomb, and Alexander Kott

(Editors' note: This appendix comprises excerpts from the following journal paper: De Lucia MJ, Newcomb A, Kott A. Features and operations of an autonomous agent for cyber defense. Journal of Cyber Security and Information Systems. 2019.)

## A.1 An Illustrative Operating Scenario

In order to illustrate how an autonomous cyber defense agent might operate, we offer a notional operating scenario. In this scenario, Blue refers to friendly forces and Red refers to the adversary. Blue-17, Blue-19, and Blue-23 are peer cyber-defense agents. Each agent is installed by a human operator on its respective device within the Blue IoBT (e.g., an Android phone) and is tasked with cyber defense of that device. Blue-C2 is the command and control (C2) node that commands, coordinates, and supports all other Blue agents, at least when communications between an agent and the Blue-C2 node are available. There is only one Red agent—Red-35—in our simple scenario.

The protagonist of our scenario is Blue-17, a cyber-defense agent that has been installed on a friendly device; it continuously monitors Blue space network and scans event logs looking for suspicious activity. The antagonist is Red-35, a malware agent successfully deployed by the Red forces on the device defended by Blue-17. The events unfold, briefly, as follows.

Blue-17 detects a hostile activity associated with Red-35 and attempts to contact the Blue-C2 for additional remediation instructions. Unfortunately, the communications are heavily contested by the adversary, and response from Blue-C2 is not coming. Therefore, Blue-17 decides to contact peer agents (Blue-19 and Blue-23) in search for relevant information. Although Blue-19 and Blue-23 receive this message from Blue-17, their responses are not arriving to Blue-17. Having heard nothing within a reasonable waiting time, Blue-17 independently formulates and executes a set of actions to defeat Red-35. However, having completed these actions, Blue-17 receives a belated reply from Blue-23. Blue-17 determines that Blue-23 is compromised because the response is suspicious. Given the extreme seriousness of this situation, Blue-17 neutralizes Blue-23 and places a copy of itself on the device that was being protected by Blue-23.

Table A-1 provides a hypothetical timeline of these events and the agents' actions. Durations are intended to merely illustrate the flow of time in the scenario and are in no way representative of execution speeds of any actual hardware or software. Following the table, we discuss each step of the scenario in more detail.

**Table A-1 Hypothetical timeline of agents' actions**

| Step | Elapsed time | Condition/event | Active software agent | Action |
|---|---|---|---|---|
| 1 | H = 0 sec | Start up | Blue-17 | Monitor network traffic and scan logs |
| 2 | H = H + 0.100 sec | Hostile software agent compromises device and network | Red-35 | Red-35 infiltrates Blue device and network. Blue-17 does not notice the infiltration. |
| 3 | H = H + 0.200 sec | Red-35 begins operations. Suspicious activity detected | Red-35 and Blue-17 | Red-35 conducts malicious activities. Blue-17 detects an activity and predicts probable compromise. |
| 4 | H = H + 0.22 sec | Compromise suspected | Blue-17 | Contacts C2 node |
| 5 | H = H + 3.00 sec | No response from C2 node | Blue-17 | Contact Blue-19 and Blue-23 agents |
| 6 | H = H + 5.00 sec | Message among Blue peer agents | Blue-19 and Blue-23 | Receive message from Blue-17 |
| 7 | H = H + 10.00 sec | Message acknowledgement time out | Blue-17 | Choose alternate course of action |
| 8 | H = H + 12.00 sec | No communication with peer defensive agents | Blue-17 and Red-35 | Block or redirect Red-35 communication. Red-35 is unable to defend itself. |
| 9 | H = H + 23.00 sec | Response received from Blue-23 | Blue-17 | Blue-17 determines that the response is invalid |
| 10 | H = H + 28.00 sec | Neutralize compromised Blue agent | Blue-17 | Quarantine or destroy Blue-23 software code |
| 11 | H = H + 28.3 sec | Replicate and overwrite | Blue-17 | Copy to device |

### Scenario Steps 1–2

In the scenario, Blue-17 passively monitors the inbound and outbound network communications using a lightweight intrusion detection system (IDS) such as FAST-D (Yu and Leslie 2018). FAST-D is a software that performs intrusion detection using far less computational resources than alternative solutions. Its algorithm uses hash kernels and byte patterns as signatures to examine the packet payload content of all network communications. Additionally, Blue-17 scans the device logs looking for indicators of compromise (privilege escalation, abnormal crashes, failed logins, etc.).

**Scenario Steps 3–4**

Blue-17 sends a message to its C2 node for further remediation instructions and verifications. A C2 node is one that is central (root) and is responsible for the management and tracking of all Blue agents. A C2 node resides in a central location that may be the tactical operations center. The message sent to the C2 node is encrypted to protect the confidentiality and integrity and is in a predefined format for agent messages. This message is split up into many small segments, is blended into normal traffic to masquerade as other legitimate traffic, and sent through different routes within the network in order to avoid an attacker from intercepting or detecting the agent message sent to the C2 node. Lastly, the address of the C2 node changes over time based on a deterministic algorithm, known to all agents to make it more difficult for Red-35 to discover its location.

**Scenario Step 5**

After some reasonable waiting time passes, and Blue-17 does not receive a reply back from the C2 node, it decides as an alternative action to send out a request to its peer agents (Blue-19 and Blue-23) for their remediation recommendations. Again, this message is sent out using an encrypted predefined format for agent messages as previously described in sending a message to the C2 node. The message is sent directly to the peers and is blended into another network traffic. The peer agents are neighbors to Blue-17 and are also be under the management of the C2 node.

**Scenario Step 6**

Both Blue-19 and Blue-23 have received the message from Blue-17. After some delay, Blue-23 sends a response and recommendation back to Blue-17 using the same method for sending a message to a peer agent.

**Scenario Steps 7–8**

Within a specified time interval, Blue-17 has not received a response from either its C2 node or its peers (Blue-19 and Blue-23). Blue-17 requested further verification of the threat before taking a destructive action against Red-35. However, since a response was not received, Blue-17 decides to take action on the perceived Red-35 malware agent threat. The Blue-17 agent first isolates the Red-35 malware agent and its communication in a honeypot to observe the actions taken by the attacker. Blue-17 has taken this action since it is not confident in its assessment of the detection of the perceived Red-35 agent.

**Scenario Step 9**

After some time has passed, and Blue-17 has already taken action, a response from Blue-23 is received. Blue-23's response contains a signature and timestamp that allows Blue-17 to determine the authenticity of the message received. However, as Blue-17 verifies the response message from Blue-23, it determines that the message signature is not valid and rejects the message. Blue-17 concludes that Blue-23 may be compromised.

**Scenario Steps 10–11**

Blue-17 has discovered that Blue-23 has been compromised. Blue-17 takes action to quarantine Blue-23. Blue-17 clones itself to create a pristine copy of the defensive agent. Blue-17 initiates the overwriting of the Blue-23 agent image with a fresh copy of a defensive agent with the initial state of Blue-17. The agent package is sent via an encrypted message from Blue-17 to the container management of Blue-23. The container management package of the agent uses cryptographic authentication, allowing the overwriting to occur. Blue-23 is restored back to a fresh agent image and is no longer infected.

## A.2 Discussion of Challenges and Requirements

Having offered a scenario—simple yet sufficiently illustrative of potential difficulties—we now have a basis for discussing the technical challenges and requirements. One of the requirements illustrated in part by the scenario is that a defensive agent must reside outside of the operating system of the device it is protecting. This arrangement avoids the possibility of the malware providing false information or changing the view of the defensive agent (i.e., Blue-17). Malware can disable processes or deceive (e.g., by providing false information) software such as the antivirus (AV) software or firewall on a device (Baliga et al. 2007). A logical separation at the hardware level between the operating system being protected and the defensive agent will protect the Blue-17 agent from being compromised by malware infection. The defensive agent will require access in a secure manner to all of the files and state from its outside view, while being protected from any threats affecting the Blue-17 operation or integrity.

Additionally, because the Blue-17 agent resides outside of the protected operating system, Red-35 will not be able to detect Blue-17's presence or any of its actions. A traditional placement alternative for an agent that resides outside of the protected operating system would be a distributed, or network-based, sensor. That configuration comes with a tradeoff: An agent (Blue-17) at the network level would not be able to monitor the file system of the protected operating system. Therefore, the Blue-17 agent must reside on the same physical device as the operating system being protected.

Also, in order for the agent to move around freely among the devices within the protected network, the agent must be agnostic of any particular operating system. It is also presumed that the container in which the agent runs has been pre-installed on the device to which agents can migrate freely to, such as in the case with Blue-17 overwriting Blue-23.

Clearly required, as illustrated in our scenario, is a fast, highly reliable and low-resource means of detecting potentially malicious activity. For example, using a low-resource intrusion detection software, Blue-17 was able to detect rapidly and with a significant degree of assurance a suspicious activity performed by a sophisticated agent Red-35. Additional solutions could be employed that use supervised machine-learning approaches, coupled with features such as network packet inter-arrival times, packet sizes, Transmission Control Protocol (TCP) flags, and such, to perform detection of malware infiltration. However, in either case it is important to understand the limitations (i.e., inability to detect malware within encrypted communications) of the intrusion detection algorithm chosen to perform detection of malicious communications. It is also important to know the possible ways an attacker could evade (fragmentation attack, encrypted attack, etc.) the IDS. Successful evasion by an attacker will result in a missed attack. It is also critical for an autonomous agent employing an IDS algorithm to have a low false-positive rate (misclassified legitimate traffic as an attack) and false- negative rate (missed attack). In a military context a false positive in an autonomous cyber-defense agent will result in an impact to the mission by denying a legitimate communication that is essential to the mission.

Another challenging requirement is the need to manage the degree of the agent's autonomy. Blue-17 could be fully autonomous or semiautonomous. In our scenario, Blue-17 is fully autonomous, evidenced by the lack of human intervention at any point. Consequently, Blue-17 must be highly confident in the detection event and its resultant course of action. The agents' actions must avoid any adverse reaction, such as degrading network performance or dropping nodes on the network as a mitigation, resulting in access denials. Alternatively, Blue-17 could act as a semiautonomous agents, with varying levels of interaction between the agent and human controllers, which present many challenges of their own (Kott and Alberts 2017). For example, Blue-17 detects a potential compromise and then defers to a human analyst (e.g., by contacting the C2 node and waiting for instruction) in a case where there is a low to moderate confidence in the detection event.

The agent will require the ability to share threat data directly with its peers (e.g., Blue-17 had the need to share data with Blue-23 and Blue-19) and orchestrate coordinated defensive actions when necessary. Additionally, the agent must be able to work in an isolated environment to make appropriate decisions independently, as Blue-17 had to do when it failed to receive response from either Blue-C2 or peers' agents. These agents will need to store pertinent information on detected attacks and outcomes (successful vs. unsuccessful) of the selected mitigation strategies. This information will need to be stored in a compressed format due to the limited resources characteristic of the various devices of the Internet of Battlefield Things (IoBT). On the other hand, when the agents return to a less-contested environment where power and bandwidth are less constrained and more reliable, the data would be uploaded to a central repository. Lessons learned (quantitative measures of outcomes) and specifics on detected attacks would be compiled to improve the process of informing other autonomous agents. This arrangement would expand and enrich the agents' knowledge and ability to learn from historical decision-making strategies.

The agent (i.e., Blue-17, Blue-19, or Blue-23) hosted within the IoBT environment will need to process and reduce the enormous amount of information produced by itself and other agents to a subset, which is relevant to the human warfighters' cognitive needs (Kott et al. 2016). For example, an enormous number of alerts may be produced by the agents, but the human warfighter cognitive requirements only include the subset of alerts to form a situational awareness of ongoing cyberattacks, which are impacting missions. Therefore, the agent (i.e., Blue-17, Blue-19, or Blue-23) will need to process and filter the alerts to a reduced subset of alerts, which are relevant to ongoing missions. Additionally, the filtered information must be relevant and trustworthy to the IoBT device and human cognitive needs, as a risk is providing information that could lead to an undesired action or outcome resulting in further impact to the mission (Kott et al. 2016). Lastly, information stored by agents on IoBT devices must be distributed and obscured from the adversary. An approach to secure the distributed agent information within an IoBT environment is to split the data into fragments and disperse them among the many devices (Kott et al. 2016a). This information will need to be obfuscated, segmented, and distributed among the many agents so that an adversary will not be able to rebuild the original information. The distribution of the segmented information among the agents will need to be performed in a way which will thwart the adversary's ability to reconstruct the information based on a number of captured segments (Kott et al., 2016a). The combination of both intelligent filtering and distributing the information among various agents will assist in informing the human warfighter cognitive needs and deceiving the adversary.

Ideally, the agents' performance would be evaluated in order to refine and share successful strategies with other agents. Performance in this context includes the agents' decision-making value, timing, and the resulting impacts of the courses of action executed (e.g., Blue-17 was successful—what factors contributed to these successes?). This further supports the need for agents to learn from their actions as well as the actions of other agents via machine-learning techniques.

The agent could employ a combination of supervised and unsupervised machine learning. The lessons learned and outcomes of the course of action taken by an agent could be used with a reinforcement-based machine-learning algorithm. For example, the successful course of action executed by Blue-17 with respect to defeating Red-35 would receive a positive reward. This approach could be used to expand the knowledge of the autonomous agents, thereby improving the agents' performance and effectiveness.

Another requirement of these agents will be trust management between devices. Each device on the network will require software-based logic to participate in the network with a full degree of trust and access. This logic can be preinstalled or can be acquired from a peer node by a device that seeks to join the network in a comply-to-connect mode of operation. Once compliance conditions are met, the agent can be transferred to other network member nodes. For example, in our scenario, Blue-17 needed a way to determine that Blue-23 is no longer trustworthy. At the same time, Blue-17 had to elicit a sufficient degree of trust from the node where Blue-23 resided in order to overwrite the Blue-23 image.

Device-to-device transfer of the agents—such as the move of a copy of Blue-17 to the node originally defended by Blue-23—necessarily raises concern for unintended propagation and behaviors beyond the intended network, as witnessed with the Morris worm (Qing and Wen 2005; Spafford 1989) and the more recent Stuxnet attack (Farwell and Rohozinski 2011). Findings from studies on limiting the spread of malware in mobile networks (Zyba et al. 2009; Li et al. 2014) could be adapted to manage the propagation of defensive agents. Another potential solution to controlling propagation is to require consensus approval of a certain number of nodes prior to enabling transfer of the agent to a new device. A suggested approach is to define boundary rules to determine whether the agent has been transferred outside its intended network. When the boundary rules evaluate to a true condition (out of bounds), mandatory removal of the agent or a self-destruct sequence would be triggered. The effects of these combined approaches to controlling propagation require additional research.

While autonomous agents should be free to learn, act, and propagate, careful thought should be given to methods that would constrain behaviors within the bounds of legal and ethical policies, as well as the chain of command. For example, it would be undesirable if Blue-17 were to learn that requests to Blue-C2 are generally fruitless and should not be attempted. An agent that is fully autonomous must be able to operate within an appropriate military C2 construct (Kott and Alberts 2017). It is imperative that a software agent be bounded in its propagation, yet capable to move around freely between authorized devices.

# Appendix B. Impact of Agent's Purpose on its Capabilities

Authors: Ryan Thomas and Martin Drašar

(Editors' note: This paper originally appeared in Kott A, Thomas R, Drašar M, Kont M, Poylisher, A, Blakely B, Théron P, Evans N, Leslie N, Singh R, Rigaki M. toward intelligent autonomous agents for cyber defense: report of the 2017 workshop by the North Atlantic Treaty Organization (NATO) Research Group IST-152-RTG; 2018. *arXiv preprint arXiv:1804.07646*.)

With the proliferation of machine-learning (ML) methods in recent years, it is likely that autonomous agents will become commonplace in day-to-day military operations. We expect a significant boost in their capabilities owing to both algorithmic advancements and adoption of purpose-built ML hardware. However, the range of agents' functions will still be, in the foreseeable future, limited by a number of factors, which we attempt to enumerate.

In this text, we recognize two types of autonomous agents as two extremes on the capability scale. At one extreme are preprogrammed heuristic agents, responding only to specified stimuli based on a set of preset actions. At the other extreme are robust intelligent systems with advanced planning and learning characteristics. Capability is then the aggregate of an agent's intelligence, awareness, connectedness, control, distributedness, level of autonomy, and adaptability. Agent's purpose prescribes specific functions and abilities, and the operational expectations place an upper bound on agents' capabilities. The following text provides a list of some limiting factors and evaluates their impact.

## B.1 Mobility

Autonomous agents deployed at stationary structures (e.g., buildings or weapon systems) should suffer the fewest limitations in their operation, as it can reasonably be expected that such agents will have enough power, processing capacity, connectivity, and other resources needed to carry out the most complicated of tasks. These systems will be restricted mostly by the ML state of the art.

Agents deployed on mobile platforms (e.g., vehicles, Soldiers, or missiles) will inevitably be limited by intermittent connectivity; power, space, and processing constraints; or even the physical implications of their actions. Furthermore, for mobile systems, it is likely that the agent will be located at a centralized point in the architecture, rather than be distributed across all subsystems. This is due to the expected difficulty in accrediting systems with robust intelligent behaviors.

## B.2 Lethality

Agents operating in systems with lethal capacity will either have to undergo much tighter scrutiny or be limited in their actions to prevent the creation of accidental or exploited killer bots. In such systems, it is easy to envision agents and humans performing as a team, with the human having the final authority for decisions with lethal implications. This will require developments in human–machine trust, interfaces, and planning.

Another option to safeguard lethality would be the use of a two-tier infrastructure, where lethal means are physically separated and thus inaccessible to even a rogue autonomous agent. The ML would control the nonlethal tier only, allowing more conventional means (or, as described previously, a human) to control the lethal tier.

## B.3 Criticality

Critical systems, whose failure has severe consequences, mostly operate with clear separation of responsibilities and are handled by rigorously trained personnel. Failures are reduced by the application of processes, which limit the impact of human error. Autonomous agents will likely introduce whole new classes of errors, so these error-controlling processes must be updated accordingly.

There are three likely approaches to this:

1) Improvements in the understanding of ML operations and performance limits will enable better scrutiny of the inner workings of autonomous agents, constraining the range of possible ML errors and formally proving the scope of exhibited behaviors.

2) Testing methodologies and testbeds will improve, allowing autonomous agents to undergo a battery of conformance tests exhaustive enough to give informal guarantees of the agent's operation with acceptable confidence.

3) Autonomous agents will be deployed redundantly, allowing for robust and resilient operations. Techniques such as voting (e.g., three implementations with a majority voting on a next action) could be used.

## B.4 Connectivity

Most mobile platforms will suffer connectivity problems or forced connection losses. Autonomous agents, which rely on communications links to enable swarm intelligence, command and control (C2), or computation offloading, would be severely impaired during connection loss. Therefore, any such ML functionality requiring connectivity must be designed with respect to the communications environment and timescale in conjunction with required ML decision accuracy. For systems in unreliable environments, which need stable communication channels to arrive at decisions quickly or require accurate and reliable decisions under all conditions, it is up to debate as to whether the presence of autonomous agents is worth the personnel training extension, related updates to operational processes, and associated certification hurdles.

## B.5 Power and Processing Constraints

Given the currently immense computation requirements for any autonomous and learning behavior, any hardware able to run sufficiently advanced agents will require nontrivial space, power, and cooling. Unless there is a significant leap in technology, this will limit the available resources for agents, especially for deployment in mobile platforms. Developers of agents and policy makers will have to carefully consider which autonomous functionalities are necessary or beneficial enough.

There is great potential in bio-inspired autonomy, assisted by mechanical and structural features on the host platforms. For instance, insects such as moths and flies are an inspiring mix of clever sensor arrays, simple processing cortexes, and advanced mechanical wing design that could enable low-power, low-processing micro-autonomous air platforms.

## B.6 Commoditization and Standardization of Agents for Environments

We expect that some standard classification of autonomous agents according to their capability and requirements is inevitable. Such classification would ease the adoption process by reducing the need to evaluate each agent in a specific context with regard to whether an agent conforms to a class specification. Military systems then could be limited to specific classes of autonomous agents, thus prescribing the level of autonomy such systems can have.

# Appendix C. "Hello, World" Autonomous Agent

Authors: Alessandro Guarino, Jana Komárková, and James Rowell

(Editors' note: This paper originally appeared in Kott A, Thomas R, Drašar M, Kont M, Poylisher, A, Blakely B, Théron P, Evans N, Leslie N, Singh R, Rigaki M. Toward intelligent autonomous agents for cyber defense: report of the 2017 workshop by the North Atlantic Treaty Organization (NATO) Research Group IST-152-RTG; 2018. *arXiv preprint arXiv:1804.07646*.)

The challenge we tackle in this section is the design of an actual autonomous agent, small and simple to implement but able to illustrate the essential functions any autonomous intelligent agent (AIA) should possess, albeit in a streamlined way. The agent proposed here is a purely software agent intended for cyber defense only.

To be a proper AIA, it should fulfill the following 6 characteristics:

1) An agent is strictly associated with its environment: an autonomous agent outside the environment it was designed for can be useless, or not even an agent at all. Franklin and Graesser (1996) have given a convincing definition of agents and the ways in which they differ from other software. The first four points in our definition draw from their definition.

2) An agent interacts with the environment, via appropriate sensors providing input from it and appropriate actuators, allowing the agent to act and influence that environment.

3) An autonomous agent acts toward a goal, or, in other words, it has an "agenda". In particular, an autonomous agent developed for warfare operations is assigned a target.

4) The activities of a truly autonomous agent are sustained "over time", so it must have a continuity of action.

5) An autonomous agent should possess an adequate internal model of its environment, including its goal—expressed possibly in terms of world states—together with some kind of performance measure or utility function that expresses its preferences.

6) An agent must possess the capability to learn new knowledge and the possibility to modify over time its model of the world and possibly also its goals and preferences.

In this section, we describe the agent and explain how it fulfills these requirements. We define its environment, task, and properties, such as sensors and actions. We also discuss possible extensions of the agent.

To make these “Hello, world” autonomous agents feasible, the design makes specific assumptions about the environment in which the agent operates, and the number and type of inputs and outputs its sensors and actuators will have. This has the aim of keeping the complexity low.

## C.1 Environment

AGENTX lives in a virtualized cloud environment that supplies some unspecified cloud-based service. We assume this platform runs three kinds of virtual machines (VMs, or virtual servers): database servers, application servers, and web servers. We also assume that a hypervisor exists to manage the platform and balance the load.

## C.2 Task

Again, for the sake of simplicity, AGENTX performs one specific function and not in an open-ended generic network defense mission. Its goal is to manage a set of honeypot (HP) virtual servers with the objective to deceive adversaries and deflect Cyberattacks against the cloud platform. Its architecture is monolithic (as opposed to a distributed, swarm-like structure) and operates at the hypervisor level of the system. To perform some of the available actions, AGENTX relies on small applets installed on each virtual server, for instance, exposing a RESTful application programming interface (API). It must be noted that in the context of this proof of concept, security measures that in a real environment would be mandatory are overlooked (e.g., encryption of communications, self-protection of the agent itself, and so on).

Since the mission of AGENTX is purely deception, it implements the capability of communicating to other autonomous agents (and/or to human supervisors) the necessity to intervene and implement active defense measures.

The agent has access to background information, such as a set of ready-made HP images, dummy process containers, and dummy files.

## C.3 Sensors

The sensors are able to access the following data and information:

i. Alerts from intrusion detection systems (IDSs) (count and severity)

ii. Integrity information of critical files on the VMs

iii. Metadata about critical files on the VMs

iv. Processes

v. Log files

vi. Metrics on the level of use of resources and system load

vii. Feedback and replies from other agents tasked with active measures

## C.4 Actions

The following actions are available to the agent:

i. Starting and stopping HP VMs.

ii. Starting and stopping actual virtual server instances (optionally).

iii. Initiating a "cry for help" message to other agents (or humans).

iv. Deploying dummy files and applications, and quarantine files (via the applets).

## C.5 Learning

The agent implements a reinforcement learning model employing an appropriate reward function:

$$R = a\frac{honey_events}{security_events} + b\frac{\Delta_resources}{total_resources} + c\frac{justified_CFH,}{CW} \qquad (1)$$

where

- *honey_events*: metric for attacks/events against the HPs;
- *security_events*: number of attacks against the real servers (detected by IDSs);
- *total_resources*: metric for the total amount of resources available;
- Δ*_resources*: resources freed or needed (example, spinning HPs) to implement an action by the agent (negative when resources are needed, positive when resources are freed);
- *justified_CFH*: "justified cries for help", number of messages (alerts) sent by the agent reacting to actual attacks; and
- *CW*: "cry wolf", number of messages sent by the agent requesting assistance for attacks that did not really happen.

The coefficients *a, b, c* state the relative importance of each factor. They shouldbe tuned beforehand or during the initial learning phase.

We consider total resources available as those actually available at the time of action, which makes the function and the agent's behavior dynamic during that time. It also means that a relatively costly action is not penalized if the system is under very light load, because the number of available resources is high and even small actions are heavily penalized if the system is utilizing almost all its resources.

Note that this function could be calculated—in a future version of AGENTX—for homogeneous groups of VMs (e.g., only the application servers), to better reflect the situation and the world state, providing AGENTX with a more granular and detailed view of its environment.

The learning is performed by implementing an anomaly detector leveraging a small set of hard-coded features (for the purposes of this section) including the following:

- Number and severity of IDS alerts
- Anti-malware software alerts
- Unauthorized accesses
- Access to HPs or dummy files
- Alerts from dummy processes
- File integrity violations
- System load (aggregate, by group, and individual)

## C.6 Testing

To validate the performance of the agent, we have to set up a testing environment, perform real attacks, and evaluate its efficiency. Since the agent is learning with each attack, we should let the evaluation continue for some time so the learning process can take place. It would also be ideal to face the agent with real attackers, not only simulated attacks.

We propose to validate the agent on defending a network with several servers in a virtualized environment with simulated "regular" traffic. The setup has the following advantage: since we know which traffic was generated by us, we can safely assume the rest of the traffic comes from the attacker; therefore, we can easily recognize the justified and unjustified cries for help. The detection part is also easily achieved in this setup. We can leave the network running for a long time with little effort. To prevent the abuse of the compromised machines, we can let the "servers" actually be high-interaction HPs.

## C.7 Additional Considerations

Future developments include, of course, the use of real-world tools to implement the autonomous agents (while this proof of concept could be developed in a scripting language like Python), the implementation of all possible security measures to secure and protect the agent, as well as the development of the cooperative agents postulated previously.

Moving away from a monolithic architecture (at the hypervisor level) to a swarm-like distributed architecture of agents living on every VM on the system is another valid possibility.

# Appendix D. The AHEAD Autonomous Agent

Authors: Fabio De Gaspari and Luigi V Mancini

(Editors' note: We view AHEAD as an agent that in many of its features is a partial instantiation of the AICA architecture.)

## D.1 Introduction

The AHEAD architecture (de Gaspari et al. 2016) defines an autonomous agent equipped with a variety of active defense tools, providing both sensing and actuating functions. The AHEAD system consists of two main components: the AHEAD Controller and the Active Defence Architecture (ADARCH) Pot (Fig. D-1). The AHEAD Controller component implements the control logic of the system and manages a group of ADARCH Pots (Al Shaer et al. 2019). The ADARCH Pots are comprised by a set of active defense tools and can dynamically change the configuration of these tools to implement the decisions taken by the controller. In the AHEAD architecture, the ADARCH Pots are integrated directly into the production systems of a target network, providing ready-to-use active defense capabilities, while at the same time safeguarding the security of the production systems themselves.

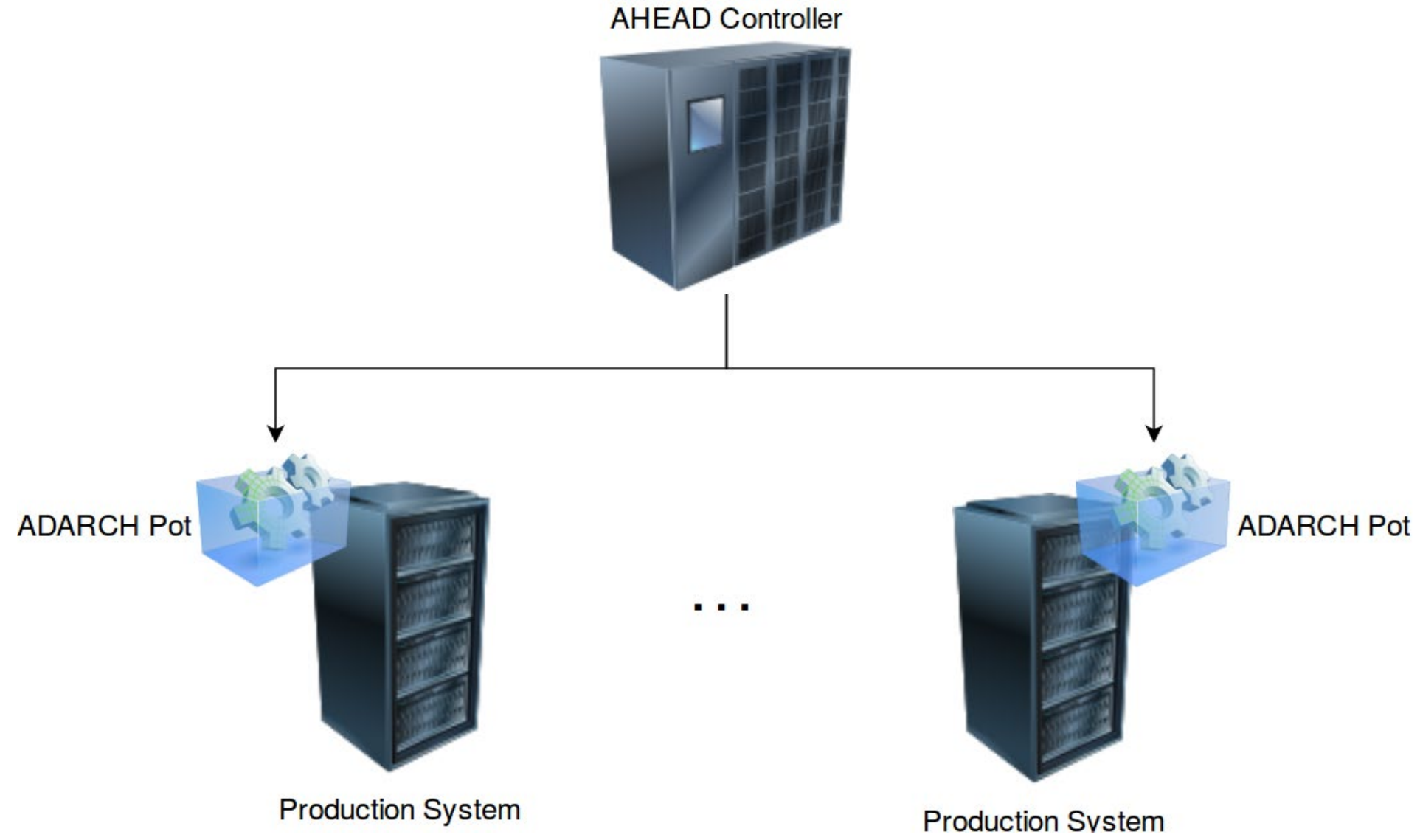

**Fig. D-1 High-level view of the AHEAD architecture**

The AHEAD system provides many functions that are also part of the AICA architecture, such as *Sensing and World State Identification*, *Planning and Action Selection*, *Action Execution*, and *Learning*. The active defense tools deployed within the ADARCH Pot provide sensing functions for the system. Every interaction with the tools is logged and forwarded to the AHEAD Controller, where the *Learning* and *Planning* components process the input and decide if and what actions are necessary based on the new world state (e.g., possible malicious activity underway). Once a decision is taken by the controller, the active defense tools of the ADARCH Pot are updated and reconfigured accordingly. Therefore, each ADARCH Pot also acts as an actuator for the system. This ability to dynamically reconfigure the Pot allows the AHEAD system to generate high amounts of accurate data on the attacker, since the controller can reconfigure the Pot to maximize the interaction with the malicious agent.

However, differently from the Autonomous Intelligent Cyber-defense Agent (AICA), the AHEAD system is designed as a single-agent system (the AHEAD controller), and does not provide for the possibility of a multiagent configuration as the AICA Reference Architecture (AICARA) does. Therefore, AHEAD does not include *Collaboration and Negotiation* functions.

## D.2 Component Comparison

In this section, we include a table showing a side-by-side comparison of AICA functions and the corresponding AHEAD module implementing it.

**Table D-1 Comparison of AICA function and the AHEAD module**

| AICA function | AHEAD |
|---|---|
| Sensing | ADARCH Pot |
| World State Identification | Controller state management component |
| Planning | Controller learning component |
| Action Selection | Controller learning component |
| Action Execution | ADARCH Pot |
| Collaboration and Negotiation | n/a |
| Learning | Controller learning component |
| Goal management | Controller learning component |
| Stealth and security | Controller/ADARCH attestation component |
| World model | Controller state management component |
| World state and history | Controller state management component |
| World dynamics | Controller learning component |
| Goals | Controller learning component |

## D.3 Technologies and Implementation

Currently the AHEAD system is only partially implemented, with the ADARCH Pot in a prototype stage and the controller still in development phase. Therefore, in this section, we focus only on the ADARCH Pot and the sensing and actuating functions of the system.

### D.3.1 System-Pot Integration

Since the Pot is designed to run alongside the real services in the production systems, it requires isolation in order to prevent malicious agents to use the active defense tools as pivot to break into the system. In ADARCH, this is accomplished by means of the Docker container technology (Docker 2019) and mandatory access control (MAC) techniques (Apparmor n.d.). Containers allow to provide separation between the Pot's environment and the real system's environment, while the mandatory access control ensures that even if the malicious agent takes over the Pot, its influence on the real system is heavily limited. Figure D-2 illustrates the integration of the Pot with the production systems.

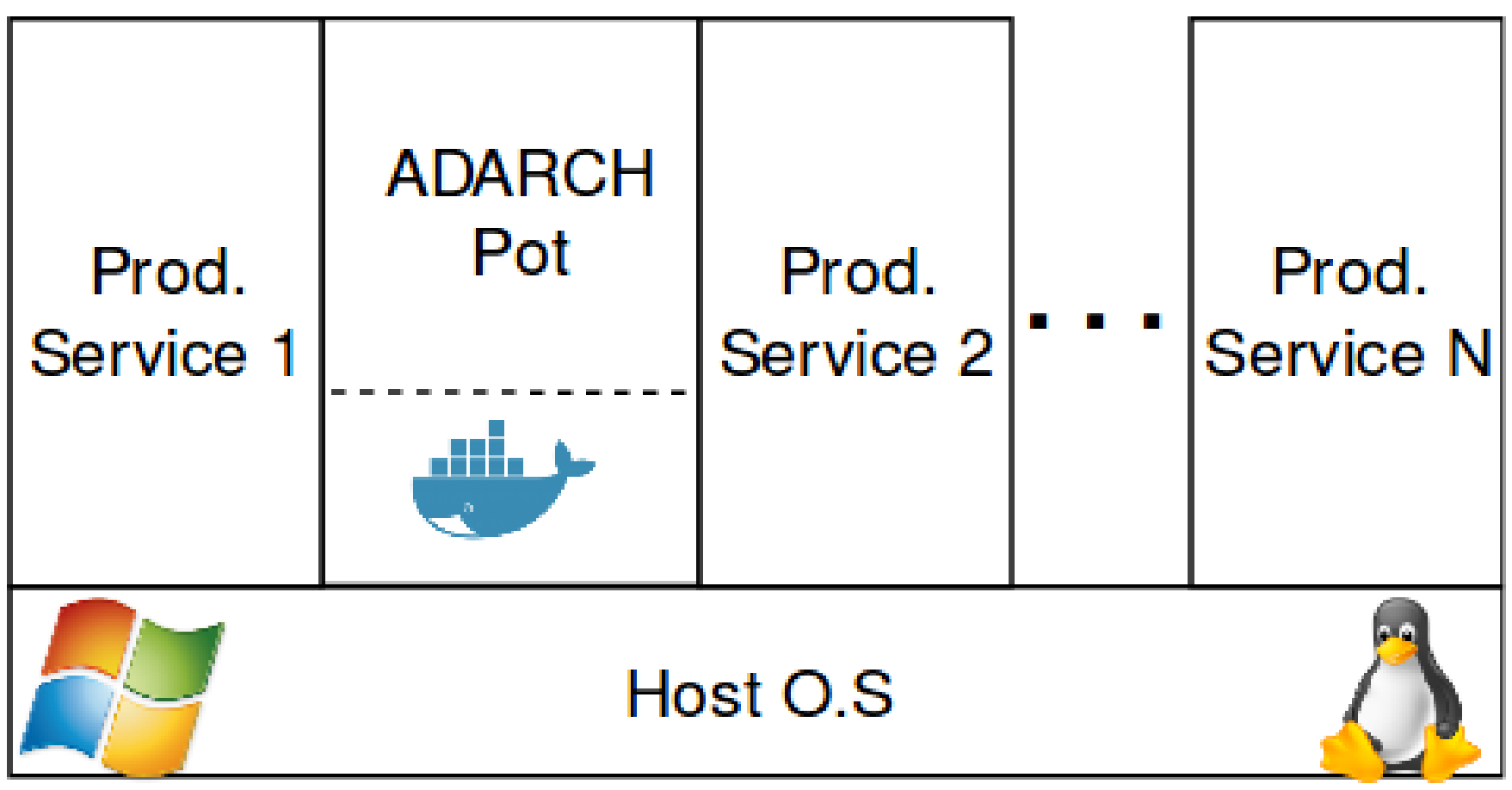

**Fig. D-2 Integration of the Pot in the production system. Docker container technology is used to isolate the Pot from the host**

The use of Docker containers provides consistency between the operating system (OS) environment and the Pot environment, preventing attacks aimed at differentiating active defense tools from legit production services.

### D.3.2 Pot Sensors and Actuators

The sensors and actuators in the AHEAD system are the active defense tools deployed within the ADARCH Pot. The ADARCH Pot provides a modular and extensible framework that allows to easily develop and configure new active defense tools, and seamlessly integrate them within the AHEAD architecture. The active defense tools are developed on top of a Python-based framework that provides common functions such as logging and networking for all the tools. The use of a common framework for all the tools allows to easily develop and integrate new active defense modules, as well as providing a common view of the underlying system to the active defense tools deployed in the Pot. In particular, having a unified system environment for all the tools provides a consistent view to malicious agents interacting with the Pot and vice versa, which in turn means it is harder for such agents to understand they are interacting with fake services. Finally, the modular nature of ADARCH makes it easy to dynamically reconfigure the services exposed based on the output of the learning and planning components of the controller.

Figure D-3 shows a high-level view of the ADARCH Pot architecture. ADARCH comprises two main components: the ADARCH core and the ADARCH interpreter. The ADARCH core is developed in C and is the core of the framework. It implements the most common functions required by a large number of active defense tools. The ADARCH interpreter encapsulates and extends a Python interpreter, and is used by external modules to transparently interact with the core framework. Active defense tools are developed as ADARCH modules written in Python, along with a corresponding configuration file. The configuration file allows tools developers to effortlessly instantiate required resources (e.g., port bindings), as well as to define callback functions associated to specific triggers. Finally, ADARCH is designed to work with container technologies to provide an additional layer of isolation to the production system.

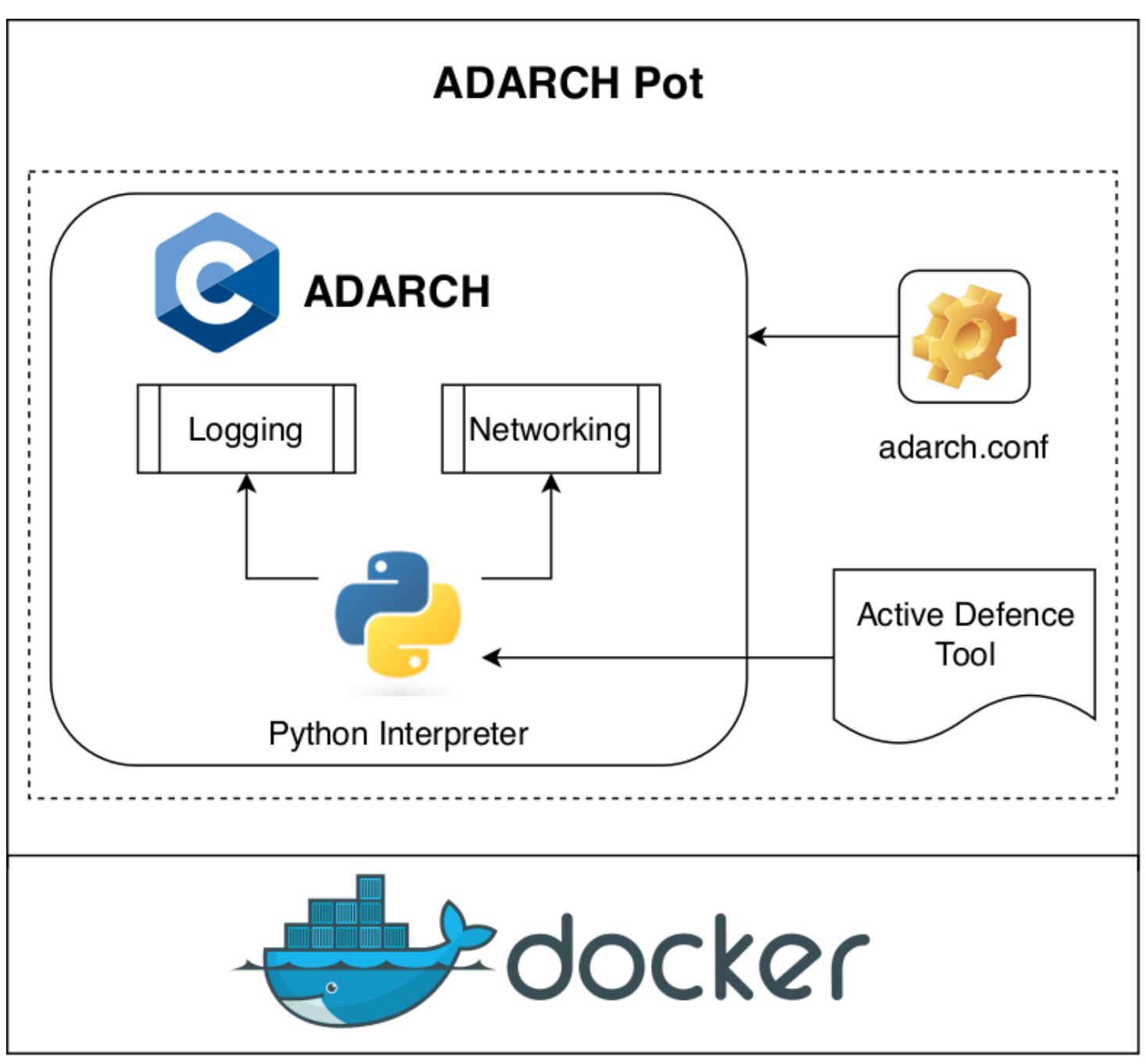

**Fig. D-3 ADARCH Pot architecture**

# Appendix E. Autonomous Cyber Deception Based on Malware Analytics

Authors: Jinpeng Wei, Ehab Al-Shaer, and Mohammed Noraden Alsaleh

(Editors' note: This appendix describes analytic framework called Autonomous Malware-centric Deception System (AMDS), which can be considered as special instance of the Autonomous Intelligent Cyber-defense Agent (AICA) with special focus on malware deception. AMDS analyzes the malware behavior, automatically extracts the deception parameters using symbolic execution, and creates cyber deception plans.)

## E.1 Introduction

*Malware as an Opportunity.* Malware is normally considered harmful and useless: when someone detects a piece of malware, their immediate reaction is to get rid of it. This is also the recommendation by state-of-the-art security solutions, including antivirus, anti-malware, intrusion detection systems, and intrusion prevention systems.

Contrary to the conventional wisdom, we argue that malware can be used to improve the effectiveness of cyber deception: malware provides a communication channel between the security defender and the attacker; therefore, it creates unique opportunities to manipulate the attacker, such as feeding misinformation back to the attacker, engaging the adversary in a deception ploy, and learning about the adversary's tactics, techniques, and procedures. In other words, we can rely on the malware to achieve our deception goal: to execute the deception plan via the malware.

Here are a few examples.

*Example one:* The malware finds FTP login credentials on the victim computer and exfiltrates them to the attacker. We can benefit from this malware by setting up a honey FTP server, creating honey accounts, and intentionally letting the malware exfiltrate the login credentials of such accounts. In other words, we can use the malware as a messenger to lure the attacker to our honey FTP server, which may be set up just for that one attacker.

*Example two:* The malware uses the victim computer to run bitcoin mining software on behalf of the attacker. We can benefit from this malware by first learning the attacker's bitcoin mining account name from the malware, and then submitting a large number of wrong mining results to the mining pool on behalf of the attacker, so that their reputation is damaged, to the extent that their account is banned.

### E.1.1 Deception Models

Cyber deception is a defense technique that deliberately introduces misinformation or misleading functionality into cyberspace in order to trick an adversary in a way that benefits the defender. Deception models allows for defining the deception goals and planning approach in order to construct an effective deception agent.

### E.1.2 Deception 4D Goals

Effective cyber deception aims to 1) *deflect* adversaries away from their goals by disrupting their progress through the kill-chain, 2) *distort* adversaries' perception of their environment by introducing doubt into the efficacy of their attacks, 3) *deplete* their financial, computing, and cognitive resources to induce biased and error-prone decisions that we influence, and4) *discover* unknown vulnerabilities and new tactics, techniques, and procedures (TTPs) of adversaries, while predicting the tactical and strategical intents of adversaries.

### E.1.3 Malware Deception Playbook: Toward Real-time Autonomous Deception of Malware

We are developing an Autonomous Malware-centric Deception System (AMDS) as shown in Fig. E-1. The main functionality of this system is to map patterns of malware behavior to prescribed sets of deceptive actions called *Deception Playbooks*. It consists of four components: the Detection Agent, the Analysis Agent, the Planning Agent, and the Deception Actuator.

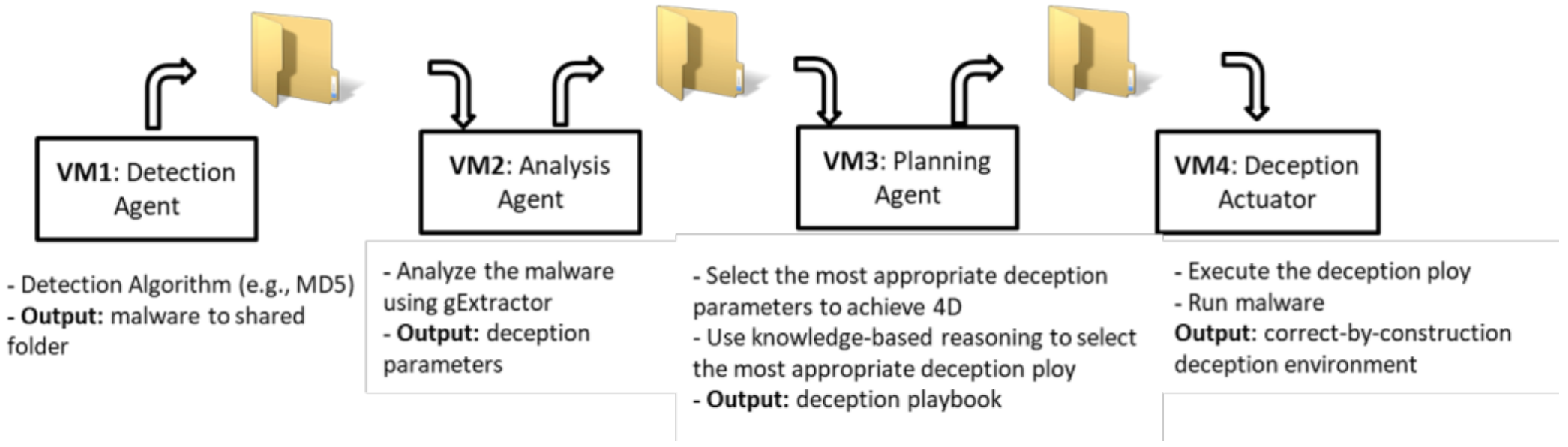

**Fig. E-1 AMDS architecture**

The Detection Agent checks whether an application is malicious, and if so notifies the Analysis Agent. The Detection Agent can be an antivirus application or an intrusion detection system.

The Analysis Agent analyzes a given malicious executable to extract *deception parameters*, which the malware depends on for reaching its goals and which can be configured/controlled by the environment to implement different deception plans.

The Planning Agent takes the deception parameters as input to find suitable Deception Playbooks for a deception ploy against the malware and the attacker behind it, based on certain rules. The Deception Playbook is actionable but it can have flexible formats (e.g., an algorithm or script template).

The Deception Actuator executes Deception Playbooks to deploy the deception plan, and it restarts the malware in the deception environment created by the deception ploy.

## E.2 Detection Agent

The Detection Agent can leverage existing malware detection techniques such as antivirus and intrusion detection systems. It scans suspicious files or detects suspicious processes in the system, using techniques such as signature matching. The signatures can be syntactic (e.g., hash code of binary files, string set, and byte sequences) or semantic (e.g., application programming interface [API] call patterns). Once a positive detection is made, the relevant file is handed over to the Analysis Agent.

## E.3 gExtractor: The Analysis Agent

To extract the complete behavior of a cyberattack, we execute its binaries (i.e., malware) symbolically and build a model that represents its behavior with respect to selected system parameters. Given that the correct set of system parameters is selected, symbolic execution can cover all relevant execution paths.

Before going through the technical steps of the symbolic malware analysis, we present the attack behavior model.

### E.3.1 Attack Behavior Model

The *attack behavior model* describes how the attack behaves based on the results of its interaction with the environment. The malware interacts with its environment through system and user library APIs characterized by their input and output arguments. Some of these arguments may be attacker-specific variables and cannot be controlled by the environment, while other parameters can be reconfigured or misrepresented. We assume that a mapping between the selected system or library APIs' arguments and the corresponding parameters in the environment, such as *files*, *registry entries*, *system time*, *processes*, *keyboard layout*, *geolocations*, *hardware ID*, *C&C*, *Internet connection*, *IP address* or *host name*, and *communication protocols*, is given. For example, the *from* argument of the *recvfrom* API can be mapped to a system parameter that represents the IP address of the sender machine.

We define the attack behavior model as a graph of *points of interaction (PoI)* nodes and *fork* nodes. The PoIs refer to the points in the malware control flow at which the malware interacts with the environment by invoking system or library APIs. The fork nodes represent the points in the control flow at which the malware makes a control decision based on the results of its interactions with the environment.

In Fig. E-2, we show an example of attack behavior model that represents a portion of the Blaster worm that delivers a copy of the worm to an exploited victim. Round nodes represent PoIs and square nodes represent fork points. The solid edges represent control dependency, while dashed ones represent data dependency. In this model, the worm first sends an instruction to a remote command shell process running on the exploited victim through the *send* library API, then it waits for a download request through the *recvfrom* API call. The attack code checks if these operations are executed successfully and terminates otherwise as depicted through the conditions shown on the outbound edges from the fork nodes 2 and 5. At node 7, the worm starts reading its executable file from the disk into a memory buffer, through *fread*, and sending the content of the buffer to the remote victim, through the *sendto* API. There is a data dependence between the third argument of the *sendto* call, which represents the number of bytes to transmit, and the return value of the *fread* call, which represents the number of bytes read from the worm file.

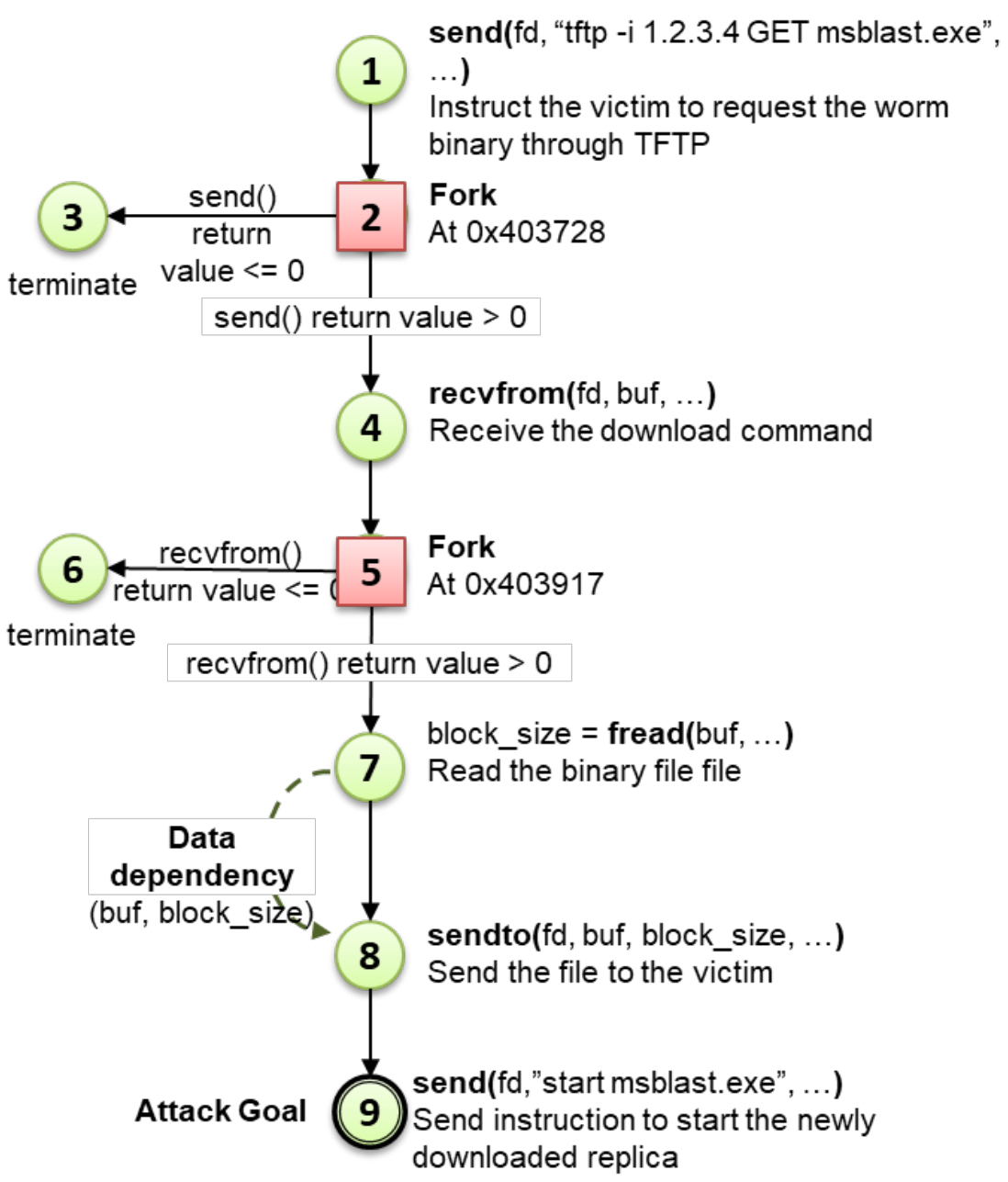

**Fig. E-2 Example of attack behavior model (Alsaleh et al. 2018)**

### E.3.2 Malware Symbolic Execution

We utilize the $S^2E$ engine to symbolically execute malware binaries. The path coverage and the progress of the executed program depends on the correct marking of symbolic variables. Since we are interested in the interactions of the malware with its environment through selected system and library APIs, we intercept these calls and mark their output arguments as symbolic. This allows us to capture the malware decisions based on those arguments and track the corresponding execution paths. In the current version of our implementation, we select about 130 APIs that cover activities related to networking, file system and registry manipulation, system information and configuration, system services control, and user interface (UI) operations.

*Marking Symbolic Variables.* To mark the appropriate symbolic variables, we take advantage of the *Annotation* plugin provided by $S^2E$, which combines monitoring and instrumentation capabilities and executes user-supplied scripts, written in LUA language, at run time when a specific annotated instruction or function call is encountered. We define an annotation entry for each API. The annotation entry consists of the module name, the address of the API within the module, and the annotation function. We identified the module names and addresses using static/dynamic code analysis tools, such as IDA and Ollydbg. The annotation function is executed at the exit of the intercepted call. It reads the addresses of the return and output arguments of the call and marks the appropriate memory locations and registers as symbolic. Note that output arguments may have different sizes and structures. Hence, we need custom scripts to mark each individual output argument of the intercepted APIs. The return values of APIs are typically held in the *EAX* register and we use special method provided by $S^2E$ to mark its value as symbolic. It should be noted that system calls and user library APIs are invoked by all applications in the environment, not only the malware process. Therefore, our annotation functions check the name of the process that invokes them and ignore calls from irrelevant processes.

*Building the Attack Behavior Model.* After preparing the appropriate annotation entries, we execute the malware using $S^2E$ to collect the execution traces. We configured the annotation functions to record the arguments, the call stack, and other metadata, such as the timestamp and the execution path number for each intercepted system and library call. By design, $S^2E$ intercepts branch statements whose conditions are based on symbolic variables and forks new states of the program for each possible branch. We collect the traces and branching conditions of all execution paths and build the attack behavior model as follows:

- We create a PoI node for each system or library API call logged by our annotation functions. Similarly, the traces contain special log entries for state forking operations. Those are used to create the *fork* nodes in our model.
- For each node in the model, we add a control dependency edge from the node preceding it in the execution path. If the preceding node is a *fork* node, the edge will be associated with a branching condition in terms of the symbolic variables.
- To capture the data dependency, we check the values of all the input arguments upon the entry of each API call. If the value is a symbolic expression, this implies that it is a transformation of previously created symbolic variables. Hence, we add a data dependency edge from the PoI nodes in which the symbols of the expression were created.

### E.3.3 Deception Parameters Extraction

Given the attack behavior model generated through symbolic execution, we extract a set of system parameters that help in designing effective deception schemes to meet the deception goals. Recall that the attack behavior model describes the complete behavior of a malware with respect to selected system parameters. However, that does not mean that every parameter in the attack behavior model is a feasible candidate for deception. That is, mutating or misrepresenting its value may not be sufficient to successfully deceive the attacker. We analyze the attack behavior model to select the appropriate set(s) of deception parameters that can help in designing deception schemes without dictating particular ones.

We present the following four criteria ($C1$ to $C4$) that must be considered to decide on which parameters are appropriate for effective deception and which are not:

- $C1$ (Goal Dependency): the selected deception parameters can directly or indirectly affect the outcomes of the attack in terms of whether the attacker can reach her goal. Hence, parameters that are used only in execution paths that do not lead to particular goals might be excluded.
- $C2$ (Resilience): in cases where multiple attack paths lead to particular goals, selected parameters must provide deception in all the paths, not only one.
- $C3$ (Consistency): the selected deception parameters must preserve the integrity of the environment from the attacker's point of view. As system parameters may be interdependent, deception schemes must take this into consideration, such that misrepresenting one parameter without misrepresenting its dependents accordingly does not disclose the deception.

- *C*4 (Cost-Effectiveness): although multiple parameters may exist in the execution paths leading to particular goals, mutating or misrepresenting different parameters may require different costs and provide different benefits from the defender's point of view. Defenders must select the most cost-effective set of parameters for deception.

#### E.3.3.1 Refining the Attack Behavior Model

The complete attack behavior model contains many execution paths that may not be relevant to our deception analysis. In this refinement step, we 1) identify the set of execution paths that are relevant to deception and 2) eliminate the don't-care symbolic variables.

*Identifying Relevant Paths.* Recall that deception is not about blocking attacks, rather, it is about misleading and forcing them to follow particular paths that serve the desired deception goals. Hence, the selection of relevant execution paths from the attack behavior model depends on the deception goal.

*Relevant Paths.* A relevant path with respect to a particular deception goal is an execution path that exhibits particular patterns of interactions with the environment that can be leveraged by the defender to achieve the deception goal.

Regardless of which deception goal is desired, it can be represented as a single call or a sequence of calls to system and library APIs leveraging existing tools that identify specific behaviors through patterns of call sequences, such as Christodorescu et al. (2007), Shankarapani et al. (2011), and Qiao et al. (2014). Then, the PoI nodes in our attack behavior model will be used to identify the execution paths that exhibit that particular sequence of calls. By pruning out all other paths that do not exhibit the desired sequence, we end up with a portion of the original behavior model that contains only the paths relevant to the deception goal.

Another simplification is to eliminate don't-care variables with respect to a particular deception goal. A don't care variable is a symbolic variable that is part of one or multiple execution path constraints and its value is irrelevant to the desired deception goal.

After eliminating the irrelevant paths and the don't-care variables, we end up with refined path constraints for the relevant paths. Any parameter extracted based on this refined model complies with $C1$ criteria.

#### E.3.3.2 Selecting Deception Parameters

Since the output of one interaction may be determined by multiple system parameters, there is no necessarily one-to-one mapping between the symbolic variables and the system parameters. Therefore, we need to map the symbolic variables to the appropriate system parameters, utilizing experts knowledge of the system and the system and library APIs. The documentation of the APIs can also be used to extract this mapping as it normally specifies the possible outputs of APIs and the cases in which each value is returned based on the system and the environment states. The result of this mapping will be a basic set of system parameters called deception parameters.

## E.4 Planning Agent

The Planning Agent uses knowledge-based reasoning to select the most appropriate deception ploy, and outputs a deception playbook.

### E.4.1 Select the Best Deception Parameters to Achieve 4D

In this step, we define a constraints optimization problem to find an optimal set of deception parameters that satisfy the following constraint: 1) at least one parameter is selected for each relevant path (to comply with $C2$), 2) if a parameter is selected, all its dependent are also selected (to comply with C3), and 3) the selected parameters incur the minimum cost on the defender (to comply with $C4$). We solve the constraints optimization problems using the Z3 solver. The result will be a set of system parameters that satisfies our four criteria to provide resilient, consistent, and cost-effective deception.

### E.4.2 Select the Most Appropriate Deception Ploy

In this step, the Planning Agent maps the chosen deception parameter to a Playbook that manipulates the deception parameter.

Each playbook has the following components: 1) a goal, which can be deflection, distortion, depletion, or discovery, 2) associated deception parameter(s), 3) preconditions, which is a predicate that must be evaluated to *true* before the actions of the playbook are enabled, 4) actions, which are concrete executable steps of a deception ploy.

The associated deception parameter is used as a key to search for playbooks. The benefit of running each playbook is reflected in its goal attribute, and depending on what the customer of deception wants to achieve, different playbooks may be selected. The actions specification can be understood by the Deception Actuator, which instantiates the action specifications into actual execution.

The actions of a deception playbook may include host-level actions and network-level actions. Host-level actions include configuration (e.g., saving of user credentials) and object manipulation (e.g., creation of honey files and honey registry entries). Network-level actions include firewall configuration and IDS configuration, such as adding a filtering rule on the firewall to allow malware to communicate with its C&C server.

Management of playbooks. All playbooks are indexed by the deception parameters that they manipulate. Our system supports the addition, query, and modification of playbooks. We provide interfaces for managing playbooks.

The query interface of the playbook manager uses the deception parameter as the input and returns a playbook. Our design allows a set of Deception Playbooks to be configured based on user demands. We leverage knowledge-based reasoning and satisfiability constraint solvers to construct a resilient deception agent.

## E.5 Deception Actuator

The Deception Actuator receives Deception Playbooks from the Planning Agent, constructs a deception environment based on specifications in the Deception Playbooks, and execute malware that it also receives from the Planning Agent, to realize the deception ploy.

To construct a deception environment, the deception actuator needs to prepare the basic OS, runtime systems, applications, and required data, and set up the network configuration, based on the specifications of the Deception Playbook. We can use a dedicated virtual machine for each deception environment, to minimize the interference from other irrelevant workloads.

## E.6 Prototype

As an illustration, we have built a prototype of AMDS. The agents are encapsulated in separate virtual machines on the same host machine, and they use shared folders on the host machine to collaborate, e.g., the Detection Agent puts malware files in the shared folder, and the Analysis Agent picks up malware from the same folder.

The Detection Agent uses hashes of know malware samples as signatures. The Analysis Agent records file- and registry-related API calls made by the given malware, including the names of files or registry entries. From the names it recognizes interesting deception parameters such as registry keys that contain saved FTP passwords (e.g., *"Software\Martin Prikryl"*).

The Planning Agent has one Deception Playbook about *"Software\Martin Prikryl"*: it belongs to WinSCP and the deception ploy includes installing WinSCP, configuring honey accounts, and saving these FTP passwords. All these steps are implemented in scripts.

Finally, the Deception Actuator runs the scripts to actually deploy the honey FTP server and prepare an execution environment for the malware (e.g., installing WinSCP and saving WinSCP passwords), and then run the malware.

## E.7 Relevance to the AICA architecture

The overall design of AMDS fits well into the AICA architecture (Kott et al. 2018). AMDS *Analysis Agent* corresponds to *Sensors* in AICA, AMDS *Planning Agent* maps to *Planner Predictor* and *Action Selector* in AICA, and AMDS *Deception Actuator* corresponds to *Action Execution* in AICA. Both AMDS and AICA leverage knowledge-based reasoning. However, AMDS can be considered as special instance of AICA for malware deception. For example, the sensing and world state identification are based on data collected from the environment but AICA does not prescribe the exact kind of data to use. However, AMDS explicitly uses live malware and its execution traces as the data type. Moreover, AICA includes a collaboration and negotiation component, which is currently not part of AMDS because the focus at this point on a single agent decision making. However, we can see that this will be in our future extension of the architecture in order to enable AMDS agents in various component of the system to share and coordinate their action to globally orchestrate deception on large-scale cyber systems such as IoT.

In summary, we consider AMDS as a concrete instantiating of AICA with a focus on malware. The Analysis Agent and the Planning Agent of AMDS can be enhanced by adopting strategies recommended in AICA, such as multiagent collaboration.

# Appendix F. Security and Trust in AICA

Authors: Edlira Dushku and Luigi Vincenzo Mancini

An Autonomous Intelligent Cyber-defense Agent (AICA) should provide real-time cyber-defense protection to other intelligent devices deployed on nearby systems within the perimeter that AICA should protect. Therefore, the security of the AICAs themselves is critical. In particular, the information shared among collaborative AICAs should be accurate and reliable regardless of the threat conditions. AICAs should be able to establish trust in a rapidly changing environment and adopt trusted collaboration mechanisms. In addition, AICAs should implement trusted learning mechanisms in order to be resilient against attack scenarios, in which a compromised AICA may maliciously influence the knowledge and the actions of other friendly AICAs.

In the following, we identify some of the security aspects that should be considered in the implementation of an AICA.

## F.1 Security

### F.1.1 Cryptographic Functions

Each AICA should use standard cryptographic functions to provide confidentiality and integrity of AICA components and operations. Cryptographic functions can be implemented in software or in a hardware accelerator. To generate unpredictable random keys, an AICA should use hardware true random number generators (TRNGs).

### F.1.2 Key Management

Key management schemes play a key role in a secure communication between AICAs. The key management process should be based on a policy and should be performed by a specific security operation or an authority. The key management schemes can be 1) centralized: only one entity manages the keys for all the agents, 2) decentralized: the agents will be organized into small subgroups, and different entities will manage the key distribution for each subgroup, and 3) distributed: the agents collaborate to generate a common key or each agent generates one key.

Alternatively, a random key pre-distribution scheme (Kahn et al. 1999; Eschenauer and Gligor 2002; Chan et al. 2003), which rely on probabilistic key sharing can be used as a lightweight key exchanging scheme between low-end devices. In these schemes, each device is initialized with $m$ keys, selected from a large pool of $S$ keys, such that two random subsets of size $m$ in $S$ will share at least one key with some probability $p$. Afterwards, devices perform shared-key-discovery to find out which of other devices they share a key with.

To protect the identity of AICAs, the private keys should be accessible only by authorized components. When necessary, the signing keys can be protected within hardware protected memory, preventing untrusted parties from using these keys. In addition, key management schemes should address the impersonation problem. For instance, an adversary can create its own cyber-defense agent (i.e., Sybil attack) which joins the Army network in order to participate in the decision-making process and influence the common goals of the friendly forces. The prominent necessity against such attacks is the development of mechanisms that can allow the legitimate agents to detect the agents with fake identity.

## F.2 Trust

### F.2.1 Trustworthiness of an AICA Agent

To provide reliable evidence about the integrity of software running on AICA, each AICA should have a hardware-based immutable root of trust, such as a Trusted Platform Module (TPM) (Trusted Computing Group 2013). Remote Attestation is a security mechanism that verifies the trustworthiness of a system or a component. In the hardware-based trust model, the trust establishment derives from the underlying cryptography based security mechanisms (e.g., digital certificates, signatures and cryptographic checksums, for instance). During the boot process, TPM measures the system's software state and stores the hash values into the TPM's Platform Configuration Registers (PCRs). However, TPM measures the software only at boot time, and it is not resilient against runtime attacks. The development of AICAs requires new security techniques that can check the integrity at runtime of each AICA component.

### F.2.2 Trusted Collaboration

The agents of friendly forces are expected to have some pre-shared cryptographic keys protected by hardware Root-of-Trust (RoT) on each agent. To guarantee a secure collaboration, all the services of the collaboration and negotiation component should use the security credentials embedded in the RoT. Since none of the distributed agents in the battlefield has a complete knowledge about the environment, it is important to construct the necessary security mechanism that could enable the legitimate to detect the agents with fake identity (e.g., a Sybil attack).

In order to deploy trust establishment mechanisms for the communication among agents, the collaboration function should be able to adopt different trust model approaches. In particular, this component should consider two possible approaches: hardware trust model and behavior-based trust model. In the hardware trust model (Jøsang et al. 2007), the trust establishment derives from the underlying cryptography based security mechanisms (e.g., digital certificates, signatures and cryptographic checksums, for instance). In the behavior-based trust model, the trust derives from external observations of agent's behavior by providing a reputation score for the agents (Zacharia and Maes 2000; Jøsang et al. 2007).

Furthermore, remote attestation can serve as a mechanism that allows an AICA to establish trust to another AICA. In addition, battlefield environment require new remote attestation schemes can also be used to verify the trustworthiness of a large group of devices in a more efficient way than attesting each of devices individually. These remote attestation schemes should be able to build trust even in the cases of disruptive networks.

Beside the attestation capabilities, the AICAs should have sufficient onboard analytics capabilities for performing local profiling of the activities of the other intelligent agents to make safe decisions when there is no connection to the friendly agents or command and control (C2). When the cyber-defense agent detects an anomalous behavior, it should be able perform further investigation on the suspicious device and then delete the malware.

### F.2.3 Trusted Learning Among AICAs

The observation of an AICA can be intentionally misleading. The C2 unit can extend the local analysis capabilities of an AICA toward an efficient security protocol that correlates the received information with different source of communications for the entire network of the agents. The verification of the received information should take in consideration not only the value of the data, but also the properties of the data reported from different agents.

Additionally, a distributed collaborative learning scheme presents a crucial threat. An adversary can compromise an AICA and make the agent to share a particular information, which will cause the other agents to be more exposed to attacks. Research has shown that malicious participants on such a learning scenario are able to effect the outcome of the learning scheme by maliciously affecting what the global model has learned (Hitaj et al. 2017).

Therefore, it is necessary to develop some new trusted learning techniques in a distributed collaborative battlefield environment.

# Appendix G. Deep Decision-Making for AICAs

Author: Paul Théron

Information technology (IT) and operations technology (OT) systems are now evolving toward autonomy and higher levels of complexity, both in the civil and military domains. From sensors and goods-delivery drones to intelligent ammunitions or augmented cognition for fire fighters through central command and control systems and maintenance and inspection things, autonomy will become a key to being successful over enemies in the increasingly fast pace of combats and to deliver better services to populations and customers.

Just like in the Internet of Battle Things, intelligent things will fight intelligent things (Kott 2018), in the Internet of Things and autonomous systems, intelligent goodware will fight intelligent malware.

With the current paradigm of cyber defense, based on the centralized monitoring of permanently connected systems, human operators who supervise cybersecurity and resolve cyberattacks will be overwhelmed by the pace, volume and complexity of cyberattacks at hand.

A bio-inspired autonomous, intelligent and trustworthy cyber-defense technology, embedded into systems, must do the job for us, at speed and scale. Multiple agents, spread across software and hardware components, will work together to monitor and defend systems when malware strikes.

Autonomous Intelligent Cyber-defense Agents (AICAs) (Fig. G-1) will monitor systems, detect attacks, design and execute tactically an appropriate response, learn and protect themselves, and report to us about their doings and circumstances.

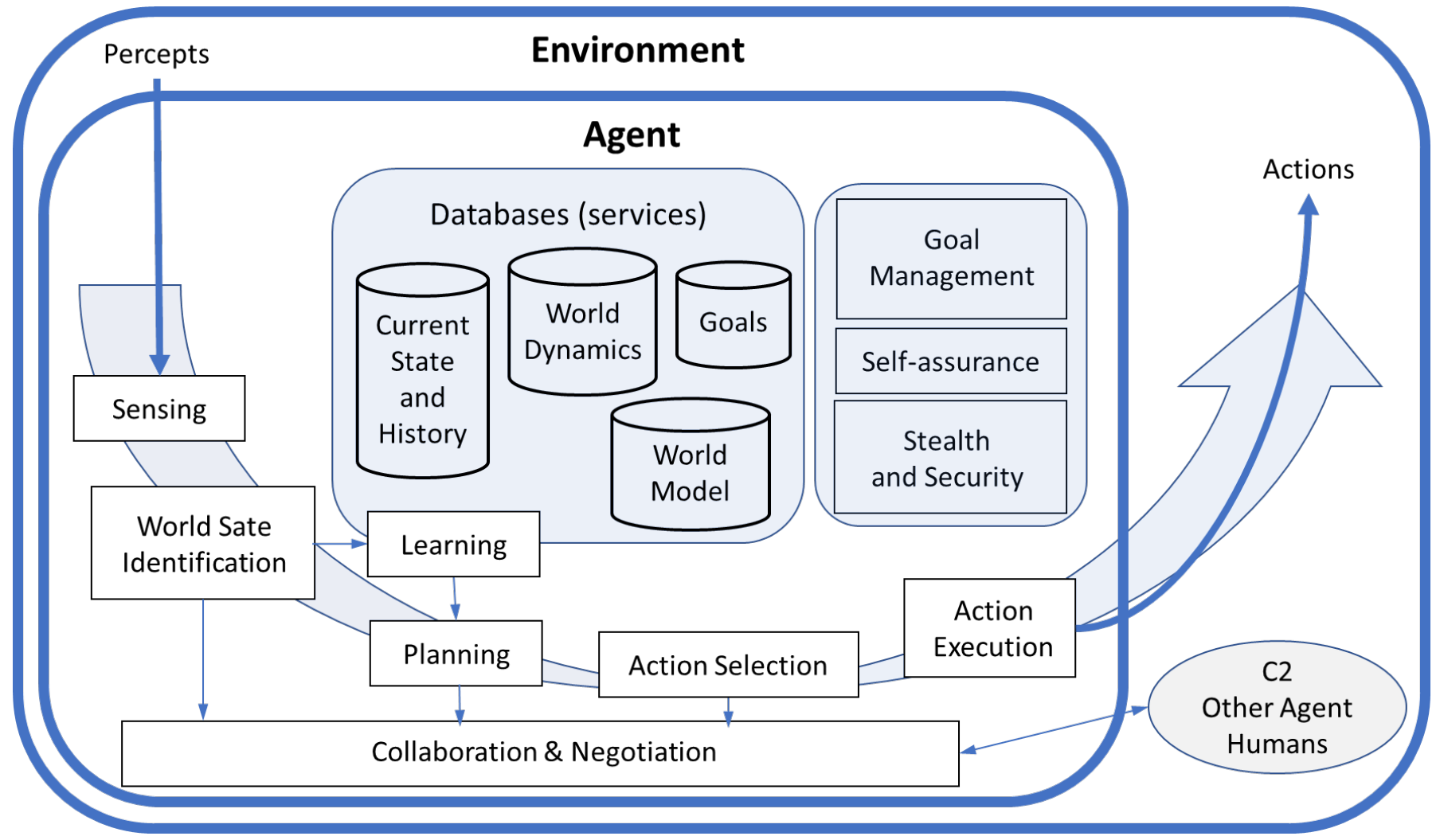

**Fig. G-1 AICA Reference Architecture (AICARA)**

## G.1 Cyber-defense Agents Must Make Smart Decisions

Agents' decision making will a key to their trustworthiness. But decision making is still today at a very early stage of development (Heinl 2014). Machine learning (ML) and reinforcement learning are regularly advocated as a pathway to the future (e.g., in Ridley [2018]), but reduce decision making to a problem-solving issue. In future highly tactical and fast-paced cyber battles, smart decisions will not be those made in the heat of a single reaction to a single state of the defended system. The adversary plans many moves and reactions to our anticipated response to each of his moves. In a tactical cyber battle, a smart decision will be the one that wins the battle, not one that counters temporarily an adversary malware but risks triggering a fatal enemy retaliation later on.

## G.2 What Makes Decisions Smart

Human decision making is smart because it builds on vigilance, vision, knowledge, experience, anticipation, wisdom, self-monitoring, deliberation, emotion, and plasticity. Instance based learning theory (IBLT) shows that five mechanisms are at play in dynamic decision making (Gonzalez et al. 2003): instance-based knowledge, recognition-based retrieval, adaptive strategies, necessity-based choice, and feedback updates. Blakely and Théron (2018) and LeBlanc et al. (2017) show that, for agents, making the right decision requires the integration of a variety of approaches. Decision making in action (DMA; Théron [2014]) suggests that the decision-making process' plasticity is an adaptive response to circumstances' characteristics and uncertainty.

## G.3 The Plasticity of Decision-Making in Action

Théron (2014) showed that the cognitive process underlying individual human DMA has plasticity to adapt to the circumstances handled by the subject in real-time episodes of action.

The following two diagrams (Théron 2014) illustrate how, during a 5-s traumatic "moment" in a fire fighter's—Lieutenant A—episode of lived experience, two successive, fast-paced, cycles of decision making involve differently shaped cognitive processes to fight the circumstances at hand (Figs. G-2 and G-3).

What these two diagrams show is that in order to escape his fate (Lieutenant A is caught in the middle of gun shots by three police officers trying to kill two Rottweiler dogs that attack), the subject struggles at a very fast pace to find margins of maneuver and ways of controlling the course of events.

Lieutenant A executes two successive cycles of decision making in a row, this representing a time span of about 5 s.

When after the first round of decision making, Lieutenant A realizes that the margins of safety and maneuver shrink, the second cycle of decision making shows Lieutenant A struggles far more, resorting on more cognitive resources to sort the situation out.

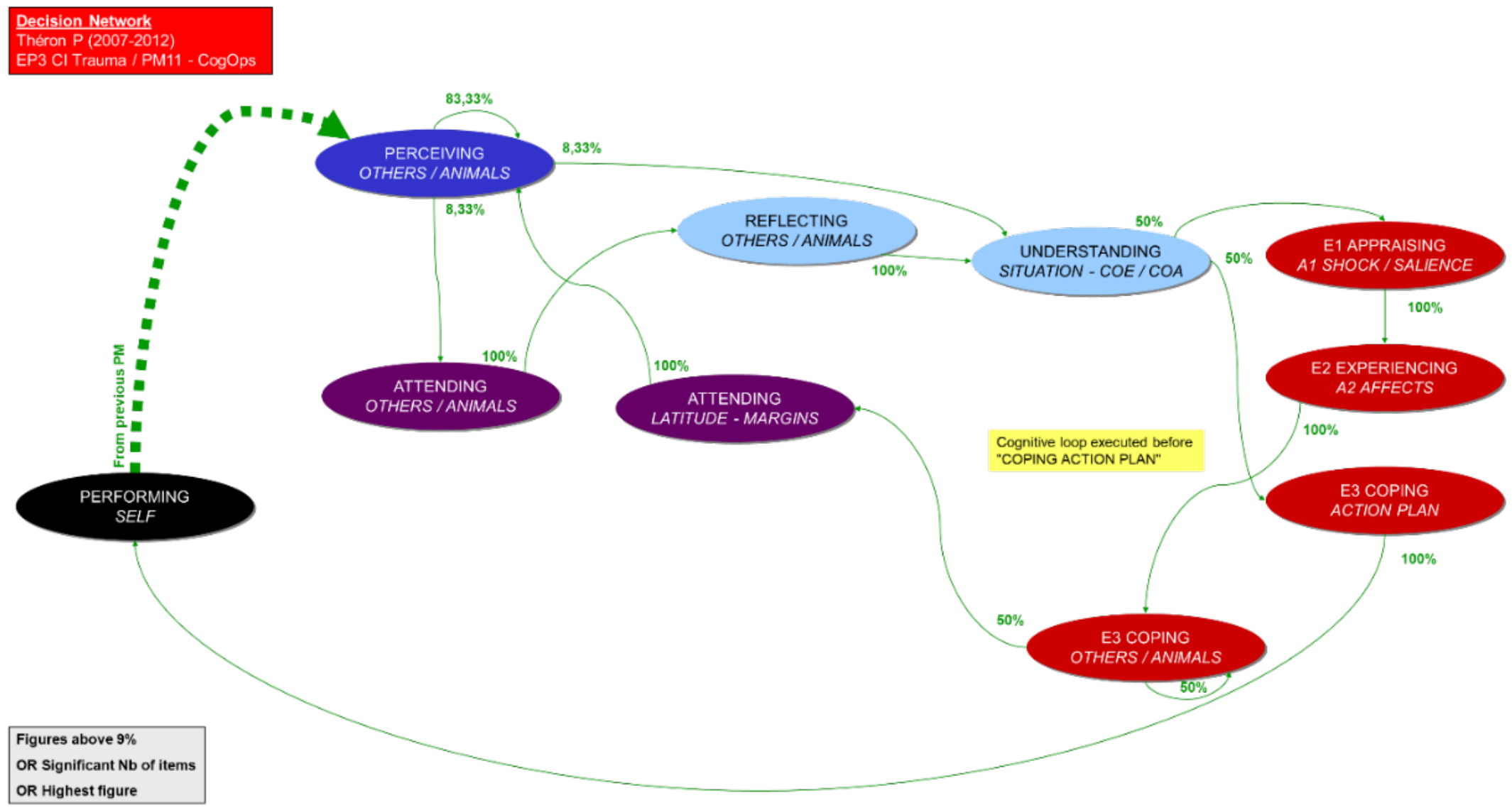

**Fig. G-2 Decision cycle in Lieutenant A's episode of experience**

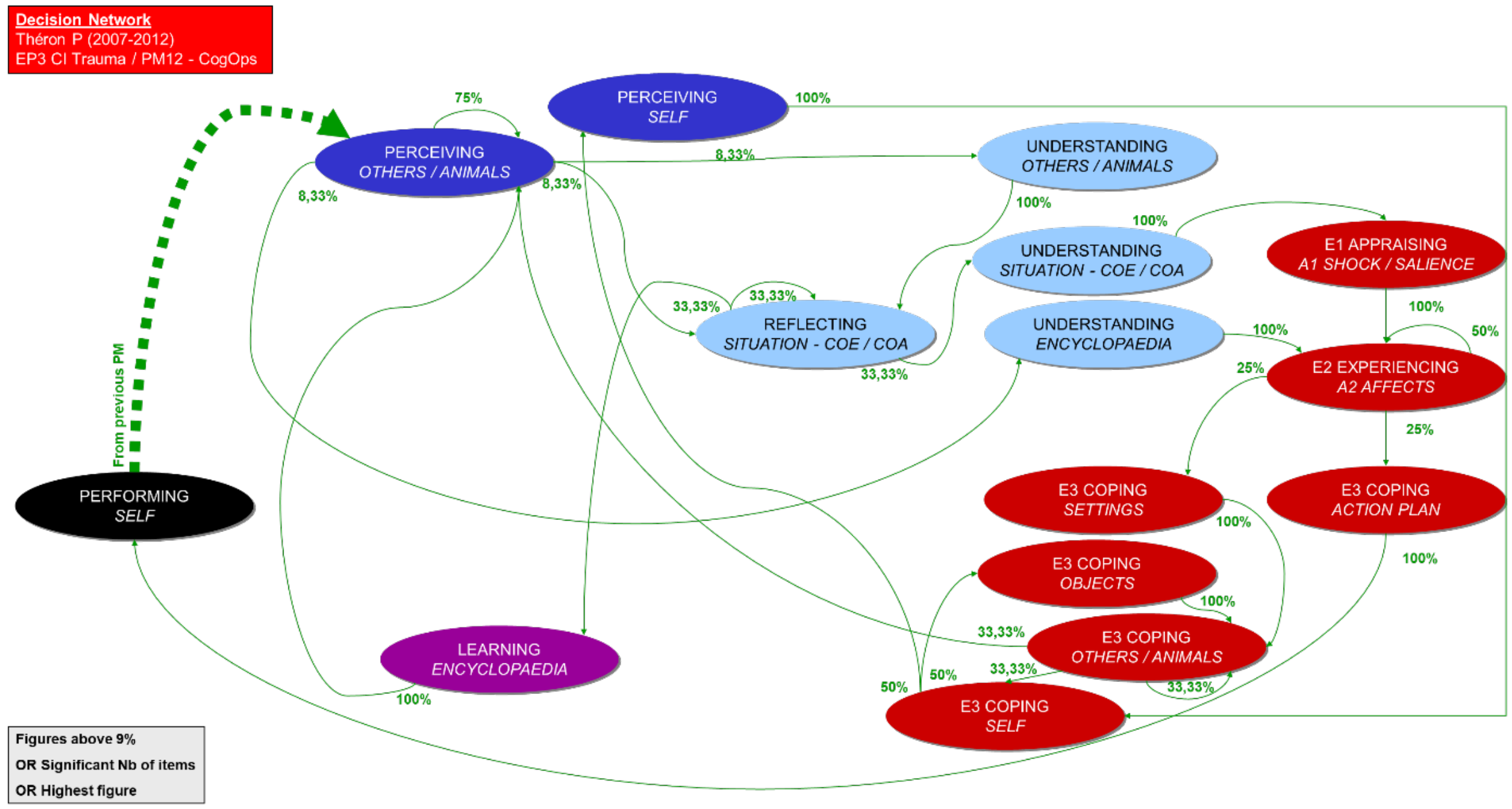

**Fig. G-3 Another decision cycle in Lieutenant A's episode of experience**

When uncertainty appears at a given stage in the DMA cognitive process, the cognitive function that identifies uncertainty calls upon other functions to resolve doubts. For instance, the “analysis” function, if it cannot make sense of the perceived situation, may call upon “long-term memory” to find in autobiographical knowledge significant world patterns, or it can refocus the “sensing” function on specific objects to acquire more data about the situation.

## G.4 DMA is applicable to AICAs

We posit that AICAs’ decision-making function will rely on a model of DMA similar to the one described in Théron (2014) and that we assume to be structured as follows (Fig. G-4):

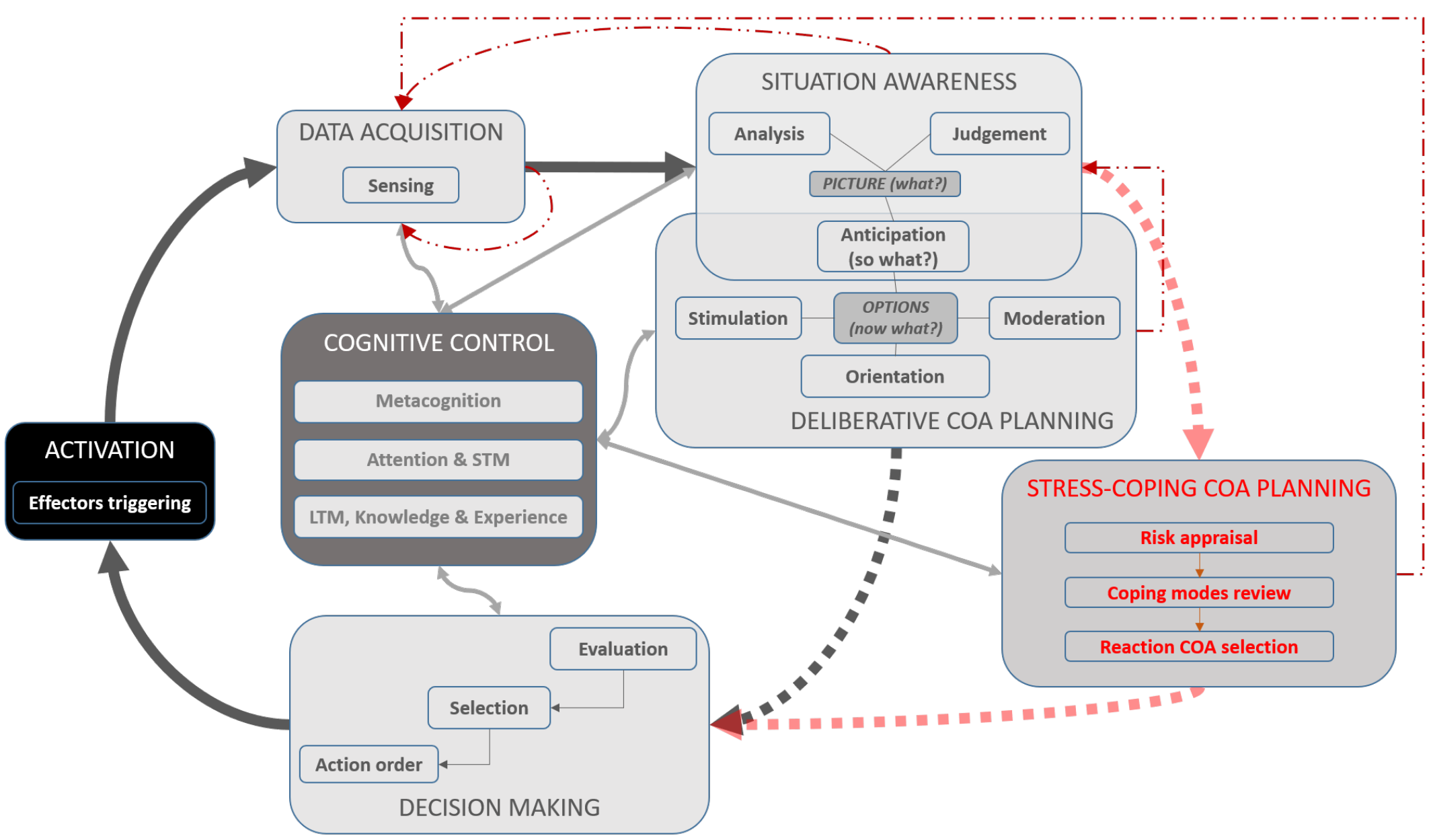

**Fig. G-4  Model of AICAs’ DMA**

- The *Sensing* cognitive component of the AICA will acquire data from its environment, from itself and form other friendly agents.
- This data set will be processed by the *World State Identification* cognitive component of the agent to provide the agent with a picture of the situation at hand.
- The agent’s *Planning* cognitive component will elaborate and the *Action Selection* cognitive component will select action options, or, alternatively in case of a major “stress”, these components of the agent will resort on a repertory of appropriate “reflex” courses of action.

- The selected course of action will be passed on to the *Action Execution* cognitive component of the agent, and the latter will activate this set of actions in its environment or within itself, or address them to other friendly agents.

Agents' DMA can be up to the challenges of fast-paced intelligent cyber battles if it is organized in a human-like plastic cognitive process.

## G.5 Deep Decision Making (DDM) in Agents

Deep decision making (DDM) is the computational model of DMA in AICAs.

In the DDM model, the agent's decision-making process will be plastic, meaning that it will activate the agent's cognitive components in a plastic variety of ways and combinations that will be dictated by circumstances at hand.

Each DDM cognitive function will itself rely upon a plastic combination of non-artificial intelligence (AI) and AI/ML techniques (e.g., genetic algorithms and classifiers or neural networks). It will be founded upon a set of models, memories, techniques, and tactics.

The specific research challenge is here to know why, when and how AICAs' cognitive components will need to trigger and communicate with one another.

This research should explore the various currents of work conducted in recent years such as cognitive architectures (Lebiere and Anderson 1993) and their use for computer games (Smart et al. 2016), naturalistic decision making (Lipshitz 1997), and DMA (Théron 2014) for they characterize the micro-cognitive processes of expert decision making, instance-based learning theory for DDM (Gonzalez et al. 2003), Agent-based modeling and simulation of cyber battles (Kotenko et al. 2012), and cyberattack graphs and models (Jajodia and Noel 2010; Noel et al. 2015) as they seek to provide models of adversaries, along with game theory, AI and ML and its current refinements.

# Appendix H. Annotated References from Game Theory Literature

Author: Martin Drašar

Game Theory primarily supports the *Planning* and *Action Selection* modules, and secondarily, the *Collaboration and Negotiation* module.

Game theory is a framework that enables us to infer future actions of agents with strategic decision-making skills. Thus, in AICA Reference Architecture (AICARA), it may form a basis for a more formal model for an advisor. Additionally, to construct a game theoretic model, one needs to reduce the complexity/ dimensionality of the issue being modeled; such a reduction helps to find the crucial elements of a problem—or study the orthogonal aspects in separation.

However, game-theoretic methods might be difficult to apply in cybersecurity. First, it is unclear how to realistically model an adversary (what are the adversary's intended actions, or their utility). Second, in military planning, game theory is seldom used, and cybersecurity might be probably even more complex than military planning.

To explore the domain, we strongly recommend these two basic textbooks on game theory:

- Osborne MJ. An introduction to game theory. Oxford University Press, 2004.
- Nisan N et al., eds. Algorithmic game theory. Cambridge University Press, 2007.

There exist some relevant game-theoretic models used to investigate security-related games. We comment on two of them:

- Korzhyk D, Yin Z, Kiekintveld C, Conitzer V, Tambe M. Stackelberg vs. Nash in security games: an extended investigation of interchangeability, equivalence, and uniqueness. J Art Int Res. 2011 May;41(2):297–327.

  In *Stackelberg's Security Game*, there are two players, the defender and the attacker. The defender has to defend some infrastructure (isolated nodes, a graph). The defender chooses a defense strategy which is the amount of effort to defend each element of the infrastructure. The attacker observes the strategy (e.g., sees how many police officers are deployed on each airport) and picks an attack strategy (what elements to attack and with which force). The utility of the defender is the value of defended infrastructure minus cost of defense. The utility of the attacker is the value of hijacked infrastructure minus the cost of the attack.

*Relevance*: we may represent the defended system as infrastructure for the Stackelberg's security game.

- Van Dijk M, Juels A, Oprea A, Rivest RL. FlipIt: the game of "stealthy takeover". J Crypto. 2013;26(4):655–713.

  In FlipIt, there are two players (the defender and the attacker) and one node (a single piece of infrastructure). The players fight for the control over the node over a certain time horizon; players can make repeated actions during that time. The attacker's action is to hijack the node. The defender's action is to reinitialize the node and thus regain its control over it (until the end of the game or the next move of the attacker). Players can't observe the state of the node without making an action. The utility of a player is the total time they control the node minus the total cost of the actions. The strategy is when to execute the actions.

  **Relevance:** FlipIt represents a system with an exploit (that cannot be permanently fixed) and a stealthy malware; a defender move represents, for example, a trusted reinitialization.

Following references are sample applications of game-theoretic reasoning to military and cybersecurity planning:

- Chatterjee S, Halappanavar M, Tipireddy R, Oster M. Game theory and uncertainty quantification for cyber defense applications. SIAM News. 2016;49(6).

  The defended cyber infrastructure is modeled by layers. Each pair of attacker–defender actions is associated with probability of penetrating each layer. Each action has a cost. The utility is the expected cost of penetrating all layers (or benefit, for the attacker); minus the cost of the action taken. The paper validates the model by a cyber-wargaming scenario involving people.

- Colbert EJ, Kott A, Knachel LP. The game-theoretic model and experimental investigation of cyber wargaming. The Journal of Defense Modeling and Simulation. 2018.

  Game-theoretic modeling commonly assumes that both sides know each other's utilities and the repertoire of actions. The paper shows how to cope with unknown attacker's strategies using reinforcement learning algorithm. The algorithm is used to tune transition probabilities between possible actions.

We indicate another interesting paper about adaptive cyber defense:

- Zhu M, Hu Z, Liu P. Reinforcement learning algorithms for adaptive cyber defense against Heartbleed. Proceedings of the First ACM Workshop on Moving Target Defense. 2014:51–58.

One crucial point to obtain autonomy for cyber-defense's agent is the learning process. The following reference deals with reinforcement learning:

- Beaudoin L, Japkowicz N, Matwin S. Autonomic computer network defence using risk state and reinforcement learning. Cryptology and Information Security Series. 2009;3:238–248.

  This paper shows an application of reinforcement learning to adapt computer network defense in order to minimize the risk (a product of the infrastructure value and the probability of a successful attack). The paper uses a simulator to learn an optimal defending strategy. The simulator assumes a queuing-theory like distribution of incoming vulnerabilities. The defended network is represented by a graph; for each component, the possible actions of the defender are fixing, patching, isolating and waiting.

  Reinforcement learning is a domain in which new important results occur any year. Research activities about neural networks always places a little further the border of what is possible to do. We choose the following reference which presents what we imagine to combine *Planning* and *Action Selection* modules in a single neural network:

- Mnih V et al. Playing Atari with deep reinforcement learning. 2013. arXiv:1312.5602 [cs].

  Cyber defense can be treated as a complex decision-making process in an environment that has a complex state that changes both stochastically and in response to the opponents' actions; and a payoff that might be delayed in time from the moment the action is taken. Such an environment is not unlike an arcade game. Reinforcement learning combined with deep learning was applied to successfully play such arcade games. This approach starts with a generic learning algorithm that is not adapted for a specific game; instead, it takes a sequence of images as an input; and joystick movements (left/right/top/bottom) as possible actions. After learning, the algorithm achieved higher scores than expert human players. It may seem that a black box consisting of a reinforcement learning algorithm could replace the combined *Planning* and *Action Selection* modules (taking along also the problem of representing the world state). However, cyber defense is more complex than an arcade game. First, there is no simulator to train on. Second, there are more possible actions. Third, the opponent is stronger than a 1980s “AI” of an arcade game. Without expertise in reinforcement learning, it is thus unclear how (and whether at all) their impressive results translate to cyber defense.

## List of Symbols, Abbreviations, and Acronyms

| | |
|---|---|
| AICA | Autonomous Intelligent Cyber-defense Agent |
| AICARA | Autonomous Intelligent Cyber-defense Agent Reference Architecture |
| AMQP | Advanced Message Queuing Protocol |
| APT | advanced persistent threat |
| ARL | Army Research Laboratory |
| BMS | battle management system |
| BUS | system bus |
| C2 | command and control |
| C4ISR | command, control, communications, computers, intelligence, surveillance, and reconnaissance |
| CAPEC | Common Attack Pattern Enumeration and Classification |
| COAP | Constrained Application Protocol |
| COMMS | communication system |
| CONOPS | Concept of Operations |
| CPS | cyber–physical system |
| CPU | central processing unit |
| CS | control systems |
| CVE | Common Vulnerability and Exposure |
| DTLS | Datagram Transport Layer Security |
| EW | electronic warfare |
| $f_a$ | set of possible plans of actions |
| $f_w$ | set of feasible actions |
| HTTP | Hypertext Transfer Protocol |
| InterCOM | internal communication system |
| IoC | indicator of compromise |
| IRM | incident response mechanisms |
| ISR | intelligence, surveillance, and reconnaissance |
| IT | information technology |
| KB | knowledge base |
| MOTS | military off-the-shelf |
| MISP | Malware Information Sharing Platform |
| MQTT | Message Queuing Telemetry Transport |
| MS | Mission Specific System |
| NATO | North Atlantic Treaty Organization |
| OPT | optoelectronic system |
| OS | operating systemP2P peer-to-peer |
| PBS | packet-based switching |
| POMDP | Partially Observable Markov Decision Process |
| RCA | root cause analysis |
| RoT | Root-of-Trust |
| RPTS | Requested Power To Send |
| RTG | Research Task Group |
| S | sensing |
| SCADA | supervisory control and data acquisition |
| SCD | service and capacity discovery |
| SHA-1 | Secure Hash Algorithm 1 |
| SNMP | Simple Network Management Protocol |
| SOAP | Simple Object Access Protocol |
| SW | switch |
| TAP | test access point |
| TCB | Trusted Computing Base |
| TCP | Transmission Control Protocol |
| TLS | Transport Layer Security |
| TTPs | tactics, techniques, and procedures |
| UAV | unmanned aerial vehicle |
| VMS | vehicle management system |
| VNS | vehicle navigation system |
| WS | weapon system |
| WSI | World State Identification |
| 1 (PDF) | DEFENSE TECHNICAL<br>INFORMATION CTR<br>DTIC OCA |
| 1 (PDF) | CCDC ARL<br>FCDD RLD CL<br>TECH LIB |
| 1 (PDF) | GOVT PRINTG OFC<br>A MALHOTRA |
| 1 (PDF) | CCDC ARL<br>FCDD RLD<br>A KOTT |